\documentclass[reprint,showpacs,twocolumn,superscriptaddress,fleqn,floatfix]{revtex4-1}
\pdfoutput=1
\usepackage{graphics}
\usepackage{graphicx}
\usepackage{amsmath,amssymb,amsthm,amsfonts,epsfig}
\usepackage{gensymb}
\usepackage{bm}
\usepackage{natbib}
\usepackage[version=3]{mhchem}
\usepackage{hyperref}
\usepackage{lipsum}
\usepackage{color}
\usepackage{gensymb}
\usepackage{longtable}
\hypersetup{%
  breaklinks=true,%
  colorlinks=true,%
  citecolor=blue,%
  filecolor=blue,%
  linkcolor=blue,%
  urlcolor=blue%
}
\newcommand{\figref}[1]{Fig.~\ref{#1}}
\newcommand{\tabref}[1]{Table~\ref{#1}}
\newcommand{\cmark}{\checkmark}
\newcommand{\xmark}{$\times$}
\begin{document}
\title{Computationally-driven, high throughput identification of CaTe and \ce{Li3Sb} as promising candidates for high mobility \emph{p}-type transparent conducting materials}
\author{Viet-Anh Ha}
\affiliation{Institute of Condensed Matter and Nanoscience (IMCN), Universit\'{e} catholique de Louvain (UCLouvain), Chemin \'{e}toiles 8, bte L7.03.01, Louvain-la-Neuve 1348, Belgium}

\author{Guodong Yu}
\email[Present address: School of Physics and Technology (SPT), Wuhan University (WHU), Wuhan 430072, China]{}
\affiliation{Institute of Condensed Matter and Nanoscience (IMCN), Universit\'{e} catholique de Louvain (UCLouvain), Chemin \'{e}toiles 8, bte L7.03.01, Louvain-la-Neuve 1348, Belgium}

\author{Francesco Ricci}
\affiliation{Institute of Condensed Matter and Nanoscience (IMCN), Universit\'{e} catholique de Louvain (UCLouvain), Chemin \'{e}toiles 8, bte L7.03.01, Louvain-la-Neuve 1348, Belgium}

\author{Diana Dahliah}
\affiliation{Institute of Condensed Matter and Nanoscience (IMCN), Universit\'{e} catholique de Louvain (UCLouvain), Chemin \'{e}toiles 8, bte L7.03.01, Louvain-la-Neuve 1348, Belgium}

\author{Michiel J. van Setten}
\email[Present address: IMEC, 75 Kapeldreef, B-3001 Leuven, Belgium]{}
\affiliation{Institute of Condensed Matter and Nanoscience (IMCN), Universit\'{e} catholique de Louvain (UCLouvain), Chemin \'{e}toiles 8, bte L7.03.01, Louvain-la-Neuve 1348, Belgium}

\author{Matteo Giantomassi}
\affiliation{Institute of Condensed Matter and Nanoscience (IMCN), Universit\'{e} catholique de Louvain (UCLouvain), Chemin \'{e}toiles 8, bte L7.03.01, Louvain-la-Neuve 1348, Belgium}

\author{Gian-Marco Rignanese}
\affiliation{Institute of Condensed Matter and Nanoscience (IMCN), Universit\'{e} catholique de Louvain (UCLouvain), Chemin \'{e}toiles 8, bte L7.03.01, Louvain-la-Neuve 1348, Belgium}

\author{Geoffroy Hautier}
\email[E-mail: ]{geoffroy.hautier@uclouvain.be}
\affiliation{Institute of Condensed Matter and Nanoscience (IMCN), Universit\'{e} catholique de Louvain (UCLouvain), Chemin \'{e}toiles 8, bte L7.03.01, Louvain-la-Neuve 1348, Belgium}

\date{\today}
\begin{abstract}
High-performance \emph{p}-type transparent conducting materials (TCMs) must exhibit a rare combination of properties including high mobility, transparency and \emph{p}-type dopability. The development of high-mobility/conductivity \emph{p}-type TCMs is necessary for many applications such as solar cells, or transparent electronic devices. Oxides have been traditionally considered as the most promising chemical space to dig out novel \emph{p}-type TCMs. However, non-oxides might perform better than traditional \emph{p}-type TCMs (oxides) in terms of mobility. We report on a high-throughput (HT) computational search for non-oxide \emph{p}-type TCMs from a large dataset of more than 30,000 compounds which identified CaTe and \ce{Li3Sb} as very good candidates for high-mobility \emph{p}-type TCMs. From our calculations, both compounds are expected to be \emph{p}-type dopable: intrinsically for \ce{Li3Sb} while CaTe would require extrinsic doping. Using electron-phonon  computations, we estimate hole mobilities at room-temperature to be about 20 and 70~cm$^2$/Vs for CaTe and \ce{Li3Sb}, respectively. The computed hole mobility for \ce{Li3Sb} is quite exceptional and comparable with the electron mobility in the best \emph{n}-type TCMs.                         
\end{abstract}
\maketitle

\section{Introduction}
Transparent conducting materials (TCMs) are necessary in many applications ranging from solar cells to transparent electronics. So far, \emph{n}-type oxides (e.g., \ce{In2O3}, \ce{SnO2} and \ce{ZnO}) are the highest performing TCMs, allowing them to be used in commercial devices~\cite{H.Ohta-MatTo04, A.Facchetti10, K.Ellmer-NatPho12, P.Barquinha12, E.Fortunato12}. On the other hand, \emph{p}-type TCMs show poorer performances, especially in terms of carrier mobility. This hinders the development of new technologies such as transparent solar cells or transistors~\cite{K.Ellmer-NatPho12, S.C.Dixon-JMChC16}. Taking advantage of the predictive power of density functional theory (DFT) calculations, we have set up a high-throughput (HT) computational framework to identify novel \emph{p}-type TCMs focusing first on oxide compounds~\cite{G.Hautier-NatCom13, J.B.Varley-PRB14, A.Bhatia-CheMat16}.

The analysis of the calculated HT data confirmed that on average \emph{p}-type oxides have inherently higher effective masses than \emph{n}-type oxides~\cite{G.Hautier-NatCom13}. This could be traced back to the strong oxygen \emph{p}-orbital character in the valence band of most oxides and has rationalized the current gap in mobility between the best \emph{p}-type and \emph{n}-type oxides. This inherent difficulty in developing high-hole-mobility oxides justifies moving towards non-oxide TCM chemistries including fluorides~\cite{H.Yanagi-APL03}, sulfides~\cite{S.Park-APL02,R.W.Robinson-AdvEMat16}, oxianions~\cite{K.Ueda-APL00}, or germanides~\cite{F.Yan-NatCom15}. Recently, we started extending our HT computing approach to search for non-oxide TCMs. Phosphides were identified to be among the lowest hole effective mass materials and more specifically boron phosphide (BP) was detected as a very promising \emph{p}-type TCM candidate~\cite{J.B.Varley-CheMat17}. We note that subsequent computational studies focusing on selected binaries and ternaries reported also on the computational screening of non-oxide TCMs~\cite{R.K.M.Raghupathy-JMCC17, K.M.Raghupathy2018}. In the present work, we extend our HT computing approach to a larger space of chemistries and investigate some selected candidates. We screen all non-oxide compounds in a large computational data set ($>$34,000 semiconductors)~\cite{F.Ricci-SciData17}. Combining DFT-based HT computations with higher accuracy methods such as $GW$, hybrid functionals and electron-phonon coupling computations (to assess the relaxation time and thus the mobility), we identify that CaTe and \ce{Li3Sb} would be of great interest as high mobility \emph{p}-type TCMs.

\section{Methods} \label{methods}

All the considered materials originate from the Inorganic Crystal Structure Database (ICSD)~\cite{ICSD13}. Their relaxed crystal structures and electronic band structures were obtained from the Materials Project database~\cite{A.Jain-APLM13, MatPro13}. These rely on DFT high-throughput computations which were performed with VASP~\cite{G.Kresse-CMS96,G.Kresse-PRB96} using the Perdew-Burke-Ernzerhof (PBE) exchange-correlation (XC) functional~\cite{J.Perdew-PRL96} within the projector augmented wave (PAW) framework~\cite{P.E.Blochl-PRB94}.

One of the first selection criteria for TCMs is their stability. Here, it is assessed by the energy above hull $E_\textrm{hull}$ in the Materials Project database~\cite{A.Jain-APLM13}. For a compound stable at 0K, $E_\mathrm{hull}=0$~meV/atom, and the stability decreases as $E_\textrm{hull}$ increases.

In the beginning of the screening procedure, the PBE band gap can be used as a filter. However, since PBE is known to systematically underestimate the band gap compared to experiments, more accurate calculations are needed in the subsequent steps (though with a limited number of materials). 
So, the fundamental and direct band gaps were also calculated with VASP for about a hundred materials using the Heyd-Scuseria-Ernzerhof (HSE) hybrid XC functional~\cite{J.Heyd03, E.N.Brothers08} and adopting the same computational parameters as for the PBE calculations.
For the final candidates (\ce{CaTe} and \ce{Li3Sb}), $G_0W_0$ calculations were performed with ABINIT~\cite{X.Gonze-Comput.Mater.Sci02,X.Gonze-Z.Kristallogr05,X.Gonze-Comput.Phys.Commun09,X.Gonze-Comput.Phys.Commun16}. In these calculations, optimized norm-conserving (NC) pseudopotentials including semi-core electrons were used which were generated with ONCVPSP~\cite{D.R.Hamann-PRB13, pseudodojo}. The kinetic cut-off energy for the wavefunctions were set to 51 and 52~Ha for \ce{CaTe} and \ce{Li3Sb} respectively, as recommended in the PseudoDojo table~\cite{pseudodojo}. The convergence of these calculations with respect to the kinetic energy cut-off $E_c$ for the dielectric function and the number of bands $N_b$ was tested using automatic $GW$ workflows~\cite{M.J.vanSetten-PRB17} based on the pymatgen~\cite{S.P.Ong-CMS13} and AbiPy packages~\cite{Abipy14,X.Gonze-Comput.Phys.Commun16}. For \ce{CaTe}, the convergence of the gap at the $\Gamma$ point (with a truncation error smaller than 0.01~eV) was obtained for $E_c=12$~Ha and $N_b=480$. In the case of \ce{Li3Sb}, the convergence is significantly faster: using $E_c=10$~Ha and $N_b=240$ guarantee a truncation error smaller than 0.01~eV. More details about the convergence tests are available in the supplementary document. For the calculations of the screening and the quasi-particle self-energy, $10\times10\times10$ and $8\times8\times8$ \textbf{k}-point meshes were used for \ce{CaTe} and \ce{Li3Sb}, respectively. The band structures are then interpolated from these \textbf{k}-point meshes using AbiPy~\cite{Abipy14,X.Gonze-Comput.Phys.Commun16}.

The point defect computations were performed using the supercell technique~\cite{C.Freysoldt-RevMP14} adopting $3\times3\times3$ supercells of the primitive cells. We calculated the defect formation energies first using the PBE XC functional but also with the more accurate HSE functional for \ce{Li3Sb} and \ce{CaTe}~\cite{J.Heyd03, E.N.Brothers08}. For the latter, the screening length and fraction of exact exchange were set to the common values of 0.2~\AA\ and 25~\% respectively. The kinetic energy cut-off for the wavefunctions was set to 19.1~Ha (520~eV) and the relaxations are stopped when the change in total energy between two ionic relaxation-steps is smaller than $3.67\times10^{-4}$~Ha (0.01~eV). The formation energy of defect $D$ in charged state $q$ can be written as~\cite{H.-P.Komsa-RRB12,S.B.Zhang-PRL91}
\begin{equation}
\begin{split}
E_f[D^q] = & E[D^q] + E_{corr}[D^q] - E[bulk] - \Sigma_{i}n_i\mu_i \\
& + q(\epsilon_\mathrm{VBM} + \Delta v + \Delta \epsilon_F)
\end{split}
\label{dfenergy}
\end{equation}
where $E[D^q]$ and $E[bulk]$ are the total energies of the supercell with a defect $D$ in the charge state $q$ and without any defects, respectively; $n_i$ is the number of atoms of type $i$ removed ($n_i<0$) or added ($n_i>0$); and, $\mu_i$ is the corresponding chemical potential. $\epsilon_\mathrm{VBM}$ is the energy of the valence band maximum (VBM), and $\Delta \epsilon_F$ is the Fermi level referenced to $\epsilon_\mathrm{VBM}$. The correction terms $E_{corr}[D^q]$ and $\Delta v$ are introduced to take care of the spurious image-charge interactions and the potential alignment for charged defects, respectively. The defect states with the charge $q$ were corrected using the extended Freysoldt's (Kumagai's) scheme~\cite{C.Freysoldt-PSSB11,Y.Kumagai-PRB14}. All defects computations were performed using the PyCDT package~\cite{D.Broberg-CPCom18}.

The effective masses were calculated with BoltzTrap (based on Boltzmann transport theory framework)~\cite{G.K.H.Madsen-CPC06} using the pymatgen~\cite{S.P.Ong-CMS13} interface and the Fireworks workflow package~\cite{A.Jain-CCPE15}. All the raw effective mass data is freely available in a separate paper which covers around 48,000 inorganic materials ~\cite{F.Ricci-SciData17}.
The mobility depends on the effective mass $m^{*}$ through $\mu = e\tau/m^{*}$ where  the relaxation time $\tau$ (inverse of the scattering rate) depends on different scattering mechanisms. Carriers can be scattered by phonons, ionized and neutral impurities, grain boundaries,... In this work, we only took into account the scattering of electrons by phonons which is likely to be an important component of scattering and is an intrinsic mechanism, difficult to control through purity and microstructure. The carriers scattering by phonons can be computed theoretically if the electron-phonon matrix elements are known. In principle, one can employ Density Functional Perturbation Theory (DFPT) to obtain all the electron-phonon matrix elements from first principles. However, converging the relevant physical properties (such as the scattering rate of electrons by phonons) often requires very dense \textbf{k}-point and \textbf{q}-point meshes for electrons and phonons respectively leading to a considerable increase of computational time.
The recently developed interpolation techniques based on Wannier functions offer a very practical and efficient solution to overcome this obstacle.
In this work, we used the EPW code~\cite{J.Noffsinger-CPC10,S.Ponce-CPC16} interfaced with Quantum ESPRESSO~\cite{P.Giannozzi-J.Phys.CondMat09,P.Giannozzi-J.Phys.CondMat17} to calculate the relaxation-time $\tau_{n\textbf{k}}$ ($n$ and \textbf{k} are band index and wave vector of a Bloch's state, respectively). More details on the theory and the implementation can be found in Ref.~\onlinecite{S.Ponce-CPC16}.
The $\tau_{n,\textbf{k}}$ were interpolated on a dense $40\times40\times40$ mesh for both \textbf{k}-points (for electrons) and \textbf{q}-points (for phonons) starting from the DFPT values on a $6\times6\times6$ mesh.
The latter (together with the structural relaxation, self-consistent, non self-consistent calculations which are needed to run EPW) were obtained using Quantum ESPRESSO with NC pseudopotentials and very stringent parameters for convergence, e.g. high cut-off energy of 40~Ha.
These $\tau_{n,\textbf{k}}$ are then used as an input to compute the carrier mobility by solving the Boltzmann transport equation by means of the BoltzTrap package~\cite{G.K.H.Madsen-CPC06}.
In the latter calculations, the DFT band-energies (computed on a finite number of \textbf{k}-points) are interpolated using star functions (see section 2 of Ref.~\onlinecite{G.K.H.Madsen-CPC06}). Here, we have implemented another interpolation for the relaxation time in BoltzTrap in order to obtain the same very dense \textbf{k}-point grid as the one used for band-energies. The physical principle for this implementation is that the symmetries of the quasi-particle energies are the same as those of band-energies~\cite{M.Giantomassi09} ($\tau_{n,\textbf{k}}$ due to the interaction with phonons can be calculated from the imaginary part of the electron-phonon self-energy).

\section{Results}

Starting from the Materials Project database, our first step was to extract materials with a low hole effective mass ($< 1~m_o$, where $m_o$ is the free electron mass) and a large enough fundamental gap ($>0.5$~eV) and direct gap ($>1.5$~eV), based on PBE calculations. Regarding the effective masses, in the most general form, they are represented by a tensor. As most TCMs are used as polycrystalline films, materials with isotropic or close to isotropic transport are easier to use in practical applications. Therefore, for the screening, we focus on the three principal values of this tensor and sort the materials based on the highest of the three principal hole effective masses. There were about 390 compounds passing through this first filter.

We then screened out very unstable materials selecting only those with an energy above hull lower than 24~meV/atoms. This threshold corresponds to the typical standard deviation of computational errors (compared with experiment) of DFT formation-energies~\cite{G.Hautier2012}. For the 107 materials passing these criteria, more accurate fundamental and direct gaps were calculated using the HSE hybrid functional. All the results of this step are presented in Table SI of the Supplemental Material~\cite{supplem}. For sake of clarity, \tabref{tableI} shows a selection of 63 materials with a direct band gap $\geq 2.8$~eV. The materials are sorted in decreasing order as a function of the computed direct band gap.

\begin{center}
\LTcapwidth=0.98\textwidth
\begin{longtable*}{@{\extracolsep{\fill}} l l r c c r c c c c l}
\caption{Formula, space group, Materials Project identification number (MP-id)~\cite{A.Jain-APLM13, MatPro13}, fundamental $E_g$ and direct gaps $E_g^d$ computed by HSE functional (in eV), energy above hull $E_\mathrm{hull}$ (in $meV$/atom), principal components $m_1$, $m_2$ and $m_3$ of the hole effective mass tensor (in atomic units), verification of the absence of toxic/rare-earth (T/RE) elements (Be, As, Cd, Yb, Hg, Pb and Th) and of the \emph{p}-type dopability (when computed here or obtained from the existing literature) for the selected compounds (see text). The materials are sorted as a function of the direct band gap in decreasing order.
} \label{tableI} \\
\hline
\multicolumn{1}{l}{Formula}    & \multicolumn{1}{l}{Space group}  & \multicolumn{1}{r}{MP-id} & \multicolumn{1}{c}{$E_g^d$}        & \multicolumn{1}{c}{$E_g$} & \multicolumn{1}{r}{$E_\mathrm{hull}$} & \multicolumn{1}{c}{$m_1$}      & \multicolumn{1}{c}{$m_2$}        & \multicolumn{1}{c}{$m_3$} & \multicolumn{1}{c}{T/RE}       & \multicolumn{1}{c}{\emph{p}-dopability}\\
\hline 
\endfirsthead

\multicolumn{11}{c}%
{{\tablename\ \thetable{} -- continued from previous page}} \\
\hline
\multicolumn{1}{l}{Formula}    & \multicolumn{1}{l}{Space group}  & \multicolumn{1}{r}{MP-id} & \multicolumn{1}{c}{$E_g^d$}        & \multicolumn{1}{c}{$E_g$} & \multicolumn{1}{r}{$E_\mathrm{hull}$} & \multicolumn{1}{c}{$m_1$}      & \multicolumn{1}{c}{$m_2$}        & \multicolumn{1}{c}{$m_3$} & \multicolumn{1}{c}{T/RE}       & \multicolumn{1}{c}{\emph{p}-dopability}\\
\hline
\endhead

\hline \multicolumn{11}{r}{{Continued on next page}} \\ \hline
\endfoot

\hline \hline
\endlastfoot
\ce{BeS}       & $F\overline{4}3m$ & 422    & 6.89 & 4.05 &  0.0 & 0.65 & 0.65 & 0.65 & \xmark & -\\
\ce{KMgH3}     & $Pm\overline{3}m$ & 23737  & 5.76 & 3.58 &  0.0 & 0.75 & 0.75 & 0.75 & \cmark & -\\
\ce{SiC}       & $F\overline{4}3m$ & 8062   & 5.75 & 2.25 &  0.7 & 0.58 & 0.58 & 0.58 & \cmark & \cmark~\cite{K.Furukawa-APL86, Y.Kondo-IEEE86, K.Shibahara-JJAP87, R.Weingartner-APL02}\\
\ce{CsPbCl3}   & $Amm2$            & 675524 & 5.69 & 5.69 &  0.0 & 0.30 & 0.32 & 0.33 & \xmark & -\\
\ce{BeSe}      & $F\overline{4}3m$ & 1541   & 5.27 & 3.36 &  0.0 & 0.55 & 0.55 & 0.55 & \xmark & -\\
\ce{BeCN2}     & $I\overline{4}2d$ & 15703  & 5.21 & 5.21 &  0.0 & 0.75 & 0.75 & 0.78 & \xmark & -\\
\ce{RbPbF3}    & $Cc$              & 674508 & 5.20 & 4.84 &  0.0 & 0.71 & 0.83 & 0.95 & \xmark & -\\
\ce{MgS}       & $Fm\overline{3}m$ & 1315   & 4.95 & 3.84 &  0.0 & 0.98 & 0.98 & 0.98 & \cmark & -\\
\ce{RbHgF3}    & $Pm\overline{3}m$ & 7482   & 4.90 & 2.11 &  0.0 & 0.93 & 0.93 & 0.93 & \xmark & -\\
\ce{AgCl}      & $Fm\overline{3}m$ & 22922  & 4.81 & 2.28 &  0.0 & 0.83 & 0.83 & 0.83 & \cmark & -\\
\ce{CsHgF3}    & $Pm\overline{3}m$ & 561947 & 4.59 & 2.20 &  0.0 & 0.89 & 0.89 & 0.89 & \xmark & -\\
\ce{Be2C}      & $Fm\overline{3}m$ & 1569   & 4.56 & 1.63 &  0.0 & 0.37 & 0.37 & 0.37 & \xmark & -\\
\ce{SrMgH4}    & $Cmc2_1$          & 643009 & 4.52 & 3.78 &  0.0 & 0.84 & 0.90 & 0.95 & \cmark & -\\
\ce{Li2Se}     & $Fm\overline{3}m$ & 2286   & 4.36 & 3.70 &  0.0 & 0.95 & 0.95 & 0.95 & \cmark & -\\
\ce{BP}        & $F\overline{4}3m$ & 1479   & 4.35 & 2.26 &  0.0 & 0.34 & 0.34 & 0.34 & \cmark & \cmark\cite{J.B.Varley-CheMat17}\\
\ce{CaS}       & $Fm\overline{3}m$ & 1672   & 4.28 & 3.34 &  0.0 & 0.88 & 0.88 & 0.88 & \cmark & -\\
\ce{LiCa4B3N6} & $Im\overline{3}m$ & 6799   & 4.25 & 3.38 &  0.0 & 0.86 & 0.86 & 0.86 & \cmark & -\\
\ce{BaSrI4}    & $R\overline{3}m$  & 754852 & 4.22 & 4.22 & 21.8 & 0.73 & 0.73 & 0.80 & \cmark & -\\
\ce{LiSr4B3N6} & $Im\overline{3}m$ & 9723   & 4.18 & 3.22 &  0.0 & 0.89 & 0.89 & 0.89 & \cmark & -\\
\ce{NaSr4B3N6} & $Im\overline{3}m$ & 10811  & 4.08 & 3.14 &  0.0 & 0.92 & 0.92 & 0.92 & \cmark & -\\
\ce{K2LiAlH6}  & $Fm\overline{3}m$ & 24411  & 4.04 & 3.70 &  9.1 & 0.65 & 0.65 & 0.65 & \cmark & -\\
\ce{BeTe}      & $F\overline{4}3m$ & 252    & 4.04 & 2.45 &  0.0 & 0.42 & 0.42 & 0.42 & \xmark & -\\
\ce{Ba3SrI8}   & $I4/mmm$          & 756235 & 4.02 & 4.02 &  7.5 & 0.70 & 0.81 & 0.81 & \cmark & -\\
\ce{CaSe}      & $Fm\overline{3}m$ & 1415   & 4.01 & 2.95 &  0.0 & 0.77 & 0.77 & 0.77 & \cmark & -\\
\ce{LiH}       & $Fm\overline{3}m$ & 23703  & 3.97 & 3.97 &  0.0 & 0.46 & 0.46 & 0.46 & \cmark & \xmark\\
\ce{AlP}       & $F\overline{4}3m$ & 1550   & 3.90 & 2.50 &  0.0 & 0.56 & 0.56 & 0.56 & \cmark & \xmark\\
\ce{YbS}       & $Fm\overline{3}m$ & 1820   & 3.76 & 2.96 &  0.0 & 0.76 & 0.76 & 0.76 & \xmark & -\\
\ce{Na2LiAlH6} & $Fm\overline{3}m$ & 644092 & 3.75 & 3.75 &  3.9 & 0.66 & 0.66 & 0.66 & \cmark & -\\
\ce{SrSe}      & $Fm\overline{3}m$ & 2758   & 3.68 & 3.03 &  0.0 & 0.83 & 0.83 & 0.83 & \cmark & -\\
\ce{BaLiH3}    & $Pm\overline{3}m$ & 23818  & 3.62 & 3.26 &  0.0 & 0.36 & 0.36 & 0.36 & \cmark & \xmark\\
\ce{CsPbF3}    & $Pm\overline{3}m$ & 5811   & 3.59 & 3.59 &  4.6 & 0.39 & 0.39 & 0.39 & \xmark & -\\
\ce{Cs3ZnH5}   & $I4/mcm$          & 643702 & 3.58 & 3.58 &  0.0 & 0.69 & 0.93 & 0.93 & \cmark & -\\
\ce{Al2CdS4}   & $Fd\overline{3}m$ & 9993   & 3.56 & 3.55 & 20.0 & 0.78 & 0.78 & 0.78 & \xmark & -\\
\ce{K2LiAlH6}  & $R\overline{3}m$  & 23774  & 3.52 & 3.52 &  0.0 & 0.68 & 0.84 & 0.84 & \cmark & -\\
\ce{BaMgH4}    & $Cmcm$            & 643718 & 3.51 & 3.26 &  4.8 & 0.48 & 0.55 & 0.70 & \cmark & -\\
\ce{CaTe}      & $Fm\overline{3}m$ & 1519   & 3.50 & 2.18 &  0.0 & 0.60 & 0.60 & 0.60 & \cmark & \cmark\\
\ce{Cs3MgH5}   & $P4/ncc$          & 23947  & 3.49 & 3.49 &  0.3 & 0.88 & 0.93 & 0.93 & \cmark & -\\
\ce{Cs3MgH5}   & $I4/mcm$          & 643895 & 3.49 & 3.49 &  0.0 & 0.83 & 0.94 & 0.94 & \cmark & -\\
\ce{YbSe}      & $Fm\overline{3}m$ & 286    & 3.48 & 2.43 &  0.0 & 0.67 & 0.67 & 0.67 & \xmark & -\\
\ce{ZnS}       & $F\overline{4}3m$ & 10695  & 3.46 & 3.46 &  0.0 & 0.81 & 0.81 & 0.81 & \cmark & \cmark\cite{R.W.Robinson-AdvEMat16}\\
\ce{TaCu3S4}   & $P\overline{4}3m$ & 10748  & 3.46 & 2.95 &  0.0 & 0.98 & 0.98 & 0.98 & \cmark & -\\
\ce{Al2ZnS4}   & $Fd\overline{3}m$ & 4842   & 3.46 & 3.43 &  0.0 & 0.66 & 0.66 & 0.66 & \cmark & \xmark\\
\ce{Li2ThN2}   & $P\overline{3}m1$ & 27487  & 3.46 & 3.33 &  0.0 & 0.85 & 0.95 & 0.95 & \xmark & -\\
\ce{Mg2B24C}   & $P\overline{4}n2$ & 568556 & 3.42 & 3.41 &  0.0 & 0.77 & 0.93 & 0.93 & \cmark & -\\
\ce{Li2GePbS4} & $I\overline{4}2m$ & 19896  & 3.33 & 3.20 &  0.0 & 0.61 & 0.61 & 0.98 & \xmark & -\\
\ce{Cs3H5Pd}   & $P4/mbm$          & 643006 & 3.32 & 3.09 &  0.0 & 0.79 & 0.83 & 0.83 & \cmark & -\\
\ce{SrTe}      & $Fm\overline{3}m$ & 1958   & 3.24 & 2.39 &  0.0 & 0.67 & 0.67 & 0.67 & \cmark & \xmark\\
\ce{MgTe}      & $F\overline{4}3m$ & 13033  & 3.24 & 3.24 &  0.9 & 0.95 & 0.95 & 0.95 & \cmark & -\\
\ce{CsTaN2}    & $I\overline{4}2d$ & 34293  & 3.22 & 3.22 &  0.0 & 0.71 & 0.71 & 0.92 & \cmark & -\\
\ce{Cs3MnH5}   & $I4/mcm$          & 643706 & 3.21 & 3.18 &  0.0 & 0.82 & 0.96 & 0.96 & \cmark & -\\
\ce{LiMgP}     & $F\overline{4}3m$ & 36111  & 3.18 & 2.00 &  0.0 & 0.65 & 0.65 & 0.65 & \cmark & -\\
\ce{BaS}       & $Fm\overline{3}m$ & 1500   & 3.17 & 3.02 &  0.0 & 0.85 & 0.85 & 0.85 & \cmark & -\\
\ce{LiAlTe2}   & $I\overline{4}2d$ & 4586   & 3.11 & 3.11 &  0.0 & 0.52 & 0.83 & 0.83 & \cmark & -\\
\ce{YbTe}      & $Fm\overline{3}m$ & 1779   & 3.09 & 1.76 &  0.0 & 0.54 & 0.54 & 0.54 & \xmark & -\\
\ce{Li3Sb}     & $Fm\overline{3}m$ & 2074   & 3.06 & 1.15 &  0.0 & 0.24 & 0.24 & 0.24 & \cmark & \cmark\\
\ce{SrAl2Te4}  & $I422$            & 37091  & 3.06 & 2.66 &  0.0 & 0.42 & 0.79 & 0.80 & \cmark & -\\
\ce{TaCu3Te4}  & $P\overline{4}3m$ & 9295   & 3.05 & 2.50 &  0.0 & 0.63 & 0.63 & 0.63 & \cmark & -\\
\ce{TaCu3Se4}  & $P\overline{4}3m$ & 4081   & 2.98 & 2.43 &  0.0 & 0.82 & 0.82 & 0.82 & \cmark & -\\
\ce{BaSe}      & $Fm\overline{3}m$ & 1253   & 2.95 & 2.59 &  0.0 & 0.76 & 0.76 & 0.76 & \cmark & -\\
\ce{KAg2PS4}   & $I\overline{4}2m$ & 12532  & 2.87 & 2.53 &  0.0 & 0.67 & 0.82 & 0.82 & \cmark & -\\
\ce{AlAs}      & $F\overline{4}3m$ & 2172   & 2.84 & 2.12 &  0.0 & 0.50 & 0.50 & 0.50 & \xmark & -\\
\ce{LiErS2}    & $I4_1/amd$        & 35591  & 2.80 & 2.80 & 10.4 & 0.62 & 0.99 & 0.99 & \cmark & -\\
\ce{GaN}       & $F\overline{4}3m$ & 830    & 2.80 & 2.80 &  5.2 & 0.94 & 0.94 & 0.94 & \cmark & -\\
\end{longtable*}
\end{center}

Among the materials at the top of the list, \ce{SiC} is a well-known wide band gap semiconductor. This material exhibits polymorphism (e.g. cubic: 3C, Rhombohedral: 15R, hexagonal: 6H, 4H, 2H)~\cite{W.J.Choyke97} and can be doped both \emph{n}- and \emph{p}-type~\cite{K.Furukawa-APL86, Y.Kondo-IEEE86, K.Shibahara-JJAP87, R.Weingartner-APL02}. A high hole mobility of 40~cm$^2$/Vs was obtained for the cubic phase~\cite{H.Morkoc-JAP94}. The indirect optical absorption of cubic phase is very weak at room temperature with a coefficient of $10^3$~cm$^{-1}$ at 3.1~eV~\cite{H.R.Philipp-PR58}. We suggest that \ce{SiC} can be considered as a good \emph{p}-type TCM. The main disadvantage of this compound is the difficulty of hole doping. Most known impurities such as Al, B, Ga and Sc create deep doping-levels leading to rather low concentrations of holes which were typically measured to be lower than $10^{18}$~cm$^{-3}$~\cite{H.Morkoc-JAP94} and is suitable for transistor applications.

Next comes a series of beryllium based compounds (\ce{BeS}, \ce{BeSe}, \ce{BeCN2}, \ce{Be2C} and \ce{BeTe}). While their computed performance in terms of band gap and hole effective masses are very attractive, the toxicity of beryllium lowers their interest for technological applications. Likewise, the many lead-based halide perovskites (\ce{CsPbCl3}, \ce{RbPbF3}, and \ce{CsPbF3}) and \ce{Li2GePbS4} also present toxicity issues. It is interesting however to see these halide perovskites being of great interest as solar absorbers when they are made in chemistries showing smaller gaps~\cite{M.Liu13, M.A.Green14}. Toxicity is also an issue with the series of arsenides, e.g. AlAs. These arsenides are also very analogous to the phosphides such as BP and AlP that were identified in a previous work~\cite{J.B.Varley-CheMat17}. Some of the materials in the list contain rare-earth elements which might present some cost issues. We consider that further assessment of all these materials in terms of dopability and mobility is not a priority. Therefore, in the penultimate column of \tabref{tableI}, the absence of toxic or rare-earth elements is verified, as indicated by a checkmark.

Continuing to explore the list of materials, many hydrides appear to be of interest with low hole effective mass and large direct band gaps for \ce{LiH}, \ce{BaLiH3} and \ce{CsH}. Unfortunately, our subsequent defect computations indicate that these hydrides have low-lying hole-killing defects especially the hydrogen vacancy making unlikely their efficient \emph{p}-type doping (see the Supplemental Material~\cite{supplem}). A few sulfides are also identified by our screening: ZnS and \ce{ZnAl2S4}. ZnS has been indeed recently studied as a good performance \emph{p}-type TCM~\cite{R.W.Robinson-AdvEMat16}. \ce{ZnAl2S4}, on the other hand, is less studied but our defect computation indicates that it is very unlikely to be \emph{p}-type dopable because Zn-Al anti-site defects form easily and act as hole-killers. \ce{Al2CdS4} is likely to present the same issues. The defect formation energies computed by DFT for \ce{ZnAl2S4} are given in the Supplemental Material~\cite{supplem}. 

Among the different materials in the table, two promising candidates, \ce{Li3Sb} and \ce{CaTe}, also attracted our attention. The rest of the paper is dedicated to the further computations that were performed for these compounds.

\begin{figure*}[!htb]
\begin{center}
\includegraphics{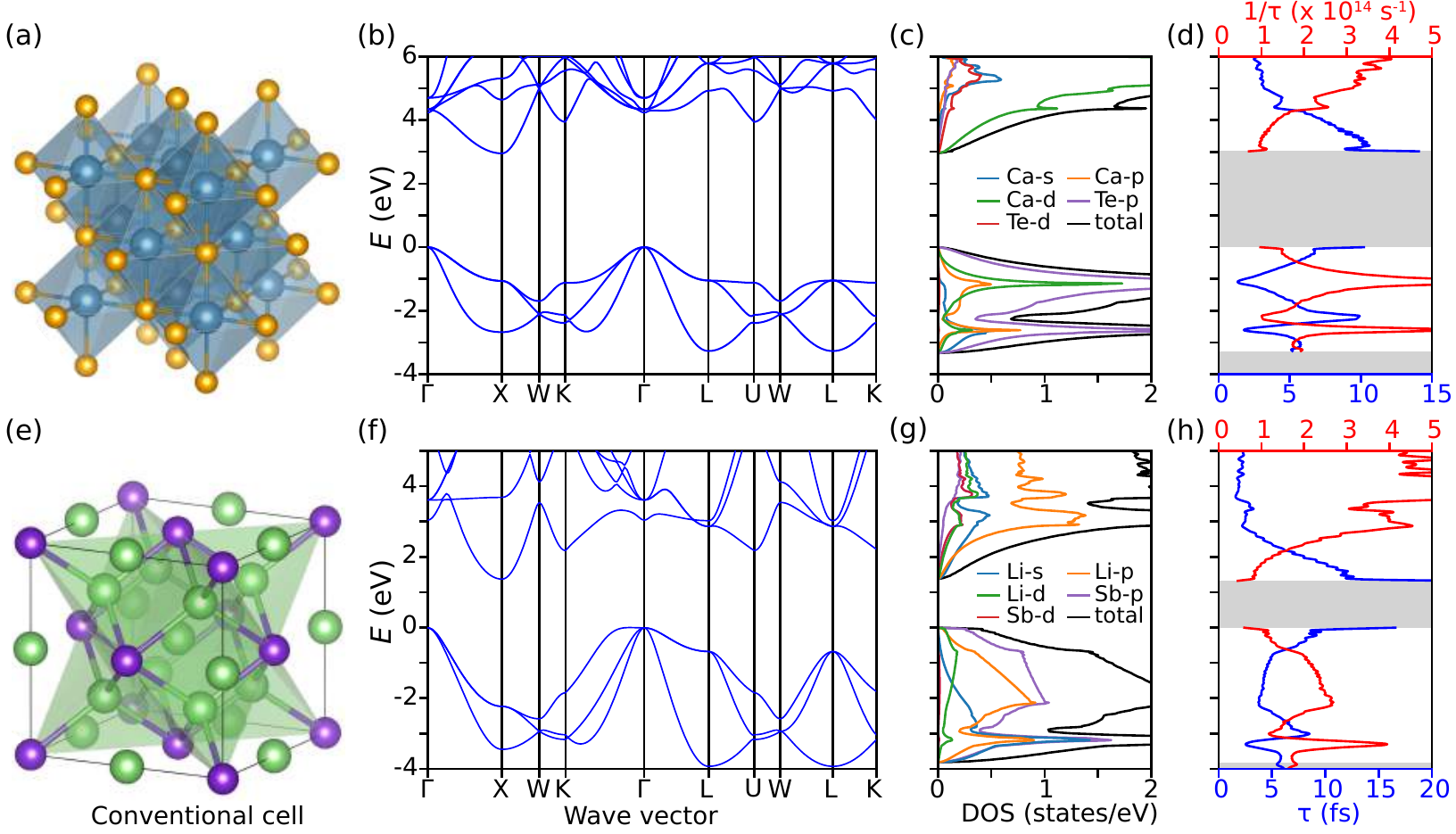}
\end{center}
\vspace{-15pt}
\caption{From the left to the right, the conventional cells, band structures, projected density of states (DOS) and relaxation time and scattering rate. Sub-figures (a)-(d) and (e)-(h) show data of CaTe and \ce{Li3Sb} respectively. The conventional cells present local environments around cations Ca (blue) and Li (green). (b) and (f) plot DFT band structures with a rigid shift of the conduction bands (scissor operator) to fit the fundamental gaps computed by $G_0W_0$. (d) and (h) show relaxation time $\tau$ (in femto-second) and scattering rate $1/\tau$ (in 1/second) as functions of energy at temperature 300~K. The projected DOS in (c) and (g) are computed by DFT. The band gaps of DOS and relaxation time are also shifted to fit $G_0W_0$ values.}
\label{joint_figs}
\end{figure*}

The conventional cells of CaTe and \ce{Li3Sb} are shown in \figref{joint_figs} (a) and (e). Ca atoms in CaTe are surrounded by six Te atoms forming an octahedral local environment. In \ce{Li3Sb}, the cation fills tetrahedral and octahedral sites. Both CaTe and \ce{Li3Sb} are cubic phases with high symmetry, which explains for their isotropy in hole effective masses ($m_1 = m_2 = m_3$). CaTe and \ce{Li3Sb} exhibit very low hole effective masses with the eigenvalues being 0.60 and 0.25 $m_o$ ($m_o$-mass of free electron), respectively.  It is worth noting that the lowest hole effective masses found so far in a computational database for a \emph{p}-type conducting oxides \ce{K2Sn2O3}~\cite{G.Hautier-NatCom13, V.-A.Ha-JMCC17} is 0.27 $m_o$. The promising non-oxide \emph{p}-type TCM reported recently~\cite{J.B.Varley-CheMat17}, BP, shows an effective mass around 0.35 $m_o$. Current Cu-based \emph{p}-type TCOs show effective masses around 1.5 to 2 $m_o$~\cite{G.Hautier-NatCom13}. The direct gaps of CaTe and \ce{Li3Sb} calculated using HSE hybrid functional are 3.5 and 3.06~eV respectively. Next to hybrid functional computations, we performed $G_0W_0$ to confirm the value of these band gaps.

%
%
%

\begin{figure}[!htb]
\begin{center}
\includegraphics{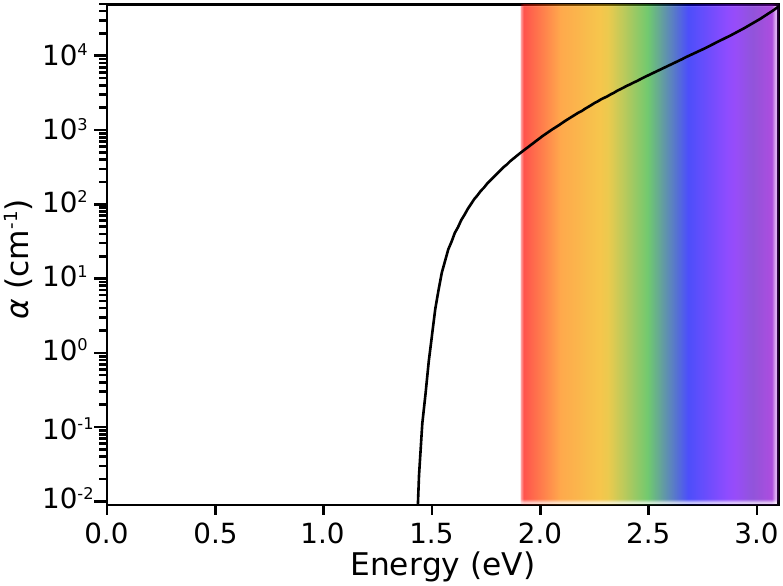}
\end{center}
\vspace{-15pt}
\caption{The indirect optical absorption of \ce{Li3Sb} due to phonon-assisted transitions.}
\label{Li3Sb_inda}
\end{figure}

\figref{joint_figs} (b) shows DFT band structure with a scissor shift to fit $G_0W_0$ fundamental gap ($G_0W_0$ band structure of CaTe is shown in Fig. S6 of the Supplemental Material~\cite{supplem}). The $G_0W_0$ fundamental gap ($\Gamma-X$) is 2.95~eV while the direct gap is located at $X$-point and has a value of 4.14~eV. The $G_0W_0$ direct gap is consistent with the optical gap of 4.1~eV measured experimentally~\cite{G.A.Saum-PR59}. We expect such a large band gap to lead to transparency in the visible region.
%
%
\ce{Li3Sb} is also an indirect semiconductor. In the same way, the DFT electronic band structure with a scissor shift is presented in \figref{joint_figs} (f) (see Fig. S7 of the Supplemental Material~\cite{supplem} for $G_0W_0$ band structure). The $G_0W_0$ band gap and direct gap are 1.37 and 3.17~eV, respectively. The $G_0W_0$ direct gap (located at the $\Gamma$-point) of 3.17~eV. This is consistent with a experimental value of 3.1~eV measured recently~\cite{T.J.Richardson-SolStaIoni03} but much lower than another experimental value of 3.9~eV reported earlier~\cite{R.Gobrecht-PSS66}. The indirect band gap is narrow and will lead to some absorption in the visible range. However, the indirect nature of the absorption makes it phonon-assisted and is expected to lead to weak absorption. To quantify this absorption, we computed the optical absorption including phonon-assisted processes using EPW~\cite{J.Noffsinger-CPC10,S.Ponce-CPC16}. Details about computational method can be found in Ref.~\onlinecite{J.Noffsinger-PRL12}. 
The result in \figref{Li3Sb_inda} shows quite weak absorption in the visible range with the average intensity about $5\times10^3$~cm$^{-1}$, which means that a 100-nm film still allows more than 70 \% of visible light energy to get through. This is suitable for applications and devices using thin-film form of \ce{Li3Sb}. The weak indirect optical absorption computed here is similar to that of established \emph{p}-type TCOs such as SnO~\cite{N.F.Quackenbush13} or recently proposed \emph{p}-type TCMs such as BP~\cite{J.B.Varley-CheMat17}.

%
%
%

CaTe and \ce{Li3Sb} show very low hole effective mass (0.60 and 0.24 $m_o$ within DFT). Indeed, both materials have threefold degeneracy at VBM ($\Gamma$ point), therefore, the transport of holes occurs in three bands with some lighter and some heavier. Our definition of effective mass takes into account the competition among these three bands and give an average value that is representative of the transport which will happen in the different bands. More details about formulas and calculation techniques can be found in Ref.~\cite{G.Hautier-CheMat14, F.Ricci-SciData17}. This should be kept in mind when comparing our results to other studies which sometimes only focus on one band when several competing bands are present ~\cite{R.K.M.Raghupathy-JMCC17,K.Kuhar-ACSEL18}. \figref{joint_figs} shows projected density of states (DOS) for (c) CaTe and (g) \ce{Li3Sb}. For both compounds, the top of valence band is mainly of anionic \emph{p}-orbital characters (\ce{Sb^{3-}} or \ce{Te^{2-}}) with some mixing from the cations. The effective masses are directly related to overlap and energy difference between orbitals~\cite{G.Hautier-CheMat14}. The lower value of hole effective masses obtained in these non-oxide compounds can be associated to both a better alignment between the anionic and cationic states than in oxides and larger anionic \emph{p}-orbitals (5\emph{p} and 4\emph{p} versus 2\emph{p} for oxides).
%
%

\begin{figure*}[!htb]
\begin{center}
\includegraphics{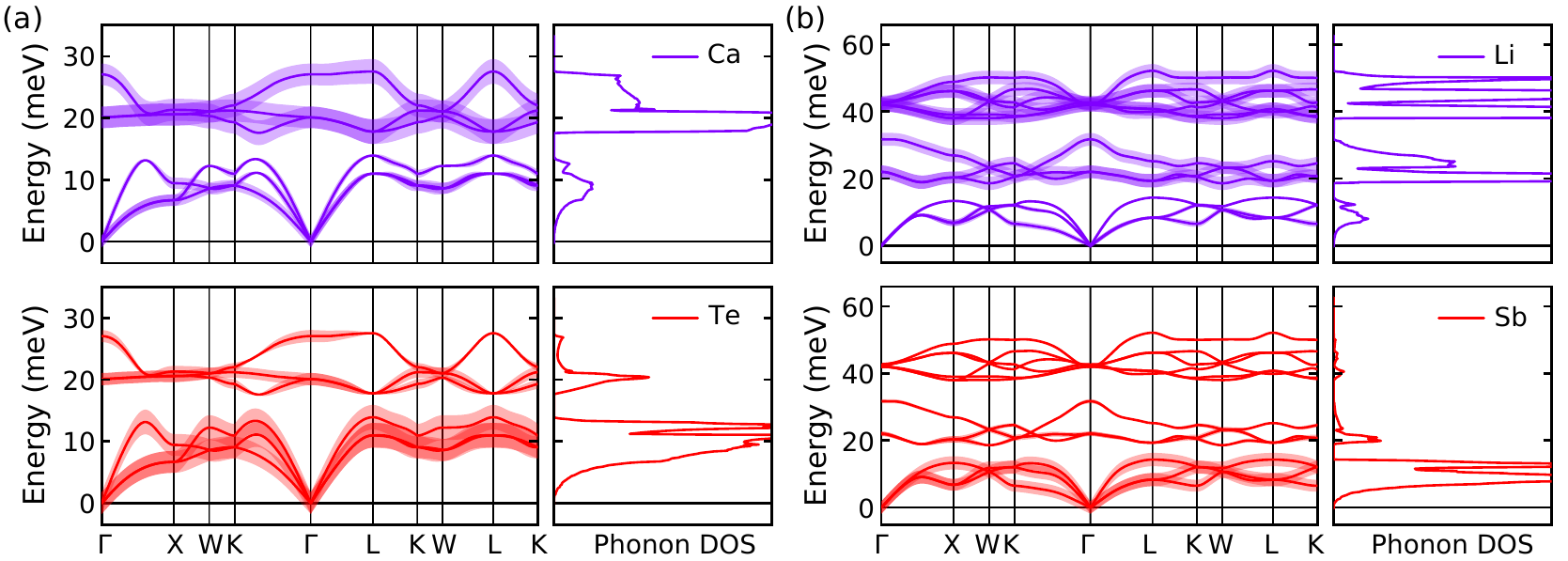}
\end{center}
\vspace{-15pt}
\caption{Phonon band structures with fat bands representing displacements of atomic vibrations. The width of fat bands gives qualitative understanding of the vibrational modes such as what are the atomic types involved in the vibrations at a given energy, their direction of oscillation and the amplitude (related to the displacement). The projected DOS of phonons on each type of atom are correspondingly shown next to the band structures. (a) CaTe and (b) \ce{Li3Sb}.}
\label{phband}
\end{figure*}

The effective mass is an important factor driving carrier mobility but not the only one. Scattering rate or relaxation time also affects the mobility. There are several mechanisms which can influence relaxation time as mentioned in \ref{methods}. Phonon scattering is the most intrinsic factor as it is not affected by purity and microstructure. The evaluation of relaxation time from phonon scattering can be performed \emph{ab initio} using electron-phonon coupling matrices obtained from DFPT phonon computations. \figref{phband} shows phonon band structures (fat bands) and projected DOS of phonons for (a) CaTe and (b) \ce{Li3Sb}. The fat bands represent qualitatively characteristics of vibrational modes including what type of atoms participates in the phonon modes at a given energy, their direction and amplitude. The absence of modes with negative (purely imaginary) frequencies show that these materials are dynamically stable at 0~K. The lighter atoms (Ca and Li) mainly contribute to the optical modes at high frequencies (3 and 9 modes in CaTe and Li3Sb, respectively) while the heavier elements (Te and Sb) play an important role in the three acoustic modes at low frequencies.
%
%

Using the DFPT phonon computations and EPW, we can extract electron-phonon coupling matrices and the relaxation time $\tau_{n\boldsymbol{k}}$ on a dense \emph{\textbf{k}}-point grid (see Eq.~S1 of the Supplemental Material~\cite{supplem}). \figref{joint_figs} (d) and (h) show the scattering rate and lifetime (inverse of scattering rate) as a function of energy at 300~K for CaTe and \ce{Li3Sb} respectively (see Eq.~S2 of the Supplemental Material~\cite{supplem}). As commonly observed, the scattering rate is proportional to the DOS. A higher DOS offer more states available for the scattered electrons. At the doping hole concentration of $10^{18}$~cm$^{-3}$, the Fermi levels are 90.5 and 120.8~meV above the VBMs for CaTe and \ce{Li3Sb}, respectively. For the highest doping of $10^{21}$~cm$^{-3}$, the Fermi levels lie below VBMs of 264.5 and 168.5~meV for CaTe and \ce{Li3Sb}, respectively. The transport of holes, therefore, takes place around VBMs ($\Gamma$-points). The DOS at $\Gamma$-point of \ce{Li3Sb} is larger than that of CaTe but the scattering rate of \ce{Li3Sb} are fairly similar (see \figref{joint_figs} (d) and (h)) indicating that a slightly weaker electron-phonon coupling is present in \ce{Li3Sb}.

%

\begin{figure}[!htb]
\begin{center}
\includegraphics{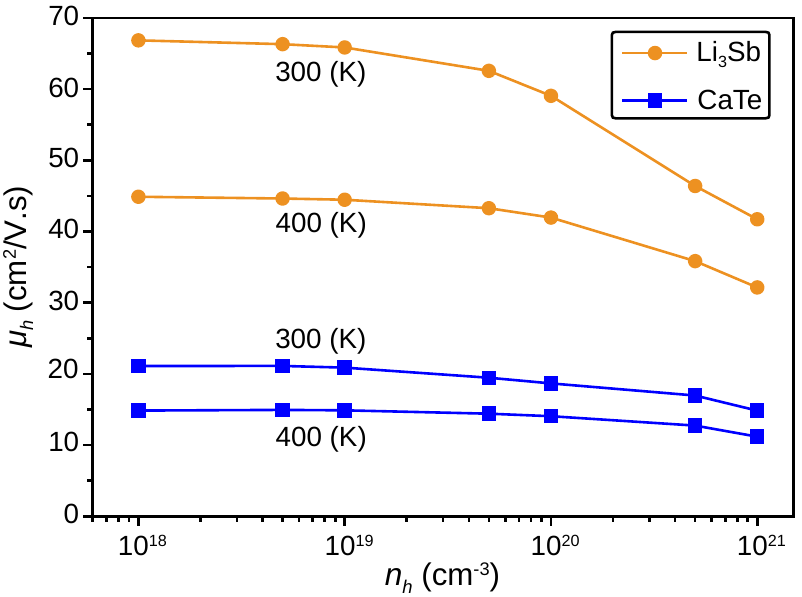}
\end{center}
\vspace{-15pt}
\caption{Hole mobilities as a function of hole concentrations of CaTe and \ce{Li3Sb} at temperatures 300 and 400~K.}
\label{mobility}
\end{figure}

We computed scattering rates at temperatures of 300 and 400~K. \figref{mobility} shows the hole mobilities as a function of hole concentrations at 300 and 400~K for both CaTe and \ce{Li3Sb}. The mobilities decreases with hole concentrations. As the Fermi levels shifts deeper below the VBMs, the DOS increases as well as the scattering rate (see \figref{joint_figs} (d) and (h)). CaTe shows values of hole mobility around 20~cm$^2$/Vs that is comparable with the mobility of \ce{Ba2BiTaO6}, a recently reported \emph{p}-type TCO~\cite{A.Bhatia-CheMat16}, and larger than mobilities of the traditional \emph{p}-type TCOs such as \ce{CuAlO2}~\cite{J.Tate-PRB09} and SnO~\cite{Y.Ogo-APL08}. \ce{Li3Sb} exhibits an exceptional hole mobility up to about 70~cm$^2$/Vs at room-temperature. This value nearly reaches the values of the electron mobilities of the best current \emph{n}-TCOs such as \ce{SnO2}, ZnO, \ce{In2O3} and \ce{Ga2O3} which are around 100~cm$^2$/Vs (see Table SII of the Supplemental Material~\cite{supplem}). It is worth noting that the mobility measured experimentally take into account other scattering processes. Our computed mobilities as they only take into account phonon scattering can be seen as an upper bound.

\begin{figure*}[!htb]
\begin{center}
\includegraphics{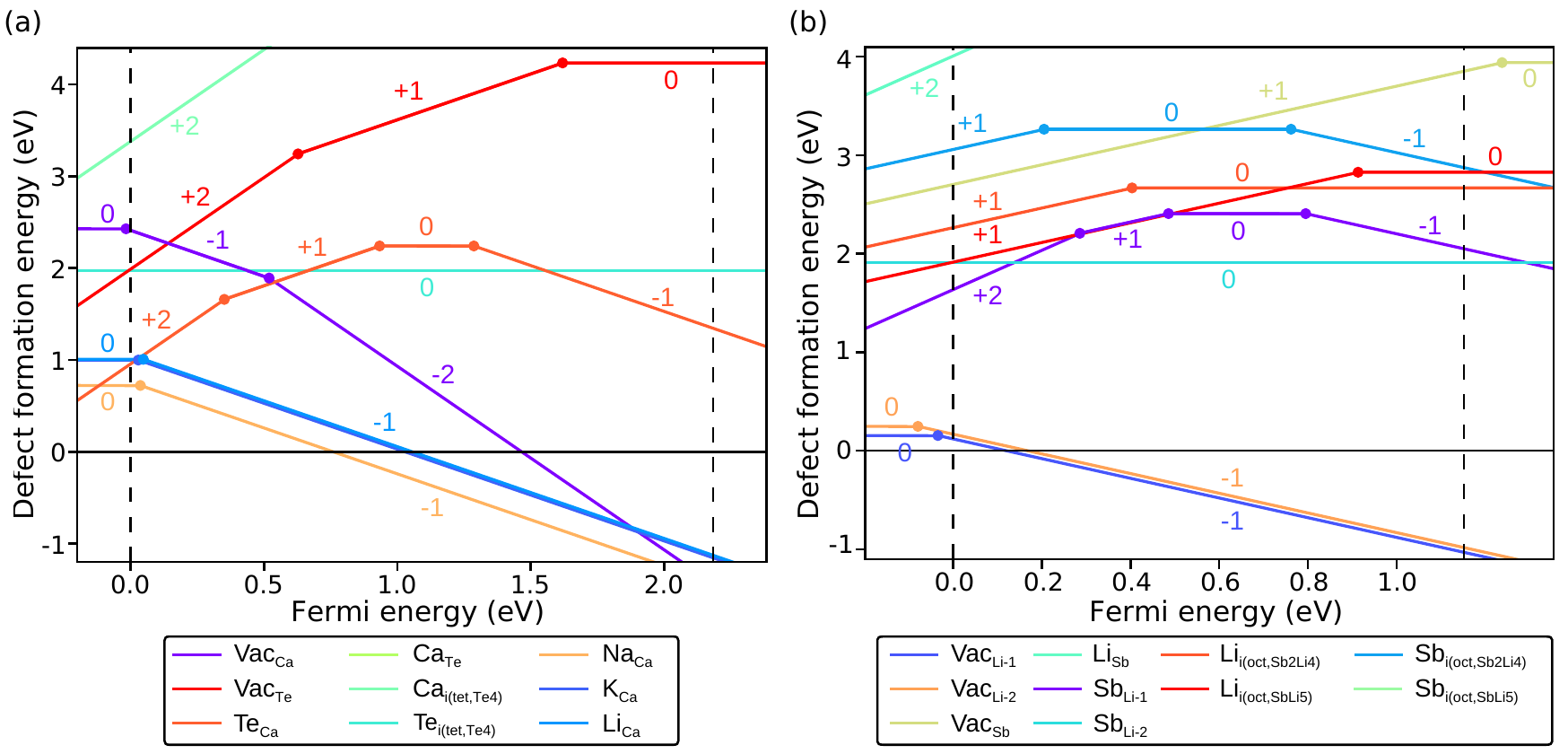}
\end{center}
\vspace{-15pt}
\caption{The defect formation energy as a function of Fermi level of intrinsic and extrinsic defects for (a) \ce{CaTe} and (b) \ce{Li3Sb}. For \ce{CaTe}, the intrinsic defects include vacancies (Vac$_\textrm{Ca}$ and Vac$_\textrm{Te}$), anti-sites (Te$_\textrm{Ca}$ and Ca$_\textrm{Te}$) and interstitial atoms inserted into the tetrahedral hollows formed by 4 Te atoms (Ca$_\textrm{i(tet,Te4)}$ and Te$_\textrm{i(tet,Te4)}$) while Na, K and Li are used as the extrinsic defects substituting Ca atoms (Na$_\textrm{Ca}$, K$_\textrm{Ca}$ and Li$_\textrm{Ca}$).
For \ce{Li3Sb}, the intrinsic defects include vacancies (Vac$_\textrm{Li-1}$, Vac$_\textrm{Li-2}$ and Vac$_\textrm{Sb}$), anti-sites (Li$_\textrm{Sb}$, Sb$_\textrm{Li-1}$ and Sb$_\textrm{Li-2}$) and interstitial atoms inserted into the octahedral hollows formed by Sb and Li atoms (Li$_\textrm{i(oct, Sb2Li4)}$, Li$_\textrm{i(oct, SbLi5)}$, Sb$_\textrm{i(oct, Sb2Li4)}$ and Sb$_\textrm{i(oct, SbLi5)}$). In both cases, the VBM is set to zero.}
\label{defects}
\end{figure*}

Our final assessment focuses on the dopability of CaTe and \ce{Li3Sb}. While we have assumed so far that the Fermi level of these two materials could be tuned to generate hole carriers, it remains to be seen if the defect chemistry is favorable to hole generation. To answer this question, we performed defect calculations using a HSE following the procedure described in section \ref{methods}. \figref{defects} (a) presents the defect formation energy for both intrinsic and extrinsic defects for each sort of defect in CaTe. The chemical potentials are chosen in conditions which lead to the most favorable \emph{p}-type doping tendency for this material. The chemical potentials corresponding to different conditions in the phase diagrams are available in Fig. S8 of the Supplemental Material~\cite{supplem}. Focusing first on intrinsic defects only including vacancies, anti-site defects and interstitial atoms, defect formation energies of these are plotted in \figref{defects} (a) with chemical potentials extracted in Te-rich condition of the phase diagram. Intrinsically, CaTe is unlikely to present \emph{p}-type doping as no intrinsic defect acts as a low lying acceptor. The vacancy of Ca will be in competition with the hole killing  vacancy of Te leading to a fermi level far from the valence band. However, the Te vacancy is not low enough in energy that it would prevent extrinsic \emph{p}-type doping. When extrinsic defects with Na, K and Li substituting onto Ca-sites are considered, we find that all these substitutions offer shallow acceptor very competitive compared to the Te vacancy. The Ca by Na substitution is the lowest in energy. Extrinsic doping by Na might therefore lead to \emph{p}-type doping in CaTe. The plots of formation energies of K$_\textrm{Ca}$, Na$_\textrm{Ca}$ and Li$_\textrm{Ca}$ in \figref{defects} (a) were achieved with chemical potentials extracted from \ce{KTe-CaTe-K2Te3}, \ce{NaTe3-CaTe-Na2Te} and \ce{Li2Te-CaTe-Te} facets of the three-element phase diagrams (see Fig. S8 of the Supplemental Material~\cite{supplem}). For \ce{Li3Sb}, \figref{defects} (b) shows an intrinsic tendency for hole doping with the lithium vacancy (Vac$_\textrm{Li}$) acting as a shallow acceptor with a very low formation energy and no competing hole-killer. This plot is produced with chemical potentials computed in \ce{Li3Sb-Li2Sb} facet of the phase diagram (see Fig. S9 of the Supplemental Material~\cite{supplem}). 

\section{Discussions} \label{discus}
The discovery of the quite unanticipated \ce{Li3Sb} with a potential for very high hole mobility demonstrates the interest of our HT screening strategy. \ce{Li3Sb} is an unexpected compound for TCM applications and would have been difficult to intuitively identify. Among other  \ce{A3B} compounds (A = Li, Na, K and Rb; and B = N, P, As and Sb), \ce{Li3Sb} is exceptional because of its very low hole effective masses (see Table SIII in the Supplemental Material~\cite{supplem}). We suggest that the energy difference between A-$ns^1$ ($n = 2, 3, 4, 5$ for Li, Na, K and Rb, respectively) and B-$np^3$ ($n = 2, 3, 4, 5$ for N, P, As and Sb, respectively) orbitals of valence electrons (A and B) might play important role here. In fact, the energy difference between Li-$2s^1$ and Sb-$5p^3$ is about 1.954 (eV)~\cite{pseudodojo} and is the smallest value among many other ones of A-$ns^1$/B-$np^3$ pairs. This leads to a small orbital-energy difference and strong $s/p$ (anti-)bonding, which results in low hole effective mass. While we only focus on CaTe and \ce{Li3Sb} as they are likely the most potential candidates, there are other interesting materials with hole effective masses from 0.6 to 1.0 $m_o$ and high direct gaps (see \tabref{tableI}) such as CaS, SrSe, SrTe, \ce{LiCa4B3N6}, \ce{LiSr4B3N6}, \ce{NaSr4B3N6}... Defects calculations for these materials have not performed in this work and we, therefore, cannot adjudge their \emph{p}-type doping-tendency.


By going beyond oxides, we identified compounds with very high hole mobility. However, several other issues also arise and need to be considered. The processing of antimonides or tellurides might be more difficult than oxides. They are, however, very common chemistries in other applications such as thermoelectrics with several exemplary compounds such as PbTe, \ce{Bi2Te3}~\cite{E.Macia-Barber15}, or more recently \ce{Mg3Sb2}~\cite{T.Kajikawa2003, C.L.Condron2006, J.Zhang2017}. The band gaps in non-oxide compounds are narrower, which lowers in average their transparency in the visible light. As we already discussed~\cite{J.B.Varley-CheMat17}, this can be overcome by exploiting the indirect gaps and weak phonon-assisted optical transitions. Lower band gaps are useful for \emph{p}-dopability though as lower band gap materials tend to be easier to dope~\cite{A.Zunger03}.

We note that the defect chemistry of non-oxide can be different than in traditional TCOs. For oxides, the cation-anion anti-site defects (replacement of anions on cations' sites and vice versa) are unlikely to be favorable energetically because of the large electronegativity difference between cations and anions. In non-oxide compounds, $e.g.$ CaTe, the cation-anion anti-sites are more likely to be present leading to potentially different hole-killing defects. While the anion (oxygen) vacancy vacancy is the most common hole-killer in oxides, we see our non-oxide materials presenting anti-sites cation-anion defects lower in energy than the anion vacancy such as in CaTe. We also identify that the hydride chemistry while offering attractive electronic structures presents dopability issues (i.e., a low lying hydrogen vacancy acting as hole killer) preventing them for further consideration in \emph{p}-type TCMs.

\section{Conclusions}
Using a large database and appropriate filtering strategies, we report on a high-throughput search for non-oxide \emph{p}-type TCMs. We identified two materials to be of interest: CaTe and \ce{Li3Sb}. We performed extensive follow-up computational investigation of these candidates, evaluating their band structure using beyond DFT techniques, their transport and phonon-assisted optical properties using electron-phonon computations as well as their defect chemistry. Both CaTe and Li3Sb present very attractive properties for \emph{p}-type TCM applications. The \ce{Li3Sb} shows a very high hole mobility of around 70~cm$^2$/Vs, which is close to electron mobility in the best \emph{n}-type TCMs. Our work motivates further experimental investigation of these two materials for TCM applications.  

\section{Acknowledgments}
V.-A.H. was funded through a grant from the FRIA. G.-M.R. is grateful to the F.R.S.-FNRS for financial support. G.H., G.-M.R., G.Y. and F.R. acknowledge the F.R.S.-FNRS project HTBaSE (contract N$^\circ$ PDR-T.1071.15) for financial support. We acknowledge access to various computational resources: the Tier-1 supercomputer of the F\'{e}d\'{e}ration Wallonie-Bruxelles funded by the Walloon Region (grant agreement N$^\circ$ 1117545), and all the facilities provided by the Universit\'{e} catholique de Louvain (CISM/UCLouvain) and by the Consortium des \'{E}quipements de Calcul Intensif en F\'{e}d\'{e}ration Wallonie Bruxelles (C\'{E}CI). The authors thank Dr. Samuel Ponc\'{e} and Professor Emmanouil Kioupakis for helpful discussions on the technical aspects of the electron-phonon computations.
\bibliographystyle{apsrev4-1}
\bibliography{biblio}

\begin{thebibliography}{75}%
\makeatletter
\providecommand \@ifxundefined [1]{%
 \@ifx{#1\undefined}
}%
\providecommand \@ifnum [1]{%
 \ifnum #1\expandafter \@firstoftwo
 \else \expandafter \@secondoftwo
 \fi
}%
\providecommand \@ifx [1]{%
 \ifx #1\expandafter \@firstoftwo
 \else \expandafter \@secondoftwo
 \fi
}%
\providecommand \natexlab [1]{#1}%
\providecommand \enquote  [1]{``#1''}%
\providecommand \bibnamefont  [1]{#1}%
\providecommand \bibfnamefont [1]{#1}%
\providecommand \citenamefont [1]{#1}%
\providecommand \href@noop [0]{\@secondoftwo}%
\providecommand \href [0]{\begingroup \@sanitize@url \@href}%
\providecommand \@href[1]{\@@startlink{#1}\@@href}%
\providecommand \@@href[1]{\endgroup#1\@@endlink}%
\providecommand \@sanitize@url [0]{\catcode `\\12\catcode `\$12\catcode
  `\&12\catcode `\#12\catcode `\^12\catcode `\_12\catcode `\%12\relax}%
\providecommand \@@startlink[1]{}%
\providecommand \@@endlink[0]{}%
\providecommand \url  [0]{\begingroup\@sanitize@url \@url }%
\providecommand \@url [1]{\endgroup\@href {#1}{\urlprefix }}%
\providecommand \urlprefix  [0]{URL }%
\providecommand \Eprint [0]{\href }%
\providecommand \doibase [0]{http://dx.doi.org/}%
\providecommand \selectlanguage [0]{\@gobble}%
\providecommand \bibinfo  [0]{\@secondoftwo}%
\providecommand \bibfield  [0]{\@secondoftwo}%
\providecommand \translation [1]{[#1]}%
\providecommand \BibitemOpen [0]{}%
\providecommand \bibitemStop [0]{}%
\providecommand \bibitemNoStop [0]{.\EOS\space}%
\providecommand \EOS [0]{\spacefactor3000\relax}%
\providecommand \BibitemShut  [1]{\csname bibitem#1\endcsname}%
\let\auto@bib@innerbib\@empty
\bibitem [{\citenamefont {Ohta}\ and\ \citenamefont
  {Hosono}(2004)}]{H.Ohta-MatTo04}%
  \BibitemOpen
  \bibfield  {author} {\bibinfo {author} {\bibfnamefont {H.}~\bibnamefont
  {Ohta}}\ and\ \bibinfo {author} {\bibfnamefont {H.}~\bibnamefont {Hosono}},\
  }\href {\doibase 10.1016/S1369-7021(04)00288-3} {\bibfield  {journal}
  {\bibinfo  {journal} {Mater. Today}\ }\textbf {\bibinfo {volume}
  {\textbf{7}}},\ \bibinfo {pages} {42} (\bibinfo {year} {2004})}\BibitemShut
  {NoStop}%
\bibitem [{\citenamefont {Facchetti}\ and\ \citenamefont
  {Marks}(2010)}]{A.Facchetti10}%
  \BibitemOpen
  \bibinfo {editor} {\bibfnamefont {A.}~\bibnamefont {Facchetti}}\ and\
  \bibinfo {editor} {\bibfnamefont {T.~J.}\ \bibnamefont {Marks}},\ eds.,\
  \enquote {\bibinfo {title} {Transparent electronics: From synthesis to
  applications},}\ \ (\bibinfo  {publisher} {Wiley},\ \bibinfo {year}
  {2010})\BibitemShut {NoStop}%
\bibitem [{\citenamefont {Ellmer}(2012)}]{K.Ellmer-NatPho12}%
  \BibitemOpen
  \bibfield  {author} {\bibinfo {author} {\bibfnamefont {K.}~\bibnamefont
  {Ellmer}},\ }\href {\doibase 10.1038/NPHOTON.2012.282} {\bibfield  {journal}
  {\bibinfo  {journal} {Nat. Photonics}\ }\textbf {\bibinfo {volume}
  {\textbf{6}}},\ \bibinfo {pages} {809} (\bibinfo {year} {2012})}\BibitemShut
  {NoStop}%
\bibitem [{\citenamefont {Barquinha}\ \emph {et~al.}(2012)\citenamefont
  {Barquinha}, \citenamefont {Martins}, \citenamefont {Pereira},\ and\
  \citenamefont {Fortunato}}]{P.Barquinha12}%
  \BibitemOpen
  \bibfield  {author} {\bibinfo {author} {\bibfnamefont {P.}~\bibnamefont
  {Barquinha}}, \bibinfo {author} {\bibfnamefont {R.}~\bibnamefont {Martins}},
  \bibinfo {author} {\bibfnamefont {L.}~\bibnamefont {Pereira}}, \ and\
  \bibinfo {author} {\bibfnamefont {E.}~\bibnamefont {Fortunato}},\ }\enquote
  {\bibinfo {title} {Transparent oxide electronics: From materials to
  devices},}\ \ (\bibinfo  {publisher} {Wiley},\ \bibinfo {year}
  {2012})\BibitemShut {NoStop}%
\bibitem [{\citenamefont {Fortunato}\ \emph {et~al.}(2012)\citenamefont
  {Fortunato}, \citenamefont {Barquinha},\ and\ \citenamefont
  {Martins}}]{E.Fortunato12}%
  \BibitemOpen
  \bibfield  {author} {\bibinfo {author} {\bibfnamefont {E.}~\bibnamefont
  {Fortunato}}, \bibinfo {author} {\bibfnamefont {P.}~\bibnamefont
  {Barquinha}}, \ and\ \bibinfo {author} {\bibfnamefont {R.}~\bibnamefont
  {Martins}},\ }\href {\doibase 10.1002/adma.201103228} {\bibfield  {journal}
  {\bibinfo  {journal} {Adv. Mater.}\ }\textbf {\bibinfo {volume}
  {\textbf{24}}},\ \bibinfo {pages} {2945} (\bibinfo {year}
  {2012})}\BibitemShut {NoStop}%
\bibitem [{\citenamefont {Dixon}\ \emph {et~al.}(2016)\citenamefont {Dixon},
  \citenamefont {Scanlon}, \citenamefont {Carmalt},\ and\ \citenamefont
  {Parkin}}]{S.C.Dixon-JMChC16}%
  \BibitemOpen
  \bibfield  {author} {\bibinfo {author} {\bibfnamefont {S.~C.}\ \bibnamefont
  {Dixon}}, \bibinfo {author} {\bibfnamefont {D.~O.}\ \bibnamefont {Scanlon}},
  \bibinfo {author} {\bibfnamefont {C.~J.}\ \bibnamefont {Carmalt}}, \ and\
  \bibinfo {author} {\bibfnamefont {I.~P.}\ \bibnamefont {Parkin}},\ }\href
  {\doibase 10.1039/c6tc01881e} {\bibfield  {journal} {\bibinfo  {journal} {J.
  Mater. Chem. C}\ }\textbf {\bibinfo {volume} {\textbf{4}}},\ \bibinfo {pages}
  {6946} (\bibinfo {year} {2016})}\BibitemShut {NoStop}%
\bibitem [{\citenamefont {Hautier}\ \emph {et~al.}(2013)\citenamefont
  {Hautier}, \citenamefont {Miglio}, \citenamefont {Ceder}, \citenamefont
  {Rignanese},\ and\ \citenamefont {Gonze}}]{G.Hautier-NatCom13}%
  \BibitemOpen
  \bibfield  {author} {\bibinfo {author} {\bibfnamefont {G.}~\bibnamefont
  {Hautier}}, \bibinfo {author} {\bibfnamefont {A.}~\bibnamefont {Miglio}},
  \bibinfo {author} {\bibfnamefont {G.}~\bibnamefont {Ceder}}, \bibinfo
  {author} {\bibfnamefont {G.-M.}\ \bibnamefont {Rignanese}}, \ and\ \bibinfo
  {author} {\bibfnamefont {X.}~\bibnamefont {Gonze}},\ }\href {\doibase
  10.1038/ncomms3292} {\bibfield  {journal} {\bibinfo  {journal} {Nat.
  Commun.}\ }\textbf {\bibinfo {volume} {\textbf{4}}},\ \bibinfo {pages} {2292}
  (\bibinfo {year} {2013})}\BibitemShut {NoStop}%
\bibitem [{\citenamefont {Varley}\ \emph {et~al.}(2014)\citenamefont {Varley},
  \citenamefont {Lordi}, \citenamefont {Miglio},\ and\ \citenamefont
  {Hautier}}]{J.B.Varley-PRB14}%
  \BibitemOpen
  \bibfield  {author} {\bibinfo {author} {\bibfnamefont {J.~B.}\ \bibnamefont
  {Varley}}, \bibinfo {author} {\bibfnamefont {V.}~\bibnamefont {Lordi}},
  \bibinfo {author} {\bibfnamefont {A.}~\bibnamefont {Miglio}}, \ and\ \bibinfo
  {author} {\bibfnamefont {G.}~\bibnamefont {Hautier}},\ }\href {\doibase
  10.1103/PhysRevB.90.045205} {\bibfield  {journal} {\bibinfo  {journal} {Phys.
  Rev. B}\ }\textbf {\bibinfo {volume} {90}},\ \bibinfo {pages} {045205}
  (\bibinfo {year} {2014})}\BibitemShut {NoStop}%
\bibitem [{\citenamefont {Bhatia}\ \emph {et~al.}(2016)\citenamefont {Bhatia},
  \citenamefont {Hautier}, \citenamefont {Nilgianskul}, \citenamefont {Miglio},
  \citenamefont {Sun}, \citenamefont {Kim}, \citenamefont {Kim}, \citenamefont
  {Chen}, \citenamefont {Rignanese}, \citenamefont {Gonze},\ and\ \citenamefont
  {Suntivich}}]{A.Bhatia-CheMat16}%
  \BibitemOpen
  \bibfield  {author} {\bibinfo {author} {\bibfnamefont {A.}~\bibnamefont
  {Bhatia}}, \bibinfo {author} {\bibfnamefont {G.}~\bibnamefont {Hautier}},
  \bibinfo {author} {\bibfnamefont {T.}~\bibnamefont {Nilgianskul}}, \bibinfo
  {author} {\bibfnamefont {A.}~\bibnamefont {Miglio}}, \bibinfo {author}
  {\bibfnamefont {J.}~\bibnamefont {Sun}}, \bibinfo {author} {\bibfnamefont
  {H.~J.}\ \bibnamefont {Kim}}, \bibinfo {author} {\bibfnamefont {K.~H.}\
  \bibnamefont {Kim}}, \bibinfo {author} {\bibfnamefont {S.}~\bibnamefont
  {Chen}}, \bibinfo {author} {\bibfnamefont {G.-M.}\ \bibnamefont {Rignanese}},
  \bibinfo {author} {\bibfnamefont {X.}~\bibnamefont {Gonze}}, \ and\ \bibinfo
  {author} {\bibfnamefont {J.}~\bibnamefont {Suntivich}},\ }\href {\doibase
  10.1021/acs.chemmater.5b03794} {\bibfield  {journal} {\bibinfo  {journal}
  {Chem. Mater.}\ }\textbf {\bibinfo {volume} {\textbf{28}}},\ \bibinfo {pages}
  {30} (\bibinfo {year} {2016})}\BibitemShut {NoStop}%
\bibitem [{\citenamefont {Yanagi}\ \emph {et~al.}(2003)\citenamefont {Yanagi},
  \citenamefont {Tate}, \citenamefont {Park}, \citenamefont {Park},\ and\
  \citenamefont {Keszler}}]{H.Yanagi-APL03}%
  \BibitemOpen
  \bibfield  {author} {\bibinfo {author} {\bibfnamefont {H.}~\bibnamefont
  {Yanagi}}, \bibinfo {author} {\bibfnamefont {J.}~\bibnamefont {Tate}},
  \bibinfo {author} {\bibfnamefont {S.}~\bibnamefont {Park}}, \bibinfo {author}
  {\bibfnamefont {C.-H.}\ \bibnamefont {Park}}, \ and\ \bibinfo {author}
  {\bibfnamefont {D.}~\bibnamefont {Keszler}},\ }\href {\doibase
  10.1063/1.1571224} {\bibfield  {journal} {\bibinfo  {journal} {Appl. Phys.
  Lett.}\ }\textbf {\bibinfo {volume} {\textbf{82}}},\ \bibinfo {pages} {2814}
  (\bibinfo {year} {2003})}\BibitemShut {NoStop}%
\bibitem [{\citenamefont {Park}\ \emph {et~al.}(2002)\citenamefont {Park},
  \citenamefont {Keszler}, \citenamefont {Valencia}, \citenamefont {Hoffman},
  \citenamefont {Bender},\ and\ \citenamefont {Wager}}]{S.Park-APL02}%
  \BibitemOpen
  \bibfield  {author} {\bibinfo {author} {\bibfnamefont {S.}~\bibnamefont
  {Park}}, \bibinfo {author} {\bibfnamefont {D.~A.}\ \bibnamefont {Keszler}},
  \bibinfo {author} {\bibfnamefont {M.~M.}\ \bibnamefont {Valencia}}, \bibinfo
  {author} {\bibfnamefont {R.~L.}\ \bibnamefont {Hoffman}}, \bibinfo {author}
  {\bibfnamefont {J.~P.}\ \bibnamefont {Bender}}, \ and\ \bibinfo {author}
  {\bibfnamefont {J.~F.}\ \bibnamefont {Wager}},\ }\href {\doibase
  10.1063/1.1485133} {\bibfield  {journal} {\bibinfo  {journal} {Appl. Phys.
  Lett.}\ }\textbf {\bibinfo {volume} {\textbf{80}}},\ \bibinfo {pages} {4393}
  (\bibinfo {year} {2002})}\BibitemShut {NoStop}%
\bibitem [{\citenamefont {Woods-Robinson}\ \emph {et~al.}(2016)\citenamefont
  {Woods-Robinson}, \citenamefont {Cooper}, \citenamefont {Xu}, \citenamefont
  {Schelhas}, \citenamefont {Pool}, \citenamefont {Faghaninia}, \citenamefont
  {Lo}, \citenamefont {Toney}, \citenamefont {Sharp},\ and\ \citenamefont
  {Ager}}]{R.W.Robinson-AdvEMat16}%
  \BibitemOpen
  \bibfield  {author} {\bibinfo {author} {\bibfnamefont {R.}~\bibnamefont
  {Woods-Robinson}}, \bibinfo {author} {\bibfnamefont {J.~K.}\ \bibnamefont
  {Cooper}}, \bibinfo {author} {\bibfnamefont {X.}~\bibnamefont {Xu}}, \bibinfo
  {author} {\bibfnamefont {L.~T.}\ \bibnamefont {Schelhas}}, \bibinfo {author}
  {\bibfnamefont {V.~L.}\ \bibnamefont {Pool}}, \bibinfo {author}
  {\bibfnamefont {A.}~\bibnamefont {Faghaninia}}, \bibinfo {author}
  {\bibfnamefont {C.~S.}\ \bibnamefont {Lo}}, \bibinfo {author} {\bibfnamefont
  {M.~F.}\ \bibnamefont {Toney}}, \bibinfo {author} {\bibfnamefont {I.~D.}\
  \bibnamefont {Sharp}}, \ and\ \bibinfo {author} {\bibfnamefont {J.~W.}\
  \bibnamefont {Ager}},\ }\href {\doibase 10.1002/aelm.201500396} {\bibfield
  {journal} {\bibinfo  {journal} {Adv. Electron. Mater.}\ }\textbf {\bibinfo
  {volume} {\textbf{2}}},\ \bibinfo {pages} {1500396} (\bibinfo {year}
  {2016})}\BibitemShut {NoStop}%
\bibitem [{\citenamefont {Ueda}\ \emph {et~al.}(2000)\citenamefont {Ueda},
  \citenamefont {Inoue}, \citenamefont {Hirose}, \citenamefont {Kawazoe},\ and\
  \citenamefont {Hosono}}]{K.Ueda-APL00}%
  \BibitemOpen
  \bibfield  {author} {\bibinfo {author} {\bibfnamefont {K.}~\bibnamefont
  {Ueda}}, \bibinfo {author} {\bibfnamefont {S.}~\bibnamefont {Inoue}},
  \bibinfo {author} {\bibfnamefont {S.}~\bibnamefont {Hirose}}, \bibinfo
  {author} {\bibfnamefont {H.}~\bibnamefont {Kawazoe}}, \ and\ \bibinfo
  {author} {\bibfnamefont {H.}~\bibnamefont {Hosono}},\ }\href {\doibase
  10.1063/1.1319507} {\bibfield  {journal} {\bibinfo  {journal} {Appl. Phys.
  Lett.}\ }\textbf {\bibinfo {volume} {\textbf{77}}},\ \bibinfo {pages} {2701}
  (\bibinfo {year} {2000})}\BibitemShut {NoStop}%
\bibitem [{\citenamefont {Yan}\ \emph {et~al.}(2015)\citenamefont {Yan},
  \citenamefont {Zhang}, \citenamefont {Yu}, \citenamefont {Yu}, \citenamefont
  {Nagaraja}, \citenamefont {Mason},\ and\ \citenamefont
  {Zunger}}]{F.Yan-NatCom15}%
  \BibitemOpen
  \bibfield  {author} {\bibinfo {author} {\bibfnamefont {F.}~\bibnamefont
  {Yan}}, \bibinfo {author} {\bibfnamefont {X.}~\bibnamefont {Zhang}}, \bibinfo
  {author} {\bibfnamefont {Y.~G.}\ \bibnamefont {Yu}}, \bibinfo {author}
  {\bibfnamefont {L.}~\bibnamefont {Yu}}, \bibinfo {author} {\bibfnamefont
  {A.}~\bibnamefont {Nagaraja}}, \bibinfo {author} {\bibfnamefont {T.~O.}\
  \bibnamefont {Mason}}, \ and\ \bibinfo {author} {\bibfnamefont
  {A.}~\bibnamefont {Zunger}},\ }\href {\doibase 10.1038/ncomms8308} {\bibfield
   {journal} {\bibinfo  {journal} {Nat. Commun.}\ }\textbf {\bibinfo {volume}
  {\textbf{6}}},\ \bibinfo {pages} {7308} (\bibinfo {year} {2015})}\BibitemShut
  {NoStop}%
\bibitem [{\citenamefont {Varley}\ \emph {et~al.}(2017)\citenamefont {Varley},
  \citenamefont {Miglio}, \citenamefont {Ha}, \citenamefont {van Setten},
  \citenamefont {Rignanese},\ and\ \citenamefont
  {Hautier}}]{J.B.Varley-CheMat17}%
  \BibitemOpen
  \bibfield  {author} {\bibinfo {author} {\bibfnamefont {J.~B.}\ \bibnamefont
  {Varley}}, \bibinfo {author} {\bibfnamefont {A.}~\bibnamefont {Miglio}},
  \bibinfo {author} {\bibfnamefont {V.-A.}\ \bibnamefont {Ha}}, \bibinfo
  {author} {\bibfnamefont {M.~J.}\ \bibnamefont {van Setten}}, \bibinfo
  {author} {\bibfnamefont {G.-M.}\ \bibnamefont {Rignanese}}, \ and\ \bibinfo
  {author} {\bibfnamefont {G.}~\bibnamefont {Hautier}},\ }\href {\doibase
  10.1021/acs.chemmater.6b04663} {\bibfield  {journal} {\bibinfo  {journal}
  {Chem. Mater.}\ }\textbf {\bibinfo {volume} {\textbf{29}}},\ \bibinfo {pages}
  {2568} (\bibinfo {year} {2017})}\BibitemShut {NoStop}%
\bibitem [{\citenamefont {Raghupathy}\ \emph
  {et~al.}(2018{\natexlab{a}})\citenamefont {Raghupathy}, \citenamefont
  {K\"uhne}, \citenamefont {Felser},\ and\ \citenamefont
  {Mirhosseini}}]{R.K.M.Raghupathy-JMCC17}%
  \BibitemOpen
  \bibfield  {author} {\bibinfo {author} {\bibfnamefont {R.~K.~M.}\
  \bibnamefont {Raghupathy}}, \bibinfo {author} {\bibfnamefont {T.~D.}\
  \bibnamefont {K\"uhne}}, \bibinfo {author} {\bibfnamefont {C.}~\bibnamefont
  {Felser}}, \ and\ \bibinfo {author} {\bibfnamefont {H.}~\bibnamefont
  {Mirhosseini}},\ }\href {\doibase 10.1039/c7tc05311h} {\bibfield  {journal}
  {\bibinfo  {journal} {J. Mater. Chem. C}\ }\textbf {\bibinfo {volume}
  {\textbf{6}}},\ \bibinfo {pages} {541} (\bibinfo {year}
  {2018}{\natexlab{a}})}\BibitemShut {NoStop}%
\bibitem [{\citenamefont {Raghupathy}\ \emph
  {et~al.}(2018{\natexlab{b}})\citenamefont {Raghupathy}, \citenamefont
  {Wiebeler}, \citenamefont {K\"{u}hne}, \citenamefont {Felser},\ and\
  \citenamefont {Mirhosseini}}]{K.M.Raghupathy2018}%
  \BibitemOpen
  \bibfield  {author} {\bibinfo {author} {\bibfnamefont {R.~K.~M.}\
  \bibnamefont {Raghupathy}}, \bibinfo {author} {\bibfnamefont
  {H.}~\bibnamefont {Wiebeler}}, \bibinfo {author} {\bibfnamefont {T.~D.}\
  \bibnamefont {K\"{u}hne}}, \bibinfo {author} {\bibfnamefont {C.}~\bibnamefont
  {Felser}}, \ and\ \bibinfo {author} {\bibfnamefont {H.}~\bibnamefont
  {Mirhosseini}},\ }\href {\doibase 10.1021/acs.chemmater.8b02719} {\bibfield
  {journal} {\bibinfo  {journal} {Chem. Mater.}\ } (\bibinfo {year}
  {2018}{\natexlab{b}}),\ 10.1021/acs.chemmater.8b02719}\BibitemShut {NoStop}%
\bibitem [{\citenamefont {Ricci}\ \emph {et~al.}(2017)\citenamefont {Ricci},
  \citenamefont {Chen}, \citenamefont {Aydemir}, \citenamefont {rey Snyder},
  \citenamefont {Rignanese}, \citenamefont {Jain},\ and\ \citenamefont
  {Hautier}}]{F.Ricci-SciData17}%
  \BibitemOpen
  \bibfield  {author} {\bibinfo {author} {\bibfnamefont {F.}~\bibnamefont
  {Ricci}}, \bibinfo {author} {\bibfnamefont {W.}~\bibnamefont {Chen}},
  \bibinfo {author} {\bibfnamefont {U.}~\bibnamefont {Aydemir}}, \bibinfo
  {author} {\bibfnamefont {G.~J.}\ \bibnamefont {rey Snyder}}, \bibinfo
  {author} {\bibfnamefont {G.-M.}\ \bibnamefont {Rignanese}}, \bibinfo {author}
  {\bibfnamefont {A.}~\bibnamefont {Jain}}, \ and\ \bibinfo {author}
  {\bibfnamefont {G.}~\bibnamefont {Hautier}},\ }\href {\doibase
  10.1038/sdata.2017.85} {\bibfield  {journal} {\bibinfo  {journal} {Sci.
  Data}\ }\textbf {\bibinfo {volume} {\textbf{4}}},\ \bibinfo {pages} {170085}
  (\bibinfo {year} {2017})}\BibitemShut {NoStop}%
\bibitem [{ICS(2013)}]{ICSD13}%
  \BibitemOpen
  \href@noop {} {\enquote {\bibinfo {title} {{Inorganic Crystal Structure
  Database}},}\ }\bibinfo {howpublished}
  {\url{https://www.fiz-karlsruhe.de/de/leistungen/kristallographie/icsd.html}}
  (\bibinfo {year} {2013}),\ \bibinfo {note} {[FIZ Karlsruhe: Karlsruhe,
  Germany, 2013]}\BibitemShut {NoStop}%
\bibitem [{\citenamefont {Jain}\ \emph {et~al.}(2013)\citenamefont {Jain},
  \citenamefont {Ong}, \citenamefont {Hautier}, \citenamefont {Chen},
  \citenamefont {Richards}, \citenamefont {Dacek}, \citenamefont {Cholia},
  \citenamefont {Gunter}, \citenamefont {Skinner}, \citenamefont {Ceder},\ and\
  \citenamefont {Persson}}]{A.Jain-APLM13}%
  \BibitemOpen
  \bibfield  {author} {\bibinfo {author} {\bibfnamefont {A.}~\bibnamefont
  {Jain}}, \bibinfo {author} {\bibfnamefont {S.~P.}\ \bibnamefont {Ong}},
  \bibinfo {author} {\bibfnamefont {G.}~\bibnamefont {Hautier}}, \bibinfo
  {author} {\bibfnamefont {W.}~\bibnamefont {Chen}}, \bibinfo {author}
  {\bibfnamefont {W.~D.}\ \bibnamefont {Richards}}, \bibinfo {author}
  {\bibfnamefont {S.}~\bibnamefont {Dacek}}, \bibinfo {author} {\bibfnamefont
  {S.}~\bibnamefont {Cholia}}, \bibinfo {author} {\bibfnamefont
  {D.}~\bibnamefont {Gunter}}, \bibinfo {author} {\bibfnamefont
  {D.}~\bibnamefont {Skinner}}, \bibinfo {author} {\bibfnamefont
  {G.}~\bibnamefont {Ceder}}, \ and\ \bibinfo {author} {\bibfnamefont {K.~A.}\
  \bibnamefont {Persson}},\ }\href {\doibase 10.1063/1.4812323} {\bibfield
  {journal} {\bibinfo  {journal} {APL Materials}\ }\textbf {\bibinfo {volume}
  {1}},\ \bibinfo {pages} {011002} (\bibinfo {year} {2013})}\BibitemShut
  {NoStop}%
\bibitem [{Mat(2013)}]{MatPro13}%
  \BibitemOpen
  \href@noop {} {\enquote {\bibinfo {title} {{The Materials Project}},}\
  }\bibinfo {howpublished} {\url{https://www.materialsproject.org/}} (\bibinfo
  {year} {2013}),\ \bibinfo {note} {[accessed September 1, 2013]}\BibitemShut
  {NoStop}%
\bibitem [{\citenamefont {Kresse}\ and\ \citenamefont
  {{Furthm\"uller}}(1996{\natexlab{a}})}]{G.Kresse-CMS96}%
  \BibitemOpen
  \bibfield  {author} {\bibinfo {author} {\bibfnamefont {G.}~\bibnamefont
  {Kresse}}\ and\ \bibinfo {author} {\bibfnamefont {J.}~\bibnamefont
  {{Furthm\"uller}}},\ }\href {\doibase 10.1016/0927-0256(96)00008-0}
  {\bibfield  {journal} {\bibinfo  {journal} {Comput. Mater. Sci.}\ }\textbf
  {\bibinfo {volume} {\textbf{6}}},\ \bibinfo {pages} {15} (\bibinfo {year}
  {1996}{\natexlab{a}})}\BibitemShut {NoStop}%
\bibitem [{\citenamefont {Kresse}\ and\ \citenamefont
  {{Furthm\"uller}}(1996{\natexlab{b}})}]{G.Kresse-PRB96}%
  \BibitemOpen
  \bibfield  {author} {\bibinfo {author} {\bibfnamefont {G.}~\bibnamefont
  {Kresse}}\ and\ \bibinfo {author} {\bibfnamefont {J.}~\bibnamefont
  {{Furthm\"uller}}},\ }\href {\doibase 10.1103/PhysRevB.54.11169} {\bibfield
  {journal} {\bibinfo  {journal} {Phys. Rev. B}\ }\textbf {\bibinfo {volume}
  {\textbf{54}}},\ \bibinfo {pages} {11169} (\bibinfo {year}
  {1996}{\natexlab{b}})}\BibitemShut {NoStop}%
\bibitem [{\citenamefont {Perdew}\ \emph {et~al.}(1996)\citenamefont {Perdew},
  \citenamefont {Burke},\ and\ \citenamefont {Ernzerhof}}]{J.Perdew-PRL96}%
  \BibitemOpen
  \bibfield  {author} {\bibinfo {author} {\bibfnamefont {J.}~\bibnamefont
  {Perdew}}, \bibinfo {author} {\bibfnamefont {K.}~\bibnamefont {Burke}}, \
  and\ \bibinfo {author} {\bibfnamefont {M.}~\bibnamefont {Ernzerhof}},\ }\href
  {\doibase 10.1103/PhysRevLett.77.3865} {\bibfield  {journal} {\bibinfo
  {journal} {Phys. Rev. Lett.}\ }\textbf {\bibinfo {volume} {\textbf{77}}},\
  \bibinfo {pages} {3865} (\bibinfo {year} {1996})}\BibitemShut {NoStop}%
\bibitem [{\citenamefont {Bl{\"o}chl}(1994)}]{P.E.Blochl-PRB94}%
  \BibitemOpen
  \bibfield  {author} {\bibinfo {author} {\bibfnamefont {P.~E.}\ \bibnamefont
  {Bl{\"o}chl}},\ }\href {\doibase 10.1103/PhysRevB.50.17953} {\bibfield
  {journal} {\bibinfo  {journal} {Phys. Rev. B}\ }\textbf {\bibinfo {volume}
  {\textbf{50}}},\ \bibinfo {pages} {17953} (\bibinfo {year}
  {1994})}\BibitemShut {NoStop}%
\bibitem [{\citenamefont {Heyd}\ \emph {et~al.}(2003)\citenamefont {Heyd},
  \citenamefont {Scuseria},\ and\ \citenamefont {Ernzerhof}}]{J.Heyd03}%
  \BibitemOpen
  \bibfield  {author} {\bibinfo {author} {\bibfnamefont {J.}~\bibnamefont
  {Heyd}}, \bibinfo {author} {\bibfnamefont {G.~E.}\ \bibnamefont {Scuseria}},
  \ and\ \bibinfo {author} {\bibfnamefont {M.}~\bibnamefont {Ernzerhof}},\
  }\href {\doibase http://dx.doi.org/10.1063/1.1564060} {\bibfield  {journal}
  {\bibinfo  {journal} {J. Chem. Phys.}\ }\textbf {\bibinfo {volume}
  {\textbf{118}}},\ \bibinfo {pages} {8207} (\bibinfo {year}
  {2003})}\BibitemShut {NoStop}%
\bibitem [{\citenamefont {Brothers}\ \emph {et~al.}(2008)\citenamefont
  {Brothers}, \citenamefont {Izmaylov}, \citenamefont {Normand}, \citenamefont
  {Barone},\ and\ \citenamefont {Scuseria}}]{E.N.Brothers08}%
  \BibitemOpen
  \bibfield  {author} {\bibinfo {author} {\bibfnamefont {E.~N.}\ \bibnamefont
  {Brothers}}, \bibinfo {author} {\bibfnamefont {A.~F.}\ \bibnamefont
  {Izmaylov}}, \bibinfo {author} {\bibfnamefont {J.~O.}\ \bibnamefont
  {Normand}}, \bibinfo {author} {\bibfnamefont {V.}~\bibnamefont {Barone}}, \
  and\ \bibinfo {author} {\bibfnamefont {G.~E.}\ \bibnamefont {Scuseria}},\
  }\href {\doibase http://dx.doi.org/10.1063/1.2955460} {\bibfield  {journal}
  {\bibinfo  {journal} {J. Chem. Phys.}\ }\textbf {\bibinfo {volume}
  {\textbf{129}}},\ \bibinfo {pages} {011102} (\bibinfo {year}
  {2008})}\BibitemShut {NoStop}%
\bibitem [{\citenamefont {Gonze}\ \emph {et~al.}(2002)\citenamefont {Gonze},
  \citenamefont {Beuken}, \citenamefont {Caracas}, \citenamefont {Detraux},
  \citenamefont {Fuchs}, \citenamefont {Rignanese}, \citenamefont {Sindic},
  \citenamefont {Verstraete}, \citenamefont {Zerah}, \citenamefont {Jollet},
  \citenamefont {Torrent}, \citenamefont {Roy}, \citenamefont {Mikami},
  \citenamefont {Ghosez}, \citenamefont {Raty},\ and\ \citenamefont
  {Allan}}]{X.Gonze-Comput.Mater.Sci02}%
  \BibitemOpen
  \bibfield  {author} {\bibinfo {author} {\bibfnamefont {X.}~\bibnamefont
  {Gonze}}, \bibinfo {author} {\bibfnamefont {J.-M.}\ \bibnamefont {Beuken}},
  \bibinfo {author} {\bibfnamefont {R.}~\bibnamefont {Caracas}}, \bibinfo
  {author} {\bibfnamefont {F.}~\bibnamefont {Detraux}}, \bibinfo {author}
  {\bibfnamefont {M.}~\bibnamefont {Fuchs}}, \bibinfo {author} {\bibfnamefont
  {G.-M.}\ \bibnamefont {Rignanese}}, \bibinfo {author} {\bibfnamefont
  {L.}~\bibnamefont {Sindic}}, \bibinfo {author} {\bibfnamefont
  {M.}~\bibnamefont {Verstraete}}, \bibinfo {author} {\bibfnamefont
  {G.}~\bibnamefont {Zerah}}, \bibinfo {author} {\bibfnamefont
  {F.}~\bibnamefont {Jollet}}, \bibinfo {author} {\bibfnamefont
  {M.}~\bibnamefont {Torrent}}, \bibinfo {author} {\bibfnamefont
  {A.}~\bibnamefont {Roy}}, \bibinfo {author} {\bibfnamefont {M.}~\bibnamefont
  {Mikami}}, \bibinfo {author} {\bibfnamefont {P.}~\bibnamefont {Ghosez}},
  \bibinfo {author} {\bibfnamefont {J.-Y.}\ \bibnamefont {Raty}}, \ and\
  \bibinfo {author} {\bibfnamefont {D.~C.}\ \bibnamefont {Allan}},\ }\href
  {\doibase 10.1016/S0927-0256(02)00325-7} {\bibfield  {journal} {\bibinfo
  {journal} {Comput. Mater. Sci.}\ }\textbf {\bibinfo {volume} {\textbf{25}}},\
  \bibinfo {pages} {478} (\bibinfo {year} {2002})}\BibitemShut {NoStop}%
\bibitem [{\citenamefont {Gonze}(2005)}]{X.Gonze-Z.Kristallogr05}%
  \BibitemOpen
  \bibfield  {author} {\bibinfo {author} {\bibfnamefont {X.}~\bibnamefont
  {Gonze}},\ }\href {\doibase 10.1524/zkri.220.5.558.65066} {\bibfield
  {journal} {\bibinfo  {journal} {Z. Kristallogr.}\ }\textbf {\bibinfo {volume}
  {\textbf{202}}},\ \bibinfo {pages} {558} (\bibinfo {year}
  {2005})}\BibitemShut {NoStop}%
\bibitem [{\citenamefont {Gonze}\ \emph {et~al.}(2009)\citenamefont {Gonze},
  \citenamefont {Amadon}, \citenamefont {Anglade}, \citenamefont {Beuken},
  \citenamefont {Bottin}, \citenamefont {Boulanger}, \citenamefont {Bruneval},
  \citenamefont {Caliste}, \citenamefont {Caracas}, \citenamefont
  {C{\^{o}}t{\'{e}}}, \citenamefont {Deutsch}, \citenamefont {Genovese},
  \citenamefont {Ghosez}, \citenamefont {Giantomassi}, \citenamefont
  {Goedecker}, \citenamefont {Hamann}, \citenamefont {Hermet}, \citenamefont
  {Jollet}, \citenamefont {Jomard}, \citenamefont {Leroux}, \citenamefont
  {Mancini}, \citenamefont {Mazevet}, \citenamefont {Oliveira}, \citenamefont
  {Onida}, \citenamefont {Pouillon}, \citenamefont {Rangel}, \citenamefont
  {Rignanese}, \citenamefont {Sangalli}, \citenamefont {Shaltaf}, \citenamefont
  {Torrent}, \citenamefont {Verstraete}, \citenamefont {Zerah},\ and\
  \citenamefont {Zwanziger}}]{X.Gonze-Comput.Phys.Commun09}%
  \BibitemOpen
  \bibfield  {author} {\bibinfo {author} {\bibfnamefont {X.}~\bibnamefont
  {Gonze}}, \bibinfo {author} {\bibfnamefont {B.}~\bibnamefont {Amadon}},
  \bibinfo {author} {\bibfnamefont {P.-M.}\ \bibnamefont {Anglade}}, \bibinfo
  {author} {\bibfnamefont {J.-M.}\ \bibnamefont {Beuken}}, \bibinfo {author}
  {\bibfnamefont {F.}~\bibnamefont {Bottin}}, \bibinfo {author} {\bibfnamefont
  {P.}~\bibnamefont {Boulanger}}, \bibinfo {author} {\bibfnamefont
  {F.}~\bibnamefont {Bruneval}}, \bibinfo {author} {\bibfnamefont
  {D.}~\bibnamefont {Caliste}}, \bibinfo {author} {\bibfnamefont
  {R.}~\bibnamefont {Caracas}}, \bibinfo {author} {\bibfnamefont
  {M.}~\bibnamefont {C{\^{o}}t{\'{e}}}}, \bibinfo {author} {\bibfnamefont
  {T.}~\bibnamefont {Deutsch}}, \bibinfo {author} {\bibfnamefont
  {L.}~\bibnamefont {Genovese}}, \bibinfo {author} {\bibfnamefont
  {P.}~\bibnamefont {Ghosez}}, \bibinfo {author} {\bibfnamefont
  {M.}~\bibnamefont {Giantomassi}}, \bibinfo {author} {\bibfnamefont
  {S.}~\bibnamefont {Goedecker}}, \bibinfo {author} {\bibfnamefont {D.~R.}\
  \bibnamefont {Hamann}}, \bibinfo {author} {\bibfnamefont {P.}~\bibnamefont
  {Hermet}}, \bibinfo {author} {\bibfnamefont {F.}~\bibnamefont {Jollet}},
  \bibinfo {author} {\bibfnamefont {G.}~\bibnamefont {Jomard}}, \bibinfo
  {author} {\bibfnamefont {S.}~\bibnamefont {Leroux}}, \bibinfo {author}
  {\bibfnamefont {M.}~\bibnamefont {Mancini}}, \bibinfo {author} {\bibfnamefont
  {S.}~\bibnamefont {Mazevet}}, \bibinfo {author} {\bibfnamefont {M.~J.~T.}\
  \bibnamefont {Oliveira}}, \bibinfo {author} {\bibfnamefont {G.}~\bibnamefont
  {Onida}}, \bibinfo {author} {\bibfnamefont {Y.}~\bibnamefont {Pouillon}},
  \bibinfo {author} {\bibfnamefont {T.}~\bibnamefont {Rangel}}, \bibinfo
  {author} {\bibfnamefont {G.-M.}\ \bibnamefont {Rignanese}}, \bibinfo {author}
  {\bibfnamefont {D.}~\bibnamefont {Sangalli}}, \bibinfo {author}
  {\bibfnamefont {R.}~\bibnamefont {Shaltaf}}, \bibinfo {author} {\bibfnamefont
  {M.}~\bibnamefont {Torrent}}, \bibinfo {author} {\bibfnamefont {M.~J.}\
  \bibnamefont {Verstraete}}, \bibinfo {author} {\bibfnamefont
  {G.}~\bibnamefont {Zerah}}, \ and\ \bibinfo {author} {\bibfnamefont {J.~W.}\
  \bibnamefont {Zwanziger}},\ }\href {\doibase 10.1016/j.cpc.2009.07.007}
  {\bibfield  {journal} {\bibinfo  {journal} {Comput. Phys. Commun.}\ }\textbf
  {\bibinfo {volume} {\textbf{180}}},\ \bibinfo {pages} {2582} (\bibinfo {year}
  {2009})}\BibitemShut {NoStop}%
\bibitem [{\citenamefont {Gonze}\ \emph {et~al.}(2016)\citenamefont {Gonze},
  \citenamefont {Jollet}, \citenamefont {Araujo}, \citenamefont {Adams},
  \citenamefont {Amadon}, \citenamefont {Applencourt}, \citenamefont {Audouze},
  \citenamefont {Beuken}, \citenamefont {Bieder}, \citenamefont {Bokhanchuk},
  \citenamefont {Bousquet}, \citenamefont {Bruneval}, \citenamefont {Caliste},
  \citenamefont {C{\^{o}}t{\'{e}}}, \citenamefont {Dahm}, \citenamefont
  {Pieve}, \citenamefont {Delaveau}, \citenamefont {Gennaro}, \citenamefont
  {Dorado}, \citenamefont {Espejo}, \citenamefont {Geneste}, \citenamefont
  {Genovese}, \citenamefont {Gerossier}, \citenamefont {Giantomassi},
  \citenamefont {Gillet}, \citenamefont {Hamann}, \citenamefont {He},
  \citenamefont {Jomard}, \citenamefont {Janssen}, \citenamefont {Roux},
  \citenamefont {Levitt}, \citenamefont {Lherbier}, \citenamefont {Liu},
  \citenamefont {Luka{\v{c}}evi{\'{c}}}, \citenamefont {Martin}, \citenamefont
  {Martins}, \citenamefont {Oliveira}, \citenamefont {Ponc{\'{e}}},
  \citenamefont {Pouillon}, \citenamefont {Rangel}, \citenamefont {Rignanese},
  \citenamefont {Romero}, \citenamefont {Rousseau}, \citenamefont {Rubel},
  \citenamefont {Shukri}, \citenamefont {Stankovski}, \citenamefont {Torrent},
  \citenamefont {Setten}, \citenamefont {Troeye}, \citenamefont {Verstraete},
  \citenamefont {Waroquiers}, \citenamefont {Wiktor}, \citenamefont {Xu},
  \citenamefont {Zhou},\ and\ \citenamefont
  {Zwanziger}}]{X.Gonze-Comput.Phys.Commun16}%
  \BibitemOpen
  \bibfield  {author} {\bibinfo {author} {\bibfnamefont {X.}~\bibnamefont
  {Gonze}}, \bibinfo {author} {\bibfnamefont {F.}~\bibnamefont {Jollet}},
  \bibinfo {author} {\bibfnamefont {F.~A.}\ \bibnamefont {Araujo}}, \bibinfo
  {author} {\bibfnamefont {D.}~\bibnamefont {Adams}}, \bibinfo {author}
  {\bibfnamefont {B.}~\bibnamefont {Amadon}}, \bibinfo {author} {\bibfnamefont
  {T.}~\bibnamefont {Applencourt}}, \bibinfo {author} {\bibfnamefont
  {C.}~\bibnamefont {Audouze}}, \bibinfo {author} {\bibfnamefont {J.-M.}\
  \bibnamefont {Beuken}}, \bibinfo {author} {\bibfnamefont {J.}~\bibnamefont
  {Bieder}}, \bibinfo {author} {\bibfnamefont {A.}~\bibnamefont {Bokhanchuk}},
  \bibinfo {author} {\bibfnamefont {E.}~\bibnamefont {Bousquet}}, \bibinfo
  {author} {\bibfnamefont {F.}~\bibnamefont {Bruneval}}, \bibinfo {author}
  {\bibfnamefont {D.}~\bibnamefont {Caliste}}, \bibinfo {author} {\bibfnamefont
  {M.}~\bibnamefont {C{\^{o}}t{\'{e}}}}, \bibinfo {author} {\bibfnamefont
  {F.}~\bibnamefont {Dahm}}, \bibinfo {author} {\bibfnamefont {F.~D.}\
  \bibnamefont {Pieve}}, \bibinfo {author} {\bibfnamefont {M.}~\bibnamefont
  {Delaveau}}, \bibinfo {author} {\bibfnamefont {M.~D.}\ \bibnamefont
  {Gennaro}}, \bibinfo {author} {\bibfnamefont {B.}~\bibnamefont {Dorado}},
  \bibinfo {author} {\bibfnamefont {C.}~\bibnamefont {Espejo}}, \bibinfo
  {author} {\bibfnamefont {G.}~\bibnamefont {Geneste}}, \bibinfo {author}
  {\bibfnamefont {L.}~\bibnamefont {Genovese}}, \bibinfo {author}
  {\bibfnamefont {A.}~\bibnamefont {Gerossier}}, \bibinfo {author}
  {\bibfnamefont {M.}~\bibnamefont {Giantomassi}}, \bibinfo {author}
  {\bibfnamefont {Y.}~\bibnamefont {Gillet}}, \bibinfo {author} {\bibfnamefont
  {D.~R.}\ \bibnamefont {Hamann}}, \bibinfo {author} {\bibfnamefont
  {L.}~\bibnamefont {He}}, \bibinfo {author} {\bibfnamefont {G.}~\bibnamefont
  {Jomard}}, \bibinfo {author} {\bibfnamefont {J.~L.}\ \bibnamefont {Janssen}},
  \bibinfo {author} {\bibfnamefont {S.~L.}\ \bibnamefont {Roux}}, \bibinfo
  {author} {\bibfnamefont {A.}~\bibnamefont {Levitt}}, \bibinfo {author}
  {\bibfnamefont {A.}~\bibnamefont {Lherbier}}, \bibinfo {author}
  {\bibfnamefont {F.}~\bibnamefont {Liu}}, \bibinfo {author} {\bibfnamefont
  {I.}~\bibnamefont {Luka{\v{c}}evi{\'{c}}}}, \bibinfo {author} {\bibfnamefont
  {A.}~\bibnamefont {Martin}}, \bibinfo {author} {\bibfnamefont
  {C.}~\bibnamefont {Martins}}, \bibinfo {author} {\bibfnamefont {M.~J.~T.}\
  \bibnamefont {Oliveira}}, \bibinfo {author} {\bibfnamefont {S.}~\bibnamefont
  {Ponc{\'{e}}}}, \bibinfo {author} {\bibfnamefont {Y.}~\bibnamefont
  {Pouillon}}, \bibinfo {author} {\bibfnamefont {T.}~\bibnamefont {Rangel}},
  \bibinfo {author} {\bibfnamefont {G.-M.}\ \bibnamefont {Rignanese}}, \bibinfo
  {author} {\bibfnamefont {A.~H.}\ \bibnamefont {Romero}}, \bibinfo {author}
  {\bibfnamefont {B.}~\bibnamefont {Rousseau}}, \bibinfo {author}
  {\bibfnamefont {O.}~\bibnamefont {Rubel}}, \bibinfo {author} {\bibfnamefont
  {A.~A.}\ \bibnamefont {Shukri}}, \bibinfo {author} {\bibfnamefont
  {M.}~\bibnamefont {Stankovski}}, \bibinfo {author} {\bibfnamefont
  {M.}~\bibnamefont {Torrent}}, \bibinfo {author} {\bibfnamefont {M.~J.~V.}\
  \bibnamefont {Setten}}, \bibinfo {author} {\bibfnamefont {B.~V.}\
  \bibnamefont {Troeye}}, \bibinfo {author} {\bibfnamefont {M.~J.}\
  \bibnamefont {Verstraete}}, \bibinfo {author} {\bibfnamefont
  {D.}~\bibnamefont {Waroquiers}}, \bibinfo {author} {\bibfnamefont
  {J.}~\bibnamefont {Wiktor}}, \bibinfo {author} {\bibfnamefont
  {B.}~\bibnamefont {Xu}}, \bibinfo {author} {\bibfnamefont {A.}~\bibnamefont
  {Zhou}}, \ and\ \bibinfo {author} {\bibfnamefont {J.~W.}\ \bibnamefont
  {Zwanziger}},\ }\href {\doibase 10.1016/j.cpc.2016.04.003} {\bibfield
  {journal} {\bibinfo  {journal} {Comput. Phys. Commun.}\ }\textbf {\bibinfo
  {volume} {\textbf{205}}},\ \bibinfo {pages} {106} (\bibinfo {year}
  {2016})}\BibitemShut {NoStop}%
\bibitem [{\citenamefont {Hamann}(2013)}]{D.R.Hamann-PRB13}%
  \BibitemOpen
  \bibfield  {author} {\bibinfo {author} {\bibfnamefont {D.~R.}\ \bibnamefont
  {Hamann}},\ }\href {\doibase 10.1103/PhysRevB.88.085117} {\bibfield
  {journal} {\bibinfo  {journal} {Phys. Rev. B}\ }\textbf {\bibinfo {volume}
  {\textbf{88}}},\ \bibinfo {pages} {085117} (\bibinfo {year}
  {2013})}\BibitemShut {NoStop}%
\bibitem [{\citenamefont {van Setten}\ \emph {et~al.}(2018)\citenamefont {van
  Setten}, \citenamefont {Giantomassi}, \citenamefont {Bousquet}, \citenamefont
  {Verstraete}, \citenamefont {Hamann}, \citenamefont {Gonze},\ and\
  \citenamefont {Rignanese}}]{pseudodojo}%
  \BibitemOpen
  \bibfield  {author} {\bibinfo {author} {\bibfnamefont {M.~J.}\ \bibnamefont
  {van Setten}}, \bibinfo {author} {\bibfnamefont {M.}~\bibnamefont
  {Giantomassi}}, \bibinfo {author} {\bibfnamefont {E.}~\bibnamefont
  {Bousquet}}, \bibinfo {author} {\bibfnamefont {M.~J.}\ \bibnamefont
  {Verstraete}}, \bibinfo {author} {\bibfnamefont {D.~R.}\ \bibnamefont
  {Hamann}}, \bibinfo {author} {\bibfnamefont {X.}~\bibnamefont {Gonze}}, \
  and\ \bibinfo {author} {\bibfnamefont {G.-M.}\ \bibnamefont {Rignanese}},\
  }\href {\doibase 10.1016/j.cpc.2018.01.012} {\bibfield  {journal} {\bibinfo
  {journal} {Comput. Phys. Commun.}\ }\textbf {\bibinfo {volume} {226}},\
  \bibinfo {pages} {39} (\bibinfo {year} {2018})}\BibitemShut {NoStop}%
\bibitem [{\citenamefont {van Setten}\ \emph {et~al.}(2017)\citenamefont {van
  Setten}, \citenamefont {Giantomassi}, \citenamefont {Gonze}, \citenamefont
  {Rignanese},\ and\ \citenamefont {Hautier}}]{M.J.vanSetten-PRB17}%
  \BibitemOpen
  \bibfield  {author} {\bibinfo {author} {\bibfnamefont {M.~J.}\ \bibnamefont
  {van Setten}}, \bibinfo {author} {\bibfnamefont {M.}~\bibnamefont
  {Giantomassi}}, \bibinfo {author} {\bibfnamefont {X.}~\bibnamefont {Gonze}},
  \bibinfo {author} {\bibfnamefont {G.-M.}\ \bibnamefont {Rignanese}}, \ and\
  \bibinfo {author} {\bibfnamefont {G.}~\bibnamefont {Hautier}},\ }\href
  {\doibase 10.1103/PhysRevB.96.155207} {\bibfield  {journal} {\bibinfo
  {journal} {Phys. Rev. B}\ }\textbf {\bibinfo {volume} {96}},\ \bibinfo
  {pages} {155207} (\bibinfo {year} {2017})}\BibitemShut {NoStop}%
\bibitem [{\citenamefont {Ong}\ \emph {et~al.}(2013)\citenamefont {Ong},
  \citenamefont {Richards}, \citenamefont {Jain}, \citenamefont {Hautier},
  \citenamefont {Kocher}, \citenamefont {Cholia}, \citenamefont {Gunter},
  \citenamefont {Chevrier}, \citenamefont {Persson},\ and\ \citenamefont
  {Ceder}}]{S.P.Ong-CMS13}%
  \BibitemOpen
  \bibfield  {author} {\bibinfo {author} {\bibfnamefont {S.~P.}\ \bibnamefont
  {Ong}}, \bibinfo {author} {\bibfnamefont {W.~D.}\ \bibnamefont {Richards}},
  \bibinfo {author} {\bibfnamefont {A.}~\bibnamefont {Jain}}, \bibinfo {author}
  {\bibfnamefont {G.}~\bibnamefont {Hautier}}, \bibinfo {author} {\bibfnamefont
  {M.}~\bibnamefont {Kocher}}, \bibinfo {author} {\bibfnamefont
  {S.}~\bibnamefont {Cholia}}, \bibinfo {author} {\bibfnamefont
  {D.}~\bibnamefont {Gunter}}, \bibinfo {author} {\bibfnamefont {V.~L.}\
  \bibnamefont {Chevrier}}, \bibinfo {author} {\bibfnamefont {K.~A.}\
  \bibnamefont {Persson}}, \ and\ \bibinfo {author} {\bibfnamefont
  {G.}~\bibnamefont {Ceder}},\ }\href {\doibase
  10.1016/j.commatsci.2012.10.028} {\bibfield  {journal} {\bibinfo  {journal}
  {Comput. Mater. Sci.}\ }\textbf {\bibinfo {volume} {\textbf{68}}},\ \bibinfo
  {pages} {314} (\bibinfo {year} {2013})}\BibitemShut {NoStop}%
\bibitem [{\citenamefont {{Giantomassi \emph{et al.}}}(2014)}]{Abipy14}%
  \BibitemOpen
  \bibfield  {author} {\bibinfo {author} {\bibfnamefont {M.}~\bibnamefont
  {{Giantomassi \emph{et al.}}}},\ }\href@noop {} {\enquote {\bibinfo {title}
  {{Open-source library for analyzing the results produced by ABINIT}},}\
  }\bibinfo {howpublished} {\url{https://github.com/abinit/abipy}} (\bibinfo
  {year} {2014})\BibitemShut {NoStop}%
\bibitem [{\citenamefont {Freysoldt}\ \emph {et~al.}(2014)\citenamefont
  {Freysoldt}, \citenamefont {Grabowski}, \citenamefont {Hickel}, \citenamefont
  {Neugebauer}, \citenamefont {Kresse}, \citenamefont {Janotti},\ and\
  \citenamefont {{Van de Walle}}}]{C.Freysoldt-RevMP14}%
  \BibitemOpen
  \bibfield  {author} {\bibinfo {author} {\bibfnamefont {C.}~\bibnamefont
  {Freysoldt}}, \bibinfo {author} {\bibfnamefont {B.}~\bibnamefont
  {Grabowski}}, \bibinfo {author} {\bibfnamefont {T.}~\bibnamefont {Hickel}},
  \bibinfo {author} {\bibfnamefont {J.}~\bibnamefont {Neugebauer}}, \bibinfo
  {author} {\bibfnamefont {G.}~\bibnamefont {Kresse}}, \bibinfo {author}
  {\bibfnamefont {A.}~\bibnamefont {Janotti}}, \ and\ \bibinfo {author}
  {\bibfnamefont {C.~G.}\ \bibnamefont {{Van de Walle}}},\ }\href {\doibase
  10.1103/RevModPhys.86.253} {\bibfield  {journal} {\bibinfo  {journal} {Rev.
  Mod. Phys.}\ }\textbf {\bibinfo {volume} {\textbf{86}}},\ \bibinfo {pages}
  {253} (\bibinfo {year} {2014})}\BibitemShut {NoStop}%
\bibitem [{\citenamefont {Komsa}\ \emph {et~al.}(2012)\citenamefont {Komsa},
  \citenamefont {Rantala},\ and\ \citenamefont
  {Pasquarello}}]{H.-P.Komsa-RRB12}%
  \BibitemOpen
  \bibfield  {author} {\bibinfo {author} {\bibfnamefont {H.-P.}\ \bibnamefont
  {Komsa}}, \bibinfo {author} {\bibfnamefont {T.~T.}\ \bibnamefont {Rantala}},
  \ and\ \bibinfo {author} {\bibfnamefont {A.}~\bibnamefont {Pasquarello}},\
  }\href {\doibase 10.1103/PhysRevB.86.045112} {\bibfield  {journal} {\bibinfo
  {journal} {Phys. Rev. B}\ }\textbf {\bibinfo {volume} {\textbf{86}}},\
  \bibinfo {pages} {045112} (\bibinfo {year} {2012})}\BibitemShut {NoStop}%
\bibitem [{\citenamefont {Zhang}\ and\ \citenamefont
  {Northrup}(1991)}]{S.B.Zhang-PRL91}%
  \BibitemOpen
  \bibfield  {author} {\bibinfo {author} {\bibfnamefont {S.~B.}\ \bibnamefont
  {Zhang}}\ and\ \bibinfo {author} {\bibfnamefont {J.~E.}\ \bibnamefont
  {Northrup}},\ }\href {\doibase 10.1103/PhysRevLett.67.2339} {\bibfield
  {journal} {\bibinfo  {journal} {Phys. Rev. Lett.}\ }\textbf {\bibinfo
  {volume} {\textbf{67}}},\ \bibinfo {pages} {2339} (\bibinfo {year}
  {1991})}\BibitemShut {NoStop}%
\bibitem [{\citenamefont {Freysoldt}\ \emph {et~al.}(2011)\citenamefont
  {Freysoldt}, \citenamefont {Neugebauer},\ and\ \citenamefont {{Van de
  Walle}}}]{C.Freysoldt-PSSB11}%
  \BibitemOpen
  \bibfield  {author} {\bibinfo {author} {\bibfnamefont {C.}~\bibnamefont
  {Freysoldt}}, \bibinfo {author} {\bibfnamefont {J.}~\bibnamefont
  {Neugebauer}}, \ and\ \bibinfo {author} {\bibfnamefont {C.~G.}\ \bibnamefont
  {{Van de Walle}}},\ }\href {\doibase 10.1002/pssb.201046289} {\bibfield
  {journal} {\bibinfo  {journal} {Phys. Status Solidi B}\ }\textbf {\bibinfo
  {volume} {\textbf{248}}},\ \bibinfo {pages} {1067} (\bibinfo {year}
  {2011})}\BibitemShut {NoStop}%
\bibitem [{\citenamefont {Kumagai}\ and\ \citenamefont
  {Oba}(2014)}]{Y.Kumagai-PRB14}%
  \BibitemOpen
  \bibfield  {author} {\bibinfo {author} {\bibfnamefont {Y.}~\bibnamefont
  {Kumagai}}\ and\ \bibinfo {author} {\bibfnamefont {F.}~\bibnamefont {Oba}},\
  }\href {\doibase 10.1103/PhysRevB.89.195205} {\bibfield  {journal} {\bibinfo
  {journal} {Phys. Rev. B}\ }\textbf {\bibinfo {volume} {\textbf{89}}},\
  \bibinfo {pages} {195205} (\bibinfo {year} {2014})}\BibitemShut {NoStop}%
\bibitem [{\citenamefont {Broberg}\ \emph {et~al.}(2018)\citenamefont
  {Broberg}, \citenamefont {Medasani}, \citenamefont {Zimmermann},
  \citenamefont {Yu}, \citenamefont {Canning}, \citenamefont {Haranczyk},
  \citenamefont {Asta},\ and\ \citenamefont {Hautier}}]{D.Broberg-CPCom18}%
  \BibitemOpen
  \bibfield  {author} {\bibinfo {author} {\bibfnamefont {D.}~\bibnamefont
  {Broberg}}, \bibinfo {author} {\bibfnamefont {B.}~\bibnamefont {Medasani}},
  \bibinfo {author} {\bibfnamefont {N.~E.}\ \bibnamefont {Zimmermann}},
  \bibinfo {author} {\bibfnamefont {G.}~\bibnamefont {Yu}}, \bibinfo {author}
  {\bibfnamefont {A.}~\bibnamefont {Canning}}, \bibinfo {author} {\bibfnamefont
  {M.}~\bibnamefont {Haranczyk}}, \bibinfo {author} {\bibfnamefont
  {M.}~\bibnamefont {Asta}}, \ and\ \bibinfo {author} {\bibfnamefont
  {G.}~\bibnamefont {Hautier}},\ }\href {\doibase 10.1016/j.cpc.2018.01.004}
  {\bibfield  {journal} {\bibinfo  {journal} {Comput. Phys. Commun}\ }\textbf
  {\bibinfo {volume} {\textbf{226}}},\ \bibinfo {pages} {165} (\bibinfo {year}
  {2018})}\BibitemShut {NoStop}%
\bibitem [{\citenamefont {Madsen}\ and\ \citenamefont
  {Singh}(2006)}]{G.K.H.Madsen-CPC06}%
  \BibitemOpen
  \bibfield  {author} {\bibinfo {author} {\bibfnamefont {G.~K.~H.}\
  \bibnamefont {Madsen}}\ and\ \bibinfo {author} {\bibfnamefont {D.~J.}\
  \bibnamefont {Singh}},\ }\href {\doibase 10.1016/j.cpc.2006.03.007}
  {\bibfield  {journal} {\bibinfo  {journal} {Comput. Phys. Commun.}\ }\textbf
  {\bibinfo {volume} {\textbf{175}}},\ \bibinfo {pages} {67} (\bibinfo {year}
  {2006})}\BibitemShut {NoStop}%
\bibitem [{\citenamefont {Jain}\ \emph {et~al.}(2015)\citenamefont {Jain},
  \citenamefont {Ong}, \citenamefont {Chen}, \citenamefont {Medasani},
  \citenamefont {Qu}, \citenamefont {Kocher}, \citenamefont {Brafman},
  \citenamefont {Petretto}, \citenamefont {Rignanese}, \citenamefont {Hautier},
  \citenamefont {Gunter},\ and\ \citenamefont {Persson}}]{A.Jain-CCPE15}%
  \BibitemOpen
  \bibfield  {author} {\bibinfo {author} {\bibfnamefont {A.}~\bibnamefont
  {Jain}}, \bibinfo {author} {\bibfnamefont {S.~P.}\ \bibnamefont {Ong}},
  \bibinfo {author} {\bibfnamefont {W.}~\bibnamefont {Chen}}, \bibinfo {author}
  {\bibfnamefont {B.}~\bibnamefont {Medasani}}, \bibinfo {author}
  {\bibfnamefont {X.}~\bibnamefont {Qu}}, \bibinfo {author} {\bibfnamefont
  {M.}~\bibnamefont {Kocher}}, \bibinfo {author} {\bibfnamefont
  {M.}~\bibnamefont {Brafman}}, \bibinfo {author} {\bibfnamefont
  {G.}~\bibnamefont {Petretto}}, \bibinfo {author} {\bibfnamefont {G.-M.}\
  \bibnamefont {Rignanese}}, \bibinfo {author} {\bibfnamefont {G.}~\bibnamefont
  {Hautier}}, \bibinfo {author} {\bibfnamefont {D.}~\bibnamefont {Gunter}}, \
  and\ \bibinfo {author} {\bibfnamefont {K.~A.}\ \bibnamefont {Persson}},\
  }\href {\doibase 10.1002/cpe.3505} {\bibfield  {journal} {\bibinfo  {journal}
  {Concurr. Comput. Pract. Exp.}\ }\textbf {\bibinfo {volume} {\textbf{27}}},\
  \bibinfo {pages} {5037} (\bibinfo {year} {2015})}\BibitemShut {NoStop}%
\bibitem [{\citenamefont {Noffsinger}\ \emph {et~al.}(2010)\citenamefont
  {Noffsinger}, \citenamefont {Giustino}, \citenamefont {Malone}, \citenamefont
  {Park}, \citenamefont {Louie},\ and\ \citenamefont
  {Cohen}}]{J.Noffsinger-CPC10}%
  \BibitemOpen
  \bibfield  {author} {\bibinfo {author} {\bibfnamefont {J.}~\bibnamefont
  {Noffsinger}}, \bibinfo {author} {\bibfnamefont {F.}~\bibnamefont
  {Giustino}}, \bibinfo {author} {\bibfnamefont {B.~D.}\ \bibnamefont
  {Malone}}, \bibinfo {author} {\bibfnamefont {C.-H.}\ \bibnamefont {Park}},
  \bibinfo {author} {\bibfnamefont {S.~G.}\ \bibnamefont {Louie}}, \ and\
  \bibinfo {author} {\bibfnamefont {M.~L.}\ \bibnamefont {Cohen}},\ }\href
  {\doibase 10.1016/j.cpc.2010.08.027} {\bibfield  {journal} {\bibinfo
  {journal} {Comput. Phys. Commun.}\ }\textbf {\bibinfo {volume}
  {\textbf{181}}},\ \bibinfo {pages} {2140} (\bibinfo {year}
  {2010})}\BibitemShut {NoStop}%
\bibitem [{\citenamefont {Ponc\'{e}}\ \emph {et~al.}(2016)\citenamefont
  {Ponc\'{e}}, \citenamefont {Margine}, \citenamefont {Verdi},\ and\
  \citenamefont {Giustino}}]{S.Ponce-CPC16}%
  \BibitemOpen
  \bibfield  {author} {\bibinfo {author} {\bibfnamefont {S.}~\bibnamefont
  {Ponc\'{e}}}, \bibinfo {author} {\bibfnamefont {E.~R.}\ \bibnamefont
  {Margine}}, \bibinfo {author} {\bibfnamefont {C.}~\bibnamefont {Verdi}}, \
  and\ \bibinfo {author} {\bibfnamefont {F.}~\bibnamefont {Giustino}},\ }\href
  {\doibase 10.1016/j.cpc.2016.07.028} {\bibfield  {journal} {\bibinfo
  {journal} {Comput. Phys. Commun.}\ }\textbf {\bibinfo {volume}
  {\textbf{209}}},\ \bibinfo {pages} {116} (\bibinfo {year}
  {2016})}\BibitemShut {NoStop}%
\bibitem [{\citenamefont {Giannozzi}\ \emph {et~al.}(2009)\citenamefont
  {Giannozzi}, \citenamefont {Baroni}, \citenamefont {Bonini}, \citenamefont
  {Calandra}, \citenamefont {Car}, \citenamefont {Cavazzoni}, \citenamefont
  {Ceresoli}, \citenamefont {Chiarotti}, \citenamefont {Cococcioni},
  \citenamefont {Dabo}, \citenamefont {{Dal Corso}}, \citenamefont {{de
  Gironcoli}}, \citenamefont {Fabris}, \citenamefont {Fratesi}, \citenamefont
  {Gebauer}, \citenamefont {Gerstmann}, \citenamefont {Gougoussis},
  \citenamefont {Kokalj}, \citenamefont {Lazzeri}, \citenamefont
  {Martin-Samos}, \citenamefont {Marzari}, \citenamefont {Mauri}, \citenamefont
  {Mazzarello}, \citenamefont {Paolini}, \citenamefont {Pasquarello},
  \citenamefont {Paulatto}, \citenamefont {Sbraccia}, \citenamefont {Scandolo},
  \citenamefont {Sclauzero}, \citenamefont {Seitsonen}, \citenamefont
  {Smogunov}, \citenamefont {Umari},\ and\ \citenamefont
  {Wentzcovitch}}]{P.Giannozzi-J.Phys.CondMat09}%
  \BibitemOpen
  \bibfield  {author} {\bibinfo {author} {\bibfnamefont {P.}~\bibnamefont
  {Giannozzi}}, \bibinfo {author} {\bibfnamefont {S.}~\bibnamefont {Baroni}},
  \bibinfo {author} {\bibfnamefont {N.}~\bibnamefont {Bonini}}, \bibinfo
  {author} {\bibfnamefont {M.}~\bibnamefont {Calandra}}, \bibinfo {author}
  {\bibfnamefont {R.}~\bibnamefont {Car}}, \bibinfo {author} {\bibfnamefont
  {C.}~\bibnamefont {Cavazzoni}}, \bibinfo {author} {\bibfnamefont
  {D.}~\bibnamefont {Ceresoli}}, \bibinfo {author} {\bibfnamefont {G.~L.}\
  \bibnamefont {Chiarotti}}, \bibinfo {author} {\bibfnamefont {M.}~\bibnamefont
  {Cococcioni}}, \bibinfo {author} {\bibfnamefont {I.}~\bibnamefont {Dabo}},
  \bibinfo {author} {\bibfnamefont {A.}~\bibnamefont {{Dal Corso}}}, \bibinfo
  {author} {\bibfnamefont {S.}~\bibnamefont {{de Gironcoli}}}, \bibinfo
  {author} {\bibfnamefont {S.}~\bibnamefont {Fabris}}, \bibinfo {author}
  {\bibfnamefont {G.}~\bibnamefont {Fratesi}}, \bibinfo {author} {\bibfnamefont
  {R.}~\bibnamefont {Gebauer}}, \bibinfo {author} {\bibfnamefont
  {U.}~\bibnamefont {Gerstmann}}, \bibinfo {author} {\bibfnamefont
  {C.}~\bibnamefont {Gougoussis}}, \bibinfo {author} {\bibfnamefont
  {A.}~\bibnamefont {Kokalj}}, \bibinfo {author} {\bibfnamefont
  {M.}~\bibnamefont {Lazzeri}}, \bibinfo {author} {\bibfnamefont
  {L.}~\bibnamefont {Martin-Samos}}, \bibinfo {author} {\bibfnamefont
  {N.}~\bibnamefont {Marzari}}, \bibinfo {author} {\bibfnamefont
  {F.}~\bibnamefont {Mauri}}, \bibinfo {author} {\bibfnamefont
  {R.}~\bibnamefont {Mazzarello}}, \bibinfo {author} {\bibfnamefont
  {S.}~\bibnamefont {Paolini}}, \bibinfo {author} {\bibfnamefont
  {A.}~\bibnamefont {Pasquarello}}, \bibinfo {author} {\bibfnamefont
  {L.}~\bibnamefont {Paulatto}}, \bibinfo {author} {\bibfnamefont
  {C.}~\bibnamefont {Sbraccia}}, \bibinfo {author} {\bibfnamefont
  {S.}~\bibnamefont {Scandolo}}, \bibinfo {author} {\bibfnamefont
  {G.}~\bibnamefont {Sclauzero}}, \bibinfo {author} {\bibfnamefont {A.~P.}\
  \bibnamefont {Seitsonen}}, \bibinfo {author} {\bibfnamefont {A.}~\bibnamefont
  {Smogunov}}, \bibinfo {author} {\bibfnamefont {P.}~\bibnamefont {Umari}}, \
  and\ \bibinfo {author} {\bibfnamefont {R.~M.}\ \bibnamefont {Wentzcovitch}},\
  }\href {\doibase 10.1088/0953-8984/21/39/395502} {\bibfield  {journal}
  {\bibinfo  {journal} {J. Phys.: Condens. Matter}\ }\textbf {\bibinfo {volume}
  {\textbf{21}}},\ \bibinfo {pages} {395502} (\bibinfo {year}
  {2009})}\BibitemShut {NoStop}%
\bibitem [{\citenamefont {Giannozzi}\ \emph {et~al.}(2017)\citenamefont
  {Giannozzi}, \citenamefont {Andreussi}, \citenamefont {Brumme}, \citenamefont
  {Bunau}, \citenamefont {{Buongiorno Nardelli}}, \citenamefont {Calandra},
  \citenamefont {Car}, \citenamefont {Cavazzoni}, \citenamefont {Ceresoli},
  \citenamefont {Cococcioni}, \citenamefont {Colonna}, \citenamefont
  {Carnimeo}, \citenamefont {{Dal Corso}}, \citenamefont {{de Gironcoli}},
  \citenamefont {Delugas}, \citenamefont {DiStasio}, \citenamefont {Ferretti},
  \citenamefont {Floris}, \citenamefont {Fratesi}, \citenamefont {Fugallo},
  \citenamefont {Gebauer}, \citenamefont {Gerstmann}, \citenamefont {Giustino},
  \citenamefont {Gorni}, \citenamefont {Jia}, \citenamefont {Kawamura},
  \citenamefont {Ko}, \citenamefont {Kokalj}, \citenamefont
  {K\"{u}{\c{c}}\"{u}kbenli}, \citenamefont {Lazzeri}, \citenamefont {Marsili},
  \citenamefont {Marzari}, \citenamefont {Mauri}, \citenamefont {Nguyen},
  \citenamefont {Nguyen}, \citenamefont {de-la Roza}, \citenamefont {Paulatto},
  \citenamefont {Ponc{\'{e}}}, \citenamefont {Rocca}, \citenamefont {Sabatini},
  \citenamefont {Santra}, \citenamefont {Schlipf}, \citenamefont {Seitsonen},
  \citenamefont {Smogunov}, \citenamefont {Timrov}, \citenamefont {Thonhauser},
  \citenamefont {Umari}, \citenamefont {Vast}, \citenamefont {Wu},\ and\
  \citenamefont {Baroni}}]{P.Giannozzi-J.Phys.CondMat17}%
  \BibitemOpen
  \bibfield  {author} {\bibinfo {author} {\bibfnamefont {P.}~\bibnamefont
  {Giannozzi}}, \bibinfo {author} {\bibfnamefont {O.}~\bibnamefont
  {Andreussi}}, \bibinfo {author} {\bibfnamefont {T.}~\bibnamefont {Brumme}},
  \bibinfo {author} {\bibfnamefont {O.}~\bibnamefont {Bunau}}, \bibinfo
  {author} {\bibfnamefont {M.}~\bibnamefont {{Buongiorno Nardelli}}}, \bibinfo
  {author} {\bibfnamefont {M.}~\bibnamefont {Calandra}}, \bibinfo {author}
  {\bibfnamefont {R.}~\bibnamefont {Car}}, \bibinfo {author} {\bibfnamefont
  {C.}~\bibnamefont {Cavazzoni}}, \bibinfo {author} {\bibfnamefont
  {D.}~\bibnamefont {Ceresoli}}, \bibinfo {author} {\bibfnamefont
  {M.}~\bibnamefont {Cococcioni}}, \bibinfo {author} {\bibfnamefont
  {N.}~\bibnamefont {Colonna}}, \bibinfo {author} {\bibfnamefont
  {I.}~\bibnamefont {Carnimeo}}, \bibinfo {author} {\bibfnamefont
  {A.}~\bibnamefont {{Dal Corso}}}, \bibinfo {author} {\bibfnamefont
  {S.}~\bibnamefont {{de Gironcoli}}}, \bibinfo {author} {\bibfnamefont
  {P.}~\bibnamefont {Delugas}}, \bibinfo {author} {\bibfnamefont {R.~A.}\
  \bibnamefont {DiStasio}}, \bibinfo {author} {\bibfnamefont {A.}~\bibnamefont
  {Ferretti}}, \bibinfo {author} {\bibfnamefont {A.}~\bibnamefont {Floris}},
  \bibinfo {author} {\bibfnamefont {G.}~\bibnamefont {Fratesi}}, \bibinfo
  {author} {\bibfnamefont {G.}~\bibnamefont {Fugallo}}, \bibinfo {author}
  {\bibfnamefont {R.}~\bibnamefont {Gebauer}}, \bibinfo {author} {\bibfnamefont
  {U.}~\bibnamefont {Gerstmann}}, \bibinfo {author} {\bibfnamefont
  {F.}~\bibnamefont {Giustino}}, \bibinfo {author} {\bibfnamefont
  {T.}~\bibnamefont {Gorni}}, \bibinfo {author} {\bibfnamefont
  {J.}~\bibnamefont {Jia}}, \bibinfo {author} {\bibfnamefont {M.}~\bibnamefont
  {Kawamura}}, \bibinfo {author} {\bibfnamefont {H.-Y.}\ \bibnamefont {Ko}},
  \bibinfo {author} {\bibfnamefont {A.}~\bibnamefont {Kokalj}}, \bibinfo
  {author} {\bibfnamefont {E.}~\bibnamefont {K\"{u}{\c{c}}\"{u}kbenli}},
  \bibinfo {author} {\bibfnamefont {M.}~\bibnamefont {Lazzeri}}, \bibinfo
  {author} {\bibfnamefont {M.}~\bibnamefont {Marsili}}, \bibinfo {author}
  {\bibfnamefont {N.}~\bibnamefont {Marzari}}, \bibinfo {author} {\bibfnamefont
  {F.}~\bibnamefont {Mauri}}, \bibinfo {author} {\bibfnamefont {N.~L.}\
  \bibnamefont {Nguyen}}, \bibinfo {author} {\bibfnamefont {H.-V.}\
  \bibnamefont {Nguyen}}, \bibinfo {author} {\bibfnamefont {A.~O.}\
  \bibnamefont {de-la Roza}}, \bibinfo {author} {\bibfnamefont
  {L.}~\bibnamefont {Paulatto}}, \bibinfo {author} {\bibfnamefont
  {S.}~\bibnamefont {Ponc{\'{e}}}}, \bibinfo {author} {\bibfnamefont
  {D.}~\bibnamefont {Rocca}}, \bibinfo {author} {\bibfnamefont
  {R.}~\bibnamefont {Sabatini}}, \bibinfo {author} {\bibfnamefont
  {B.}~\bibnamefont {Santra}}, \bibinfo {author} {\bibfnamefont
  {M.}~\bibnamefont {Schlipf}}, \bibinfo {author} {\bibfnamefont {A.~P.}\
  \bibnamefont {Seitsonen}}, \bibinfo {author} {\bibfnamefont {A.}~\bibnamefont
  {Smogunov}}, \bibinfo {author} {\bibfnamefont {I.}~\bibnamefont {Timrov}},
  \bibinfo {author} {\bibfnamefont {T.}~\bibnamefont {Thonhauser}}, \bibinfo
  {author} {\bibfnamefont {P.}~\bibnamefont {Umari}}, \bibinfo {author}
  {\bibfnamefont {N.}~\bibnamefont {Vast}}, \bibinfo {author} {\bibfnamefont
  {X.}~\bibnamefont {Wu}}, \ and\ \bibinfo {author} {\bibfnamefont
  {S.}~\bibnamefont {Baroni}},\ }\href {\doibase 10.1088/1361-648X/aa8f79}
  {\bibfield  {journal} {\bibinfo  {journal} {J. Phys.: Condens. Matter}\
  }\textbf {\bibinfo {volume} {\textbf{29}}},\ \bibinfo {pages} {465901}
  (\bibinfo {year} {2017})}\BibitemShut {NoStop}%
\bibitem [{\citenamefont {Giantomassi}(2009)}]{M.Giantomassi09}%
  \BibitemOpen
  \bibfield  {author} {\bibinfo {author} {\bibfnamefont {M.}~\bibnamefont
  {Giantomassi}},\ }\emph {\bibinfo {title} {Core-electrons and
  self-consistency in the \emph{GW} approximation from a PAW perspective}},\
  \href@noop {} {Ph.D. thesis},\ \bibinfo  {school} {Universit\'{e} catholique
  de Louvain} (\bibinfo {year} {2009}),\ \bibinfo {note} {chapter 5 and
  appendix B}\BibitemShut {NoStop}%
\bibitem [{\citenamefont {Hautier}\ \emph {et~al.}(2012)\citenamefont
  {Hautier}, \citenamefont {Ong}, \citenamefont {Jain}, \citenamefont {Moore},\
  and\ \citenamefont {Ceder}}]{G.Hautier2012}%
  \BibitemOpen
  \bibfield  {author} {\bibinfo {author} {\bibfnamefont {G.}~\bibnamefont
  {Hautier}}, \bibinfo {author} {\bibfnamefont {S.~P.}\ \bibnamefont {Ong}},
  \bibinfo {author} {\bibfnamefont {A.}~\bibnamefont {Jain}}, \bibinfo {author}
  {\bibfnamefont {C.~J.}\ \bibnamefont {Moore}}, \ and\ \bibinfo {author}
  {\bibfnamefont {G.}~\bibnamefont {Ceder}},\ }\href {\doibase
  10.1103/physrevb.85.155208} {\bibfield  {journal} {\bibinfo  {journal} {Phys.
  Rev. B}\ }\textbf {\bibinfo {volume} {85}} (\bibinfo {year} {2012}),\
  10.1103/physrevb.85.155208}\BibitemShut {NoStop}%
\bibitem [{sup()}]{supplem}%
  \BibitemOpen
  \href@noop {} {\bibinfo  {journal} {See See Supplemental Material at [URL
  will be inserted by publisher]}\ }\BibitemShut {NoStop}%
\bibitem [{\citenamefont {Furukawa}\ \emph {et~al.}(1986)\citenamefont
  {Furukawa}, \citenamefont {Uemoto}, \citenamefont {Shigeta}, \citenamefont
  {Suzuki},\ and\ \citenamefont {Nakajima}}]{K.Furukawa-APL86}%
  \BibitemOpen
\bibfield  {journal} {  }\bibfield  {author} {\bibinfo {author} {\bibfnamefont
  {K.}~\bibnamefont {Furukawa}}, \bibinfo {author} {\bibfnamefont
  {A.}~\bibnamefont {Uemoto}}, \bibinfo {author} {\bibfnamefont
  {M.}~\bibnamefont {Shigeta}}, \bibinfo {author} {\bibfnamefont
  {A.}~\bibnamefont {Suzuki}}, \ and\ \bibinfo {author} {\bibfnamefont
  {S.}~\bibnamefont {Nakajima}},\ }\href {\doibase 10.1063/1.96860} {\bibfield
  {journal} {\bibinfo  {journal} {Appl. Phys. Lett.}\ }\textbf {\bibinfo
  {volume} {48}},\ \bibinfo {pages} {1536} (\bibinfo {year}
  {1986})}\BibitemShut {NoStop}%
\bibitem [{\citenamefont {Kondo}\ \emph {et~al.}(1986)\citenamefont {Kondo},
  \citenamefont {Takahashi}, \citenamefont {Ishii}, \citenamefont {Hayashi},
  \citenamefont {Sakuma}, \citenamefont {Misawa}, \citenamefont {Daimon},
  \citenamefont {Yamanaka},\ and\ \citenamefont {Yoshida}}]{Y.Kondo-IEEE86}%
  \BibitemOpen
  \bibfield  {author} {\bibinfo {author} {\bibfnamefont {Y.}~\bibnamefont
  {Kondo}}, \bibinfo {author} {\bibfnamefont {T.}~\bibnamefont {Takahashi}},
  \bibinfo {author} {\bibfnamefont {K.}~\bibnamefont {Ishii}}, \bibinfo
  {author} {\bibfnamefont {Y.}~\bibnamefont {Hayashi}}, \bibinfo {author}
  {\bibfnamefont {E.}~\bibnamefont {Sakuma}}, \bibinfo {author} {\bibfnamefont
  {S.~.}\ \bibnamefont {Misawa}}, \bibinfo {author} {\bibfnamefont
  {H.}~\bibnamefont {Daimon}}, \bibinfo {author} {\bibfnamefont
  {M.}~\bibnamefont {Yamanaka}}, \ and\ \bibinfo {author} {\bibfnamefont
  {S.~.}\ \bibnamefont {Yoshida}},\ }\href {\doibase 10.1109/EDL.1986.26417}
  {\bibfield  {journal} {\bibinfo  {journal} {IEEE Electron Device Lett.}\
  }\textbf {\bibinfo {volume} {7}},\ \bibinfo {pages} {404} (\bibinfo {year}
  {1986})}\BibitemShut {NoStop}%
\bibitem [{\citenamefont {Shibahara}\ \emph {et~al.}(1987)\citenamefont
  {Shibahara}, \citenamefont {Kuroda}, \citenamefont {Nishino},\ and\
  \citenamefont {Matsunami}}]{K.Shibahara-JJAP87}%
  \BibitemOpen
  \bibfield  {author} {\bibinfo {author} {\bibfnamefont {K.}~\bibnamefont
  {Shibahara}}, \bibinfo {author} {\bibfnamefont {N.}~\bibnamefont {Kuroda}},
  \bibinfo {author} {\bibfnamefont {S.}~\bibnamefont {Nishino}}, \ and\
  \bibinfo {author} {\bibfnamefont {H.}~\bibnamefont {Matsunami}},\ }\href
  {\doibase 10.1143/JJAP.26.L1815} {\bibfield  {journal} {\bibinfo  {journal}
  {Jpn. J. Appl. Phys.}\ }\textbf {\bibinfo {volume} {26}},\ \bibinfo {pages}
  {1815} (\bibinfo {year} {1987})}\BibitemShut {NoStop}%
\bibitem [{\citenamefont {Weing\"artner}\ \emph {et~al.}(2002)\citenamefont
  {Weing\"artner}, \citenamefont {Wellmann}, \citenamefont {Bickermann},
  \citenamefont {Hofmann}, \citenamefont {Straubinger},\ and\ \citenamefont
  {Winnacker}}]{R.Weingartner-APL02}%
  \BibitemOpen
  \bibfield  {author} {\bibinfo {author} {\bibfnamefont {R.}~\bibnamefont
  {Weing\"artner}}, \bibinfo {author} {\bibfnamefont {P.~J.}\ \bibnamefont
  {Wellmann}}, \bibinfo {author} {\bibfnamefont {M.}~\bibnamefont
  {Bickermann}}, \bibinfo {author} {\bibfnamefont {D.}~\bibnamefont {Hofmann}},
  \bibinfo {author} {\bibfnamefont {T.~L.}\ \bibnamefont {Straubinger}}, \ and\
  \bibinfo {author} {\bibfnamefont {A.}~\bibnamefont {Winnacker}},\ }\href
  {\doibase 10.1063/1.1430262} {\bibfield  {journal} {\bibinfo  {journal}
  {Appl. Phys. Lett.}\ }\textbf {\bibinfo {volume} {80}},\ \bibinfo {pages}
  {70} (\bibinfo {year} {2002})}\BibitemShut {NoStop}%
\bibitem [{\citenamefont {Choyke}\ and\ \citenamefont
  {Pensl}(1997)}]{W.J.Choyke97}%
  \BibitemOpen
  \bibfield  {author} {\bibinfo {author} {\bibfnamefont {W.~J.}\ \bibnamefont
  {Choyke}}\ and\ \bibinfo {author} {\bibfnamefont {G.}~\bibnamefont {Pensl}},\
  }\href {\doibase 10.1557/S0883769400032723} {\bibfield  {journal} {\bibinfo
  {journal} {MRS Bull.}\ }\textbf {\bibinfo {volume} {22}},\ \bibinfo {pages}
  {25} (\bibinfo {year} {1997})}\BibitemShut {NoStop}%
\bibitem [{\citenamefont {Morko\c{c}}\ \emph {et~al.}(1994)\citenamefont
  {Morko\c{c}}, \citenamefont {Strite}, \citenamefont {Gao}, \citenamefont
  {Lin}, \citenamefont {Sverdlov},\ and\ \citenamefont
  {Burns}}]{H.Morkoc-JAP94}%
  \BibitemOpen
  \bibfield  {author} {\bibinfo {author} {\bibfnamefont {H.}~\bibnamefont
  {Morko\c{c}}}, \bibinfo {author} {\bibfnamefont {S.}~\bibnamefont {Strite}},
  \bibinfo {author} {\bibfnamefont {G.~B.}\ \bibnamefont {Gao}}, \bibinfo
  {author} {\bibfnamefont {M.~E.}\ \bibnamefont {Lin}}, \bibinfo {author}
  {\bibfnamefont {B.}~\bibnamefont {Sverdlov}}, \ and\ \bibinfo {author}
  {\bibfnamefont {M.}~\bibnamefont {Burns}},\ }\href {\doibase
  10.1063/1.358463} {\bibfield  {journal} {\bibinfo  {journal} {J. Appl.
  Phys.}\ }\textbf {\bibinfo {volume} {\textbf{76}}},\ \bibinfo {pages} {1363}
  (\bibinfo {year} {1994})}\BibitemShut {NoStop}%
\bibitem [{\citenamefont {Philipp}(1958)}]{H.R.Philipp-PR58}%
  \BibitemOpen
  \bibfield  {author} {\bibinfo {author} {\bibfnamefont {H.~R.}\ \bibnamefont
  {Philipp}},\ }\href {\doibase 10.1103/PhysRev.111.440} {\bibfield  {journal}
  {\bibinfo  {journal} {Phys. Rev.}\ }\textbf {\bibinfo {volume}
  {\textbf{111}}},\ \bibinfo {pages} {440} (\bibinfo {year}
  {1958})}\BibitemShut {NoStop}%
\bibitem [{\citenamefont {Liu}\ \emph {et~al.}(2013)\citenamefont {Liu},
  \citenamefont {Johnston},\ and\ \citenamefont {Snaith}}]{M.Liu13}%
  \BibitemOpen
  \bibfield  {author} {\bibinfo {author} {\bibfnamefont {M.}~\bibnamefont
  {Liu}}, \bibinfo {author} {\bibfnamefont {M.~B.}\ \bibnamefont {Johnston}}, \
  and\ \bibinfo {author} {\bibfnamefont {H.~J.}\ \bibnamefont {Snaith}},\
  }\href {\doibase 10.1038/nature12509} {\bibfield  {journal} {\bibinfo
  {journal} {Nature}\ }\textbf {\bibinfo {volume} {501}},\ \bibinfo {pages}
  {395} (\bibinfo {year} {2013})}\BibitemShut {NoStop}%
\bibitem [{\citenamefont {Green}\ \emph {et~al.}(2014)\citenamefont {Green},
  \citenamefont {Ho-Baillie},\ and\ \citenamefont {Snaith}}]{M.A.Green14}%
  \BibitemOpen
  \bibfield  {author} {\bibinfo {author} {\bibfnamefont {M.~A.}\ \bibnamefont
  {Green}}, \bibinfo {author} {\bibfnamefont {A.}~\bibnamefont {Ho-Baillie}}, \
  and\ \bibinfo {author} {\bibfnamefont {H.~J.}\ \bibnamefont {Snaith}},\
  }\href {\doibase 10.1038/nphoton.2014.134} {\bibfield  {journal} {\bibinfo
  {journal} {Nat. Photonics}\ }\textbf {\bibinfo {volume} {8}},\ \bibinfo
  {pages} {506} (\bibinfo {year} {2014})}\BibitemShut {NoStop}%
\bibitem [{\citenamefont {Ha}\ \emph {et~al.}(2017)\citenamefont {Ha},
  \citenamefont {Ricci}, \citenamefont {Rignanese},\ and\ \citenamefont
  {Hautier}}]{V.-A.Ha-JMCC17}%
  \BibitemOpen
  \bibfield  {author} {\bibinfo {author} {\bibfnamefont {V.-A.}\ \bibnamefont
  {Ha}}, \bibinfo {author} {\bibfnamefont {F.}~\bibnamefont {Ricci}}, \bibinfo
  {author} {\bibfnamefont {G.-M.}\ \bibnamefont {Rignanese}}, \ and\ \bibinfo
  {author} {\bibfnamefont {G.}~\bibnamefont {Hautier}},\ }\href {\doibase
  10.1039/c7tc00528h} {\bibfield  {journal} {\bibinfo  {journal} {J. Mater.
  Chem. C}\ }\textbf {\bibinfo {volume} {\textbf{5}}},\ \bibinfo {pages} {5772}
  (\bibinfo {year} {2017})}\BibitemShut {NoStop}%
\bibitem [{\citenamefont {Saum}\ and\ \citenamefont
  {Hensley}(1959)}]{G.A.Saum-PR59}%
  \BibitemOpen
  \bibfield  {author} {\bibinfo {author} {\bibfnamefont {G.~A.}\ \bibnamefont
  {Saum}}\ and\ \bibinfo {author} {\bibfnamefont {E.~B.}\ \bibnamefont
  {Hensley}},\ }\href {\doibase 10.1103/PhysRev.113.1019} {\bibfield  {journal}
  {\bibinfo  {journal} {Phys. Rev.}\ }\textbf {\bibinfo {volume}
  {\textbf{7}}},\ \bibinfo {pages} {1019} (\bibinfo {year} {1959})}\BibitemShut
  {NoStop}%
\bibitem [{\citenamefont {Richardson}(2003)}]{T.J.Richardson-SolStaIoni03}%
  \BibitemOpen
  \bibfield  {author} {\bibinfo {author} {\bibfnamefont {T.~J.}\ \bibnamefont
  {Richardson}},\ }\href {\doibase 10.1016/j.ssi.2003.08.047} {\bibfield
  {journal} {\bibinfo  {journal} {Solid State Ionics}\ }\textbf {\bibinfo
  {volume} {\textbf{165}}},\ \bibinfo {pages} {305} (\bibinfo {year}
  {2003})}\BibitemShut {NoStop}%
\bibitem [{\citenamefont {Gobrecht}(1966)}]{R.Gobrecht-PSS66}%
  \BibitemOpen
  \bibfield  {author} {\bibinfo {author} {\bibfnamefont {R.}~\bibnamefont
  {Gobrecht}},\ }\href {\doibase 10.1002/pssb.19660130215} {\bibfield
  {journal} {\bibinfo  {journal} {Phys. Status Solidi}\ }\textbf {\bibinfo
  {volume} {\textbf{13}}},\ \bibinfo {pages} {429} (\bibinfo {year}
  {1966})}\BibitemShut {NoStop}%
\bibitem [{\citenamefont {Noffsinger}\ \emph {et~al.}(2012)\citenamefont
  {Noffsinger}, \citenamefont {Kioupakis}, \citenamefont {{Van de Walle}},
  \citenamefont {Louie},\ and\ \citenamefont {Cohen}}]{J.Noffsinger-PRL12}%
  \BibitemOpen
  \bibfield  {author} {\bibinfo {author} {\bibfnamefont {J.}~\bibnamefont
  {Noffsinger}}, \bibinfo {author} {\bibfnamefont {E.}~\bibnamefont
  {Kioupakis}}, \bibinfo {author} {\bibfnamefont {C.~G.}\ \bibnamefont {{Van de
  Walle}}}, \bibinfo {author} {\bibfnamefont {S.~G.}\ \bibnamefont {Louie}}, \
  and\ \bibinfo {author} {\bibfnamefont {M.~L.}\ \bibnamefont {Cohen}},\ }\href
  {\doibase 10.1103/PhysRevLett.108.167402} {\bibfield  {journal} {\bibinfo
  {journal} {Phys. Rev. Lett.}\ }\textbf {\bibinfo {volume} {\textbf{108}}},\
  \bibinfo {pages} {167402} (\bibinfo {year} {2012})}\BibitemShut {NoStop}%
\bibitem [{\citenamefont {Quackenbush}\ \emph {et~al.}(2013)\citenamefont
  {Quackenbush}, \citenamefont {Allen}, \citenamefont {Scanlon}, \citenamefont
  {Sallis}, \citenamefont {Hewlett}, \citenamefont {Nandur}, \citenamefont
  {Chen}, \citenamefont {Smith}, \citenamefont {Weiland}, \citenamefont
  {Fischer}, \citenamefont {Woicik}, \citenamefont {White}, \citenamefont
  {Watson},\ and\ \citenamefont {Piper}}]{N.F.Quackenbush13}%
  \BibitemOpen
  \bibfield  {author} {\bibinfo {author} {\bibfnamefont {N.~F.}\ \bibnamefont
  {Quackenbush}}, \bibinfo {author} {\bibfnamefont {J.~P.}\ \bibnamefont
  {Allen}}, \bibinfo {author} {\bibfnamefont {D.~O.}\ \bibnamefont {Scanlon}},
  \bibinfo {author} {\bibfnamefont {S.}~\bibnamefont {Sallis}}, \bibinfo
  {author} {\bibfnamefont {J.~A.}\ \bibnamefont {Hewlett}}, \bibinfo {author}
  {\bibfnamefont {A.~S.}\ \bibnamefont {Nandur}}, \bibinfo {author}
  {\bibfnamefont {B.}~\bibnamefont {Chen}}, \bibinfo {author} {\bibfnamefont
  {K.~E.}\ \bibnamefont {Smith}}, \bibinfo {author} {\bibfnamefont
  {C.}~\bibnamefont {Weiland}}, \bibinfo {author} {\bibfnamefont {D.~A.}\
  \bibnamefont {Fischer}}, \bibinfo {author} {\bibfnamefont {J.~C.}\
  \bibnamefont {Woicik}}, \bibinfo {author} {\bibfnamefont {B.~E.}\
  \bibnamefont {White}}, \bibinfo {author} {\bibfnamefont {G.~W.}\ \bibnamefont
  {Watson}}, \ and\ \bibinfo {author} {\bibfnamefont {L.~F.~J.}\ \bibnamefont
  {Piper}},\ }\href {\doibase 10.1021/cm401343a} {\bibfield  {journal}
  {\bibinfo  {journal} {Chem. Mater.}\ }\textbf {\bibinfo {volume}
  {\textbf{25}}},\ \bibinfo {pages} {3114} (\bibinfo {year}
  {2013})}\BibitemShut {NoStop}%
\bibitem [{\citenamefont {Hautier}\ \emph {et~al.}(2014)\citenamefont
  {Hautier}, \citenamefont {Miglio}, \citenamefont {Waroquiers}, \citenamefont
  {Rignanese},\ and\ \citenamefont {Gonze}}]{G.Hautier-CheMat14}%
  \BibitemOpen
  \bibfield  {author} {\bibinfo {author} {\bibfnamefont {G.}~\bibnamefont
  {Hautier}}, \bibinfo {author} {\bibfnamefont {A.}~\bibnamefont {Miglio}},
  \bibinfo {author} {\bibfnamefont {D.}~\bibnamefont {Waroquiers}}, \bibinfo
  {author} {\bibfnamefont {G.-M.}\ \bibnamefont {Rignanese}}, \ and\ \bibinfo
  {author} {\bibfnamefont {X.}~\bibnamefont {Gonze}},\ }\href {\doibase
  10.1021/cm404079a} {\bibfield  {journal} {\bibinfo  {journal} {Chem. Mater.}\
  }\textbf {\bibinfo {volume} {\textbf{26}}},\ \bibinfo {pages} {5447}
  (\bibinfo {year} {2014})}\BibitemShut {NoStop}%
\bibitem [{\citenamefont {Kuhar}\ \emph {et~al.}(2018)\citenamefont {Kuhar},
  \citenamefont {Pandey}, \citenamefont {Thygesen},\ and\ \citenamefont
  {Jacobsen}}]{K.Kuhar-ACSEL18}%
  \BibitemOpen
  \bibfield  {author} {\bibinfo {author} {\bibfnamefont {K.}~\bibnamefont
  {Kuhar}}, \bibinfo {author} {\bibfnamefont {M.}~\bibnamefont {Pandey}},
  \bibinfo {author} {\bibfnamefont {K.~S.}\ \bibnamefont {Thygesen}}, \ and\
  \bibinfo {author} {\bibfnamefont {K.~W.}\ \bibnamefont {Jacobsen}},\ }\href
  {\doibase 10.1021/acsenergylett.7b01312} {\bibfield  {journal} {\bibinfo
  {journal} {ACS Energy Lett.}\ }\textbf {\bibinfo {volume} {\textbf{3}}},\
  \bibinfo {pages} {436} (\bibinfo {year} {2018})}\BibitemShut {NoStop}%
\bibitem [{\citenamefont {Tate}\ \emph {et~al.}(2009)\citenamefont {Tate},
  \citenamefont {Ju}, \citenamefont {Moon}, \citenamefont {Zakutayev},
  \citenamefont {Richard}, \citenamefont {Russell},\ and\ \citenamefont
  {McIntyre}}]{J.Tate-PRB09}%
  \BibitemOpen
  \bibfield  {author} {\bibinfo {author} {\bibfnamefont {J.}~\bibnamefont
  {Tate}}, \bibinfo {author} {\bibfnamefont {H.~L.}\ \bibnamefont {Ju}},
  \bibinfo {author} {\bibfnamefont {J.~C.}\ \bibnamefont {Moon}}, \bibinfo
  {author} {\bibfnamefont {A.}~\bibnamefont {Zakutayev}}, \bibinfo {author}
  {\bibfnamefont {A.~P.}\ \bibnamefont {Richard}}, \bibinfo {author}
  {\bibfnamefont {J.}~\bibnamefont {Russell}}, \ and\ \bibinfo {author}
  {\bibfnamefont {D.~H.}\ \bibnamefont {McIntyre}},\ }\href {\doibase
  10.1103/PhysRevB.80.165206} {\bibfield  {journal} {\bibinfo  {journal} {Phys.
  Rev. B}\ }\textbf {\bibinfo {volume} {80}},\ \bibinfo {pages} {165206}
  (\bibinfo {year} {2009})}\BibitemShut {NoStop}%
\bibitem [{\citenamefont {Ogo}\ \emph {et~al.}(2008)\citenamefont {Ogo},
  \citenamefont {Hiramatsu}, \citenamefont {Nomura}, \citenamefont {Yanagi},
  \citenamefont {Kamiya}, \citenamefont {Hirano},\ and\ \citenamefont
  {Hosono}}]{Y.Ogo-APL08}%
  \BibitemOpen
  \bibfield  {author} {\bibinfo {author} {\bibfnamefont {Y.}~\bibnamefont
  {Ogo}}, \bibinfo {author} {\bibfnamefont {H.}~\bibnamefont {Hiramatsu}},
  \bibinfo {author} {\bibfnamefont {K.}~\bibnamefont {Nomura}}, \bibinfo
  {author} {\bibfnamefont {H.}~\bibnamefont {Yanagi}}, \bibinfo {author}
  {\bibfnamefont {T.}~\bibnamefont {Kamiya}}, \bibinfo {author} {\bibfnamefont
  {M.}~\bibnamefont {Hirano}}, \ and\ \bibinfo {author} {\bibfnamefont
  {H.}~\bibnamefont {Hosono}},\ }\href {\doibase 10.1063/1.2964197} {\bibfield
  {journal} {\bibinfo  {journal} {Appl. Phys. Lett.}\ }\textbf {\bibinfo
  {volume} {\textbf{93}}},\ \bibinfo {pages} {032113} (\bibinfo {year}
  {2008})}\BibitemShut {NoStop}%
\bibitem [{\citenamefont {Maci\'{a}-Barber}(2015)}]{E.Macia-Barber15}%
  \BibitemOpen
  \bibfield  {author} {\bibinfo {author} {\bibfnamefont {E.}~\bibnamefont
  {Maci\'{a}-Barber}},\ }\enquote {\bibinfo {title} {Thermoelectric materials:
  Advances and applications},}\ \ (\bibinfo  {publisher} {Pan Stanford},\
  \bibinfo {year} {2015})\BibitemShut {NoStop}%
\bibitem [{\citenamefont {Kajikawa}\ \emph {et~al.}(2003)\citenamefont
  {Kajikawa}, \citenamefont {Kimura},\ and\ \citenamefont
  {Yokoyama}}]{T.Kajikawa2003}%
  \BibitemOpen
  \bibfield  {author} {\bibinfo {author} {\bibfnamefont {T.}~\bibnamefont
  {Kajikawa}}, \bibinfo {author} {\bibfnamefont {N.}~\bibnamefont {Kimura}}, \
  and\ \bibinfo {author} {\bibfnamefont {T.}~\bibnamefont {Yokoyama}},\ }in\
  \href {\doibase 10.1109/ict.2003.1287510} {\emph {\bibinfo {booktitle}
  {Proceedings {ICT'}03. 22nd International Conference on Thermoelectrics
  ({IEEE} Cat. No.03TH8726)}}}\ (\bibinfo {year} {2003})\BibitemShut {NoStop}%
\bibitem [{\citenamefont {Condron}\ \emph {et~al.}(2006)\citenamefont
  {Condron}, \citenamefont {Kauzlarich}, \citenamefont {Gascoin},\ and\
  \citenamefont {Snyder}}]{C.L.Condron2006}%
  \BibitemOpen
  \bibfield  {author} {\bibinfo {author} {\bibfnamefont {C.~L.}\ \bibnamefont
  {Condron}}, \bibinfo {author} {\bibfnamefont {S.~M.}\ \bibnamefont
  {Kauzlarich}}, \bibinfo {author} {\bibfnamefont {F.}~\bibnamefont {Gascoin}},
  \ and\ \bibinfo {author} {\bibfnamefont {G.~J.}\ \bibnamefont {Snyder}},\
  }\href {\doibase 10.1016/j.jssc.2006.01.034} {\bibfield  {journal} {\bibinfo
  {journal} {J. Solid State Chem.}\ }\textbf {\bibinfo {volume} {179}},\
  \bibinfo {pages} {2252} (\bibinfo {year} {2006})}\BibitemShut {NoStop}%
\bibitem [{\citenamefont {Zhang}\ \emph {et~al.}(2017)\citenamefont {Zhang},
  \citenamefont {Song}, \citenamefont {Mamakhel}, \citenamefont
  {J{\o}rgensen},\ and\ \citenamefont {Iversen}}]{J.Zhang2017}%
  \BibitemOpen
  \bibfield  {author} {\bibinfo {author} {\bibfnamefont {J.}~\bibnamefont
  {Zhang}}, \bibinfo {author} {\bibfnamefont {L.}~\bibnamefont {Song}},
  \bibinfo {author} {\bibfnamefont {A.}~\bibnamefont {Mamakhel}}, \bibinfo
  {author} {\bibfnamefont {M.~R.~V.}\ \bibnamefont {J{\o}rgensen}}, \ and\
  \bibinfo {author} {\bibfnamefont {B.~B.}\ \bibnamefont {Iversen}},\ }\href
  {\doibase 10.1021/acs.chemmater.7b01746} {\bibfield  {journal} {\bibinfo
  {journal} {Chem. Mater.}\ }\textbf {\bibinfo {volume} {29}},\ \bibinfo
  {pages} {5371} (\bibinfo {year} {2017})}\BibitemShut {NoStop}%
\bibitem [{\citenamefont {Zunger}(2003)}]{A.Zunger03}%
  \BibitemOpen
  \bibfield  {author} {\bibinfo {author} {\bibfnamefont {A.}~\bibnamefont
  {Zunger}},\ }\href {\doibase 10.1063/1.1584074} {\bibfield  {journal}
  {\bibinfo  {journal} {Appl. Phys. Lett.}\ }\textbf {\bibinfo {volume}
  {\textbf{83}}},\ \bibinfo {pages} {57} (\bibinfo {year} {2003})}\BibitemShut
  {NoStop}%
\end{thebibliography}%


\begin{thebibliography}{37}%
\makeatletter
\providecommand \@ifxundefined [1]{%
 \@ifx{#1\undefined}
}%
\providecommand \@ifnum [1]{%
 \ifnum #1\expandafter \@firstoftwo
 \else \expandafter \@secondoftwo
 \fi
}%
\providecommand \@ifx [1]{%
 \ifx #1\expandafter \@firstoftwo
 \else \expandafter \@secondoftwo
 \fi
}%
\providecommand \natexlab [1]{#1}%
\providecommand \enquote  [1]{``#1''}%
\providecommand \bibnamefont  [1]{#1}%
\providecommand \bibfnamefont [1]{#1}%
\providecommand \citenamefont [1]{#1}%
\providecommand \href@noop [0]{\@secondoftwo}%
\providecommand \href [0]{\begingroup \@sanitize@url \@href}%
\providecommand \@href[1]{\@@startlink{#1}\@@href}%
\providecommand \@@href[1]{\endgroup#1\@@endlink}%
\providecommand \@sanitize@url [0]{\catcode `\\12\catcode `\$12\catcode
  `\&12\catcode `\#12\catcode `\^12\catcode `\_12\catcode `\%12\relax}%
\providecommand \@@startlink[1]{}%
\providecommand \@@endlink[0]{}%
\providecommand \url  [0]{\begingroup\@sanitize@url \@url }%
\providecommand \@url [1]{\endgroup\@href {#1}{\urlprefix }}%
\providecommand \urlprefix  [0]{URL }%
\providecommand \Eprint [0]{\href }%
\providecommand \doibase [0]{http://dx.doi.org/}%
\providecommand \selectlanguage [0]{\@gobble}%
\providecommand \bibinfo  [0]{\@secondoftwo}%
\providecommand \bibfield  [0]{\@secondoftwo}%
\providecommand \translation [1]{[#1]}%
\providecommand \BibitemOpen [0]{}%
\providecommand \bibitemStop [0]{}%
\providecommand \bibitemNoStop [0]{.\EOS\space}%
\providecommand \EOS [0]{\spacefactor3000\relax}%
\providecommand \BibitemShut  [1]{\csname bibitem#1\endcsname}%
\let\auto@bib@innerbib\@empty
\bibitem [{\citenamefont {Ponc\'{e}}\ \emph {et~al.}(2016)\citenamefont
  {Ponc\'{e}}, \citenamefont {Margine}, \citenamefont {Verdi},\ and\
  \citenamefont {Giustino}}]{S.Ponce-Comput.Phys.Commun16}%
  \BibitemOpen
  \bibfield  {author} {\bibinfo {author} {\bibfnamefont {S.}~\bibnamefont
  {Ponc\'{e}}}, \bibinfo {author} {\bibfnamefont {E.~R.}\ \bibnamefont
  {Margine}}, \bibinfo {author} {\bibfnamefont {C.}~\bibnamefont {Verdi}}, \
  and\ \bibinfo {author} {\bibfnamefont {F.}~\bibnamefont {Giustino}},\ }\href
  {\doibase 10.1016/j.cpc.2016.07.028} {\bibfield  {journal} {\bibinfo
  {journal} {Comput. Phys. Commun.}\ }\textbf {\bibinfo {volume}
  {\textbf{209}}},\ \bibinfo {pages} {116} (\bibinfo {year}
  {2016})}\BibitemShut {NoStop}%
\bibitem [{\citenamefont {Giustino}(2017)}]{F.Giustino-RMP17}%
  \BibitemOpen
  \bibfield  {author} {\bibinfo {author} {\bibfnamefont {F.}~\bibnamefont
  {Giustino}},\ }\href {\doibase 10.1103/RevModPhys.89.015003} {\bibfield
  {journal} {\bibinfo  {journal} {Rev. Mod. Phys.}\ }\textbf {\bibinfo {volume}
  {\textbf{89}}},\ \bibinfo {pages} {015003} (\bibinfo {year}
  {2017})}\BibitemShut {NoStop}%
\bibitem [{\citenamefont {Madsen}\ and\ \citenamefont
  {Singh}(2006)}]{G.K.H.Madsen-CPC06}%
  \BibitemOpen
  \bibfield  {author} {\bibinfo {author} {\bibfnamefont {G.~K.~H.}\
  \bibnamefont {Madsen}}\ and\ \bibinfo {author} {\bibfnamefont {D.~J.}\
  \bibnamefont {Singh}},\ }\href {\doibase 10.1016/j.cpc.2006.03.007}
  {\bibfield  {journal} {\bibinfo  {journal} {Comput. Phys. Commun.}\ }\textbf
  {\bibinfo {volume} {\textbf{175}}},\ \bibinfo {pages} {67} (\bibinfo {year}
  {2006})}\BibitemShut {NoStop}%
\bibitem [{\citenamefont {Giantomassi}(2009)}]{M.Giantomassi09}%
  \BibitemOpen
  \bibfield  {author} {\bibinfo {author} {\bibfnamefont {M.}~\bibnamefont
  {Giantomassi}},\ }\emph {\bibinfo {title} {Core-electrons and
  self-consistency in the \emph{GW} approximation from a PAW perspective}},\
  \href@noop {} {Ph.D. thesis},\ \bibinfo  {school} {Universit\'{e} catholique
  de Louvain} (\bibinfo {year} {2009}),\ \bibinfo {note} {chapter 5 and
  appendix B}\BibitemShut {NoStop}%
\bibitem [{\citenamefont {Jain}\ \emph {et~al.}(2013)\citenamefont {Jain},
  \citenamefont {Ong}, \citenamefont {Hautier}, \citenamefont {Chen},
  \citenamefont {Richards}, \citenamefont {Dacek}, \citenamefont {Cholia},
  \citenamefont {Gunter}, \citenamefont {Skinner}, \citenamefont {Ceder},\ and\
  \citenamefont {Persson}}]{A.Jain-APLM13}%
  \BibitemOpen
  \bibfield  {author} {\bibinfo {author} {\bibfnamefont {A.}~\bibnamefont
  {Jain}}, \bibinfo {author} {\bibfnamefont {S.~P.}\ \bibnamefont {Ong}},
  \bibinfo {author} {\bibfnamefont {G.}~\bibnamefont {Hautier}}, \bibinfo
  {author} {\bibfnamefont {W.}~\bibnamefont {Chen}}, \bibinfo {author}
  {\bibfnamefont {W.~D.}\ \bibnamefont {Richards}}, \bibinfo {author}
  {\bibfnamefont {S.}~\bibnamefont {Dacek}}, \bibinfo {author} {\bibfnamefont
  {S.}~\bibnamefont {Cholia}}, \bibinfo {author} {\bibfnamefont
  {D.}~\bibnamefont {Gunter}}, \bibinfo {author} {\bibfnamefont
  {D.}~\bibnamefont {Skinner}}, \bibinfo {author} {\bibfnamefont
  {G.}~\bibnamefont {Ceder}}, \ and\ \bibinfo {author} {\bibfnamefont {K.~A.}\
  \bibnamefont {Persson}},\ }\href {\doibase 10.1063/1.4812323} {\bibfield
  {journal} {\bibinfo  {journal} {APL Materials}\ }\textbf {\bibinfo {volume}
  {1}},\ \bibinfo {pages} {011002} (\bibinfo {year} {2013})}\BibitemShut
  {NoStop}%
\bibitem [{Mat(2013)}]{MatPro13}%
  \BibitemOpen
  \href@noop {} {\enquote {\bibinfo {title} {{The Materials Project}},}\
  }\bibinfo {howpublished} {\url{https://www.materialsproject.org/}} (\bibinfo
  {year} {2013}),\ \bibinfo {note} {[accessed September 1, 2013]}\BibitemShut
  {NoStop}%
\bibitem [{\citenamefont {Ricci}\ \emph {et~al.}(2017)\citenamefont {Ricci},
  \citenamefont {Chen}, \citenamefont {Aydemir}, \citenamefont {rey Snyder},
  \citenamefont {Rignanese}, \citenamefont {Jain},\ and\ \citenamefont
  {Hautier}}]{F.Ricci-SciData17}%
  \BibitemOpen
  \bibfield  {author} {\bibinfo {author} {\bibfnamefont {F.}~\bibnamefont
  {Ricci}}, \bibinfo {author} {\bibfnamefont {W.}~\bibnamefont {Chen}},
  \bibinfo {author} {\bibfnamefont {U.}~\bibnamefont {Aydemir}}, \bibinfo
  {author} {\bibfnamefont {G.~J.}\ \bibnamefont {rey Snyder}}, \bibinfo
  {author} {\bibfnamefont {G.-M.}\ \bibnamefont {Rignanese}}, \bibinfo {author}
  {\bibfnamefont {A.}~\bibnamefont {Jain}}, \ and\ \bibinfo {author}
  {\bibfnamefont {G.}~\bibnamefont {Hautier}},\ }\href {\doibase
  10.1038/sdata.2017.85} {\bibfield  {journal} {\bibinfo  {journal} {Sci.
  Data}\ }\textbf {\bibinfo {volume} {\textbf{4}}},\ \bibinfo {pages} {170085}
  (\bibinfo {year} {2017})}\BibitemShut {NoStop}%
\bibitem [{\citenamefont {Heyd}\ \emph {et~al.}(2003)\citenamefont {Heyd},
  \citenamefont {Scuseria},\ and\ \citenamefont {Ernzerhof}}]{J.Heyd03}%
  \BibitemOpen
  \bibfield  {author} {\bibinfo {author} {\bibfnamefont {J.}~\bibnamefont
  {Heyd}}, \bibinfo {author} {\bibfnamefont {G.~E.}\ \bibnamefont {Scuseria}},
  \ and\ \bibinfo {author} {\bibfnamefont {M.}~\bibnamefont {Ernzerhof}},\
  }\href {\doibase http://dx.doi.org/10.1063/1.1564060} {\bibfield  {journal}
  {\bibinfo  {journal} {J. Chem. Phys.}\ }\textbf {\bibinfo {volume}
  {\textbf{118}}},\ \bibinfo {pages} {8207} (\bibinfo {year}
  {2003})}\BibitemShut {NoStop}%
\bibitem [{\citenamefont {Brothers}\ \emph {et~al.}(2008)\citenamefont
  {Brothers}, \citenamefont {Izmaylov}, \citenamefont {Normand}, \citenamefont
  {Barone},\ and\ \citenamefont {Scuseria}}]{E.N.Brothers08}%
  \BibitemOpen
  \bibfield  {author} {\bibinfo {author} {\bibfnamefont {E.~N.}\ \bibnamefont
  {Brothers}}, \bibinfo {author} {\bibfnamefont {A.~F.}\ \bibnamefont
  {Izmaylov}}, \bibinfo {author} {\bibfnamefont {J.~O.}\ \bibnamefont
  {Normand}}, \bibinfo {author} {\bibfnamefont {V.}~\bibnamefont {Barone}}, \
  and\ \bibinfo {author} {\bibfnamefont {G.~E.}\ \bibnamefont {Scuseria}},\
  }\href {\doibase http://dx.doi.org/10.1063/1.2955460} {\bibfield  {journal}
  {\bibinfo  {journal} {J. Chem. Phys.}\ }\textbf {\bibinfo {volume}
  {\textbf{129}}},\ \bibinfo {pages} {011102} (\bibinfo {year}
  {2008})}\BibitemShut {NoStop}%
\bibitem [{\citenamefont {Fonstad}\ and\ \citenamefont
  {Rediker}(1971)}]{C.G.Fonstad71}%
  \BibitemOpen
  \bibfield  {author} {\bibinfo {author} {\bibfnamefont {C.~G.}\ \bibnamefont
  {Fonstad}}\ and\ \bibinfo {author} {\bibfnamefont {R.~H.}\ \bibnamefont
  {Rediker}},\ }\href {\doibase 10.1063/1.1660648} {\bibfield  {journal}
  {\bibinfo  {journal} {J. Appl. Phys.}\ }\textbf {\bibinfo {volume}
  {\textbf{42}}},\ \bibinfo {pages} {2911} (\bibinfo {year}
  {1971})}\BibitemShut {NoStop}%
\bibitem [{\citenamefont {Nakao}\ \emph {et~al.}(2010)\citenamefont {Nakao},
  \citenamefont {Yamada}, \citenamefont {Hitosugi}, \citenamefont {Hirose},
  \citenamefont {Shimada},\ and\ \citenamefont {Hasegawa}}]{S.Nakao10}%
  \BibitemOpen
  \bibfield  {author} {\bibinfo {author} {\bibfnamefont {S.}~\bibnamefont
  {Nakao}}, \bibinfo {author} {\bibfnamefont {N.}~\bibnamefont {Yamada}},
  \bibinfo {author} {\bibfnamefont {T.}~\bibnamefont {Hitosugi}}, \bibinfo
  {author} {\bibfnamefont {Y.}~\bibnamefont {Hirose}}, \bibinfo {author}
  {\bibfnamefont {T.}~\bibnamefont {Shimada}}, \ and\ \bibinfo {author}
  {\bibfnamefont {T.}~\bibnamefont {Hasegawa}},\ }\href {\doibase
  10.1143/APEX.3.031102} {\bibfield  {journal} {\bibinfo  {journal} {Appl.
  Phys. Express}\ }\textbf {\bibinfo {volume} {\textbf{3}}},\ \bibinfo {pages}
  {031102} (\bibinfo {year} {2010})}\BibitemShut {NoStop}%
\bibitem [{\citenamefont {Dominguez}\ \emph {et~al.}(2002)\citenamefont
  {Dominguez}, \citenamefont {Fu},\ and\ \citenamefont
  {Pan}}]{J.E.Dominguez02}%
  \BibitemOpen
  \bibfield  {author} {\bibinfo {author} {\bibfnamefont {J.~E.}\ \bibnamefont
  {Dominguez}}, \bibinfo {author} {\bibfnamefont {L.}~\bibnamefont {Fu}}, \
  and\ \bibinfo {author} {\bibfnamefont {X.~Q.}\ \bibnamefont {Pan}},\ }\href
  {\doibase 10.1063/1.1530745} {\bibfield  {journal} {\bibinfo  {journal}
  {Appl. Phys. Lett.}\ }\textbf {\bibinfo {volume} {\textbf{81}}},\ \bibinfo
  {pages} {5168} (\bibinfo {year} {2002})}\BibitemShut {NoStop}%
\bibitem [{\citenamefont {Look}\ \emph {et~al.}(1998)\citenamefont {Look},
  \citenamefont {Reynolds}, \citenamefont {Sizelove}, \citenamefont {Jones},
  \citenamefont {Litton}, \citenamefont {Cantwell},\ and\ \citenamefont
  {Harsch}}]{D.C.Look98}%
  \BibitemOpen
  \bibfield  {author} {\bibinfo {author} {\bibfnamefont {D.~C.}\ \bibnamefont
  {Look}}, \bibinfo {author} {\bibfnamefont {D.~C.}\ \bibnamefont {Reynolds}},
  \bibinfo {author} {\bibfnamefont {J.~R.}\ \bibnamefont {Sizelove}}, \bibinfo
  {author} {\bibfnamefont {R.~L.}\ \bibnamefont {Jones}}, \bibinfo {author}
  {\bibfnamefont {C.~W.}\ \bibnamefont {Litton}}, \bibinfo {author}
  {\bibfnamefont {G.}~\bibnamefont {Cantwell}}, \ and\ \bibinfo {author}
  {\bibfnamefont {W.~C.}\ \bibnamefont {Harsch}},\ }\href {\doibase
  10.1016/S0038-1098(97)10145-4} {\bibfield  {journal} {\bibinfo  {journal}
  {Solid State Commun.}\ }\textbf {\bibinfo {volume} {\textbf{105}}},\ \bibinfo
  {pages} {399} (\bibinfo {year} {1998})}\BibitemShut {NoStop}%
\bibitem [{\citenamefont {Ellmer}(2012)}]{K.Ellmer12}%
  \BibitemOpen
  \bibfield  {author} {\bibinfo {author} {\bibfnamefont {K.}~\bibnamefont
  {Ellmer}},\ }\href {\doibase 10.1038/NPHOTON.2012.282} {\bibfield  {journal}
  {\bibinfo  {journal} {Nature Photon.}\ }\textbf {\bibinfo {volume}
  {\textbf{6}}},\ \bibinfo {pages} {809} (\bibinfo {year} {2012})}\BibitemShut
  {NoStop}%
\bibitem [{\citenamefont {Ginley}\ \emph {et~al.}(2010)\citenamefont {Ginley},
  \citenamefont {Hosono},\ and\ \citenamefont {Paine}}]{D.S.Ginley10}%
  \BibitemOpen
  \bibinfo {editor} {\bibfnamefont {D.~S.}\ \bibnamefont {Ginley}}, \bibinfo
  {editor} {\bibfnamefont {H.}~\bibnamefont {Hosono}}, \ and\ \bibinfo {editor}
  {\bibfnamefont {D.~C.}\ \bibnamefont {Paine}},\ eds.,\ \enquote {\bibinfo
  {title} {Handbook of transparent conductors},}\ \ (\bibinfo  {publisher}
  {Springer},\ \bibinfo {year} {2010})\BibitemShut {NoStop}%
\bibitem [{\citenamefont {Kaidashev}\ \emph {et~al.}(2003)\citenamefont
  {Kaidashev}, \citenamefont {Lorenz}, \citenamefont {von Wenckstern},
  \citenamefont {Rahm}, \citenamefont {Semmelhack}, \citenamefont {Han},
  \citenamefont {Benndorf}, \citenamefont {Bundesmann}, \citenamefont
  {Hochmuth},\ and\ \citenamefont {Grundmann}}]{E.M.Kaidashev03}%
  \BibitemOpen
  \bibfield  {author} {\bibinfo {author} {\bibfnamefont {E.~M.}\ \bibnamefont
  {Kaidashev}}, \bibinfo {author} {\bibfnamefont {M.}~\bibnamefont {Lorenz}},
  \bibinfo {author} {\bibfnamefont {H.}~\bibnamefont {von Wenckstern}},
  \bibinfo {author} {\bibfnamefont {A.}~\bibnamefont {Rahm}}, \bibinfo {author}
  {\bibfnamefont {H.-C.}\ \bibnamefont {Semmelhack}}, \bibinfo {author}
  {\bibfnamefont {K.-H.}\ \bibnamefont {Han}}, \bibinfo {author} {\bibfnamefont
  {G.}~\bibnamefont {Benndorf}}, \bibinfo {author} {\bibfnamefont
  {C.}~\bibnamefont {Bundesmann}}, \bibinfo {author} {\bibfnamefont
  {H.}~\bibnamefont {Hochmuth}}, \ and\ \bibinfo {author} {\bibfnamefont
  {M.}~\bibnamefont {Grundmann}},\ }\href {\doibase 10.1063/1.1578694}
  {\bibfield  {journal} {\bibinfo  {journal} {Appl. Phys. Lett.}\ }\textbf
  {\bibinfo {volume} {\textbf{82}}},\ \bibinfo {pages} {3901} (\bibinfo {year}
  {2003})}\BibitemShut {NoStop}%
\bibitem [{\citenamefont {Agashe}\ \emph {et~al.}(2004)\citenamefont {Agashe},
  \citenamefont {Kluth}, \citenamefont {H{\"u}pkes}, \citenamefont {Zastrow},
  \citenamefont {Rech},\ and\ \citenamefont {Wuttig}}]{C.Agashe04}%
  \BibitemOpen
  \bibfield  {author} {\bibinfo {author} {\bibfnamefont {C.}~\bibnamefont
  {Agashe}}, \bibinfo {author} {\bibfnamefont {O.}~\bibnamefont {Kluth}},
  \bibinfo {author} {\bibfnamefont {J.}~\bibnamefont {H{\"u}pkes}}, \bibinfo
  {author} {\bibfnamefont {U.}~\bibnamefont {Zastrow}}, \bibinfo {author}
  {\bibfnamefont {B.}~\bibnamefont {Rech}}, \ and\ \bibinfo {author}
  {\bibfnamefont {M.}~\bibnamefont {Wuttig}},\ }\href {\doibase
  10.1063/1.1641524} {\bibfield  {journal} {\bibinfo  {journal} {J. Appl.
  Phys.}\ }\textbf {\bibinfo {volume} {\textbf{95}}},\ \bibinfo {pages} {1911}
  (\bibinfo {year} {2004})}\BibitemShut {NoStop}%
\bibitem [{\citenamefont {Ellmer}\ \emph {et~al.}(1994)\citenamefont {Ellmer},
  \citenamefont {Kudella}, \citenamefont {Mientus}, \citenamefont {Schieck},\
  and\ \citenamefont {Fiechter}}]{K.Ellmer94}%
  \BibitemOpen
  \bibfield  {author} {\bibinfo {author} {\bibfnamefont {K.}~\bibnamefont
  {Ellmer}}, \bibinfo {author} {\bibfnamefont {F.}~\bibnamefont {Kudella}},
  \bibinfo {author} {\bibfnamefont {R.}~\bibnamefont {Mientus}}, \bibinfo
  {author} {\bibfnamefont {R.}~\bibnamefont {Schieck}}, \ and\ \bibinfo
  {author} {\bibfnamefont {S.}~\bibnamefont {Fiechter}},\ }\href {\doibase
  10.1016/0040-6090(94)90470-7} {\bibfield  {journal} {\bibinfo  {journal}
  {Semicond. Sci. Technol.}\ }\textbf {\bibinfo {volume} {\textbf{247}}},\
  \bibinfo {pages} {15} (\bibinfo {year} {1994})}\BibitemShut {NoStop}%
\bibitem [{\citenamefont {Ellmer}(2000)}]{K.Ellmer00}%
  \BibitemOpen
  \bibfield  {author} {\bibinfo {author} {\bibfnamefont {K.}~\bibnamefont
  {Ellmer}},\ }\href {\doibase 10.1088/0022-3727/33/4/201} {\bibfield
  {journal} {\bibinfo  {journal} {J. Phys. D: Appl. Phys.}\ }\textbf {\bibinfo
  {volume} {\textbf{33}}},\ \bibinfo {pages} {17} (\bibinfo {year}
  {2000})}\BibitemShut {NoStop}%
\bibitem [{\citenamefont {Weiher}(1962)}]{R.L.Weiher62}%
  \BibitemOpen
  \bibfield  {author} {\bibinfo {author} {\bibfnamefont {R.~L.}\ \bibnamefont
  {Weiher}},\ }\href {\doibase 10.1063/1.1702560} {\bibfield  {journal}
  {\bibinfo  {journal} {J. Appl. Phys.}\ }\textbf {\bibinfo {volume}
  {\textbf{33}}},\ \bibinfo {pages} {2834} (\bibinfo {year}
  {1962})}\BibitemShut {NoStop}%
\bibitem [{\citenamefont {Groth}(1966)}]{R.Groth66}%
  \BibitemOpen
  \bibfield  {author} {\bibinfo {author} {\bibfnamefont {R.}~\bibnamefont
  {Groth}},\ }\href {\doibase 10.1002/pssb.19660140104} {\bibfield  {journal}
  {\bibinfo  {journal} {Phys. stat. sol.}\ }\textbf {\bibinfo {volume}
  {\textbf{14}}},\ \bibinfo {pages} {69} (\bibinfo {year} {1966})}\BibitemShut
  {NoStop}%
\bibitem [{\citenamefont {M{\"u}ller}(1968)}]{H.K.Muller68}%
  \BibitemOpen
  \bibfield  {author} {\bibinfo {author} {\bibfnamefont {H.~K.}\ \bibnamefont
  {M{\"u}ller}},\ }\href {\doibase 10.1002/pssb.19680270229} {\bibfield
  {journal} {\bibinfo  {journal} {Phys. Status Solidi}\ }\textbf {\bibinfo
  {volume} {\textbf{27}}},\ \bibinfo {pages} {723} (\bibinfo {year}
  {1968})}\BibitemShut {NoStop}%
\bibitem [{\citenamefont {Noguchi}\ and\ \citenamefont
  {Sakata}(1980)}]{S.Noguchi80}%
  \BibitemOpen
  \bibfield  {author} {\bibinfo {author} {\bibfnamefont {S.}~\bibnamefont
  {Noguchi}}\ and\ \bibinfo {author} {\bibfnamefont {H.}~\bibnamefont
  {Sakata}},\ }\href {\doibase 10.1088/0022-3727/13/6/023} {\bibfield
  {journal} {\bibinfo  {journal} {J. Phys. D : Appl. Phys.}\ }\textbf {\bibinfo
  {volume} {\textbf{13}}},\ \bibinfo {pages} {1129} (\bibinfo {year}
  {1980})}\BibitemShut {NoStop}%
\bibitem [{\citenamefont {Pan}\ and\ \citenamefont {Ma}(1981)}]{C.A.Pan81}%
  \BibitemOpen
  \bibfield  {author} {\bibinfo {author} {\bibfnamefont {C.~A.}\ \bibnamefont
  {Pan}}\ and\ \bibinfo {author} {\bibfnamefont {T.~P.}\ \bibnamefont {Ma}},\
  }\href {\doibase 10.1007/BF02654901} {\bibfield  {journal} {\bibinfo
  {journal} {J. Electron. Mater.}\ }\textbf {\bibinfo {volume} {\textbf{10}}},\
  \bibinfo {pages} {43} (\bibinfo {year} {1981})}\BibitemShut {NoStop}%
\bibitem [{\citenamefont {Hamberg}\ and\ \citenamefont
  {Granqvist}(1986)}]{I.Hamberg86}%
  \BibitemOpen
  \bibfield  {author} {\bibinfo {author} {\bibfnamefont {I.}~\bibnamefont
  {Hamberg}}\ and\ \bibinfo {author} {\bibfnamefont {C.~G.}\ \bibnamefont
  {Granqvist}},\ }\href {\doibase 10.1063/1.337534} {\bibfield  {journal}
  {\bibinfo  {journal} {J. Appl. Phys.}\ }\textbf {\bibinfo {volume}
  {\textbf{60}}},\ \bibinfo {pages} {123} (\bibinfo {year} {1986})}\BibitemShut
  {NoStop}%
\bibitem [{\citenamefont {Wen}\ \emph {et~al.}(1992)\citenamefont {Wen},
  \citenamefont {Couturier}, \citenamefont {Chaminade}, \citenamefont
  {Marquestaut}, \citenamefont {Claverie},\ and\ \citenamefont
  {Hagenmuller}}]{S.J.Wen92}%
  \BibitemOpen
  \bibfield  {author} {\bibinfo {author} {\bibfnamefont {S.~J.}\ \bibnamefont
  {Wen}}, \bibinfo {author} {\bibfnamefont {G.}~\bibnamefont {Couturier}},
  \bibinfo {author} {\bibfnamefont {J.~P.}\ \bibnamefont {Chaminade}}, \bibinfo
  {author} {\bibfnamefont {E.}~\bibnamefont {Marquestaut}}, \bibinfo {author}
  {\bibfnamefont {J.}~\bibnamefont {Claverie}}, \ and\ \bibinfo {author}
  {\bibfnamefont {P.}~\bibnamefont {Hagenmuller}},\ }\href {\doibase
  10.1016/0022-4596(92)90176-V} {\bibfield  {journal} {\bibinfo  {journal} {J.
  Solid State Chem.}\ }\textbf {\bibinfo {volume} {\textbf{101}}},\ \bibinfo
  {pages} {203} (\bibinfo {year} {1992})}\BibitemShut {NoStop}%
\bibitem [{\citenamefont {Sawada}\ and\ \citenamefont
  {Higuchi}(1998)}]{M.Sawada98}%
  \BibitemOpen
  \bibfield  {author} {\bibinfo {author} {\bibfnamefont {M.}~\bibnamefont
  {Sawada}}\ and\ \bibinfo {author} {\bibfnamefont {M.}~\bibnamefont
  {Higuchi}},\ }\href {\doibase 10.1016/S0040-6090(97)00513-0} {\bibfield
  {journal} {\bibinfo  {journal} {Thin Solid Films}\ }\textbf {\bibinfo
  {volume} {\textbf{317}}},\ \bibinfo {pages} {157} (\bibinfo {year}
  {1998})}\BibitemShut {NoStop}%
\bibitem [{\citenamefont {Meng}\ \emph {et~al.}(2001)\citenamefont {Meng},
  \citenamefont {Yang}, \citenamefont {Chen}, \citenamefont {Shen},
  \citenamefont {Jiang}, \citenamefont {Zhang},\ and\ \citenamefont
  {Hua}}]{Y.Meng01}%
  \BibitemOpen
  \bibfield  {author} {\bibinfo {author} {\bibfnamefont {Y.}~\bibnamefont
  {Meng}}, \bibinfo {author} {\bibfnamefont {X.-L.}\ \bibnamefont {Yang}},
  \bibinfo {author} {\bibfnamefont {H.-X.}\ \bibnamefont {Chen}}, \bibinfo
  {author} {\bibfnamefont {J.}~\bibnamefont {Shen}}, \bibinfo {author}
  {\bibfnamefont {Y.-M.}\ \bibnamefont {Jiang}}, \bibinfo {author}
  {\bibfnamefont {Z.-J.}\ \bibnamefont {Zhang}}, \ and\ \bibinfo {author}
  {\bibfnamefont {Z.-Y.}\ \bibnamefont {Hua}},\ }\href {\doibase
  10.1016/S0040-6090(01)01142-7} {\bibfield  {journal} {\bibinfo  {journal}
  {Thin Solid Films}\ }\textbf {\bibinfo {volume} {\textbf{394}}},\ \bibinfo
  {pages} {219} (\bibinfo {year} {2001})}\BibitemShut {NoStop}%
\bibitem [{\citenamefont {Warmsingh}\ \emph {et~al.}(2004)\citenamefont
  {Warmsingh}, \citenamefont {Yoshida}, \citenamefont {Readey}, \citenamefont
  {Teplin}, \citenamefont {Perkins}, \citenamefont {Parilla}, \citenamefont
  {Gedvilas}, \citenamefont {Keyes},\ and\ \citenamefont
  {Ginley}}]{C.Warmsingh04}%
  \BibitemOpen
  \bibfield  {author} {\bibinfo {author} {\bibfnamefont {C.}~\bibnamefont
  {Warmsingh}}, \bibinfo {author} {\bibfnamefont {Y.}~\bibnamefont {Yoshida}},
  \bibinfo {author} {\bibfnamefont {D.~W.}\ \bibnamefont {Readey}}, \bibinfo
  {author} {\bibfnamefont {C.~W.}\ \bibnamefont {Teplin}}, \bibinfo {author}
  {\bibfnamefont {J.~D.}\ \bibnamefont {Perkins}}, \bibinfo {author}
  {\bibfnamefont {P.~A.}\ \bibnamefont {Parilla}}, \bibinfo {author}
  {\bibfnamefont {L.~M.}\ \bibnamefont {Gedvilas}}, \bibinfo {author}
  {\bibfnamefont {B.~M.}\ \bibnamefont {Keyes}}, \ and\ \bibinfo {author}
  {\bibfnamefont {D.~S.}\ \bibnamefont {Ginley}},\ }\href {\doibase
  10.1063/1.1646468} {\bibfield  {journal} {\bibinfo  {journal} {J. Appl.
  Phys.}\ }\textbf {\bibinfo {volume} {\textbf{95}}},\ \bibinfo {pages} {3831}
  (\bibinfo {year} {2004})}\BibitemShut {NoStop}%
\bibitem [{\citenamefont {Koida}\ \emph {et~al.}(2007)\citenamefont {Koida},
  \citenamefont {Fujiwara},\ and\ \citenamefont {Kondo}}]{T.Koida07}%
  \BibitemOpen
  \bibfield  {author} {\bibinfo {author} {\bibfnamefont {T.}~\bibnamefont
  {Koida}}, \bibinfo {author} {\bibfnamefont {H.}~\bibnamefont {Fujiwara}}, \
  and\ \bibinfo {author} {\bibfnamefont {M.}~\bibnamefont {Kondo}},\ }\href
  {\doibase 10.1143/JJAP.46.L685} {\bibfield  {journal} {\bibinfo  {journal}
  {Jpn. J. Appl. Phys.}\ }\textbf {\bibinfo {volume} {\textbf{46}}},\ \bibinfo
  {pages} {685} (\bibinfo {year} {2007})}\BibitemShut {NoStop}%
\bibitem [{\citenamefont {Koida}\ \emph {et~al.}(2009)\citenamefont {Koida},
  \citenamefont {Fujiwara},\ and\ \citenamefont {Kondo}}]{T.Koida09}%
  \BibitemOpen
  \bibfield  {author} {\bibinfo {author} {\bibfnamefont {T.}~\bibnamefont
  {Koida}}, \bibinfo {author} {\bibfnamefont {H.}~\bibnamefont {Fujiwara}}, \
  and\ \bibinfo {author} {\bibfnamefont {M.}~\bibnamefont {Kondo}},\ }\href
  {\doibase 10.1016/j.solmat.2008.09.047} {\bibfield  {journal} {\bibinfo
  {journal} {Sol. Energy Mater Sol. Cells}\ }\textbf {\bibinfo {volume}
  {\textbf{93}}},\ \bibinfo {pages} {851} (\bibinfo {year} {2009})}\BibitemShut
  {NoStop}%
\bibitem [{\citenamefont {Oka}\ \emph {et~al.}(2012)\citenamefont {Oka},
  \citenamefont {Kawase},\ and\ \citenamefont {Shigesato}}]{N.Oka12}%
  \BibitemOpen
  \bibfield  {author} {\bibinfo {author} {\bibfnamefont {N.}~\bibnamefont
  {Oka}}, \bibinfo {author} {\bibfnamefont {Y.}~\bibnamefont {Kawase}}, \ and\
  \bibinfo {author} {\bibfnamefont {Y.}~\bibnamefont {Shigesato}},\ }\href
  {\doibase 10.1016/j.tsf.2011.06.063} {\bibfield  {journal} {\bibinfo
  {journal} {Thin Solid Films}\ }\textbf {\bibinfo {volume} {\textbf{520}}},\
  \bibinfo {pages} {4101} (\bibinfo {year} {2012})}\BibitemShut {NoStop}%
\bibitem [{\citenamefont {Lorenz}\ and\ \citenamefont
  {Gambino}(1967)}]{M.R.Lorenz67}%
  \BibitemOpen
  \bibfield  {author} {\bibinfo {author} {\bibfnamefont {M.~R.}\ \bibnamefont
  {Lorenz}}\ and\ \bibinfo {author} {\bibfnamefont {J.~F. W. R.~J.}\
  \bibnamefont {Gambino}},\ }\href {\doibase 10.1016/0022-3697(67)90305-8}
  {\bibfield  {journal} {\bibinfo  {journal} {J. Phys. Chem. Solids}\ }\textbf
  {\bibinfo {volume} {\textbf{28}}},\ \bibinfo {pages} {403} (\bibinfo {year}
  {1967})}\BibitemShut {NoStop}%
\bibitem [{\citenamefont {Ueda}\ \emph {et~al.}(1997)\citenamefont {Ueda},
  \citenamefont {Hosono}, \citenamefont {Waseda},\ and\ \citenamefont
  {Kawazoe}}]{N.Ueda97}%
  \BibitemOpen
  \bibfield  {author} {\bibinfo {author} {\bibfnamefont {N.}~\bibnamefont
  {Ueda}}, \bibinfo {author} {\bibfnamefont {H.}~\bibnamefont {Hosono}},
  \bibinfo {author} {\bibfnamefont {R.}~\bibnamefont {Waseda}}, \ and\ \bibinfo
  {author} {\bibfnamefont {H.}~\bibnamefont {Kawazoe}},\ }\href {\doibase
  10.1063/1.119693} {\bibfield  {journal} {\bibinfo  {journal} {Appl. Phys.
  Lett.}\ }\textbf {\bibinfo {volume} {\textbf{71}}},\ \bibinfo {pages} {933}
  (\bibinfo {year} {1997})}\BibitemShut {NoStop}%
\bibitem [{\citenamefont {Oishi}\ \emph {et~al.}(2015)\citenamefont {Oishi},
  \citenamefont {Koga}, \citenamefont {Harada},\ and\ \citenamefont
  {Kasu}}]{T.Oishi15}%
  \BibitemOpen
  \bibfield  {author} {\bibinfo {author} {\bibfnamefont {T.}~\bibnamefont
  {Oishi}}, \bibinfo {author} {\bibfnamefont {Y.}~\bibnamefont {Koga}},
  \bibinfo {author} {\bibfnamefont {K.}~\bibnamefont {Harada}}, \ and\ \bibinfo
  {author} {\bibfnamefont {M.}~\bibnamefont {Kasu}},\ }\href {\doibase
  10.7567/APEX.8.031101} {\bibfield  {journal} {\bibinfo  {journal} {Appl.
  Phys. Express}\ }\textbf {\bibinfo {volume} {\textbf{8}}},\ \bibinfo {pages}
  {031101} (\bibinfo {year} {2015})}\BibitemShut {NoStop}%
\bibitem [{\citenamefont {Ma}\ \emph {et~al.}(2016)\citenamefont {Ma},
  \citenamefont {Tanen}, \citenamefont {Verma}, \citenamefont {Guo},
  \citenamefont {Luo}, \citenamefont {Xing},\ and\ \citenamefont
  {Jena}}]{N.Ma16}%
  \BibitemOpen
  \bibfield  {author} {\bibinfo {author} {\bibfnamefont {N.}~\bibnamefont
  {Ma}}, \bibinfo {author} {\bibfnamefont {N.}~\bibnamefont {Tanen}}, \bibinfo
  {author} {\bibfnamefont {A.}~\bibnamefont {Verma}}, \bibinfo {author}
  {\bibfnamefont {Z.}~\bibnamefont {Guo}}, \bibinfo {author} {\bibfnamefont
  {T.}~\bibnamefont {Luo}}, \bibinfo {author} {\bibfnamefont {H.}~\bibnamefont
  {Xing}}, \ and\ \bibinfo {author} {\bibfnamefont {D.}~\bibnamefont {Jena}},\
  }\href {\doibase 10.1063/1.4968550} {\bibfield  {journal} {\bibinfo
  {journal} {Appl. Phys. Lett.}\ }\textbf {\bibinfo {volume} {\textbf{109}}},\
  \bibinfo {pages} {212101} (\bibinfo {year} {2016})}\BibitemShut {NoStop}%
\bibitem [{\citenamefont {Peng}\ \emph {et~al.}(2013)\citenamefont {Peng},
  \citenamefont {Scanlon}, \citenamefont {Stevanovic}, \citenamefont {Vidal},
  \citenamefont {Watson},\ and\ \citenamefont {Lany}}]{Peng2013}%
  \BibitemOpen
  \bibfield  {author} {\bibinfo {author} {\bibfnamefont {H.}~\bibnamefont
  {Peng}}, \bibinfo {author} {\bibfnamefont {D.~O.}\ \bibnamefont {Scanlon}},
  \bibinfo {author} {\bibfnamefont {V.}~\bibnamefont {Stevanovic}}, \bibinfo
  {author} {\bibfnamefont {J.}~\bibnamefont {Vidal}}, \bibinfo {author}
  {\bibfnamefont {G.~W.}\ \bibnamefont {Watson}}, \ and\ \bibinfo {author}
  {\bibfnamefont {S.}~\bibnamefont {Lany}},\ }\href {\doibase
  10.1103/PhysRevB.88.115201} {\bibfield  {journal} {\bibinfo  {journal} {Phys.
  Rev. B}\ }\textbf {\bibinfo {volume} {\textbf{88}}},\ \bibinfo {pages}
  {115201} (\bibinfo {year} {2013})}\BibitemShut {NoStop}%
\end{thebibliography}%
\end{document}